\documentclass[journal=jacsat,manuscript=article]{achemso}
\usepackage{achemso}
\usepackage[version=3]{mhchem}
\usepackage[english]{babel}
\usepackage{filecontents}
\usepackage{graphicx}
\graphicspath{{./Figures/}}
\usepackage{courier}
\usepackage{hyperref,url}
\usepackage{amssymb,amsmath,amsthm,mathrsfs,comment,afterpage,accents}
\usepackage[capitalise]{cleveref}
\hypersetup{breaklinks,colorlinks=true,linkcolor=blue,citecolor=blue,urlcolor=blue,filecolor=blue}
\usepackage{mathtools}
\usepackage{braket}
\usepackage{mathrsfs}
\usepackage{courier}
\usepackage{multirow}
\usepackage{tabu}
\usepackage{color}
\usepackage{caption}
\usepackage{subcaption}
\usepackage{physics}

\usepackage[table,xcdraw]{xcolor}
\usepackage[numbers]{natbib}
\usepackage{xr} %External referencing
\usepackage{soul}
\usepackage[normalem]{ulem} %For StrikeThrough text

\newcommand{\remove}[1]{}  %Does not show changes

\makeatletter
\def\@dotsep{4.5}
\makeatother
\usepackage{dcolumn}% Align table columns on decimal point
\usepackage{bm}% bold maths
\renewcommand\vec\mathbf
\setkeys{acs}{maxauthors = 0}

\title{A Non-Perturbative Pairwise-Additive Analysis of Charge Transfer Contributions to Intermolecular Interaction Energies}

\author{Srimukh Prasad Veccham}
\author{Joonho Lee}
\author{Yuezhi Mao}
\author{Paul R. Horn}
\author{Martin Head-Gordon}
\email{mhg@cchem.berkeley.edu}
\affiliation{
Department of Chemistry, University of California, Berkeley, California 94720, USA
Chemical Sciences Division, Lawrence Berkeley National Laboratory, Berkeley, California 94720, USA
}

\begin{tocentry}
\includegraphics[width=8.5cm]{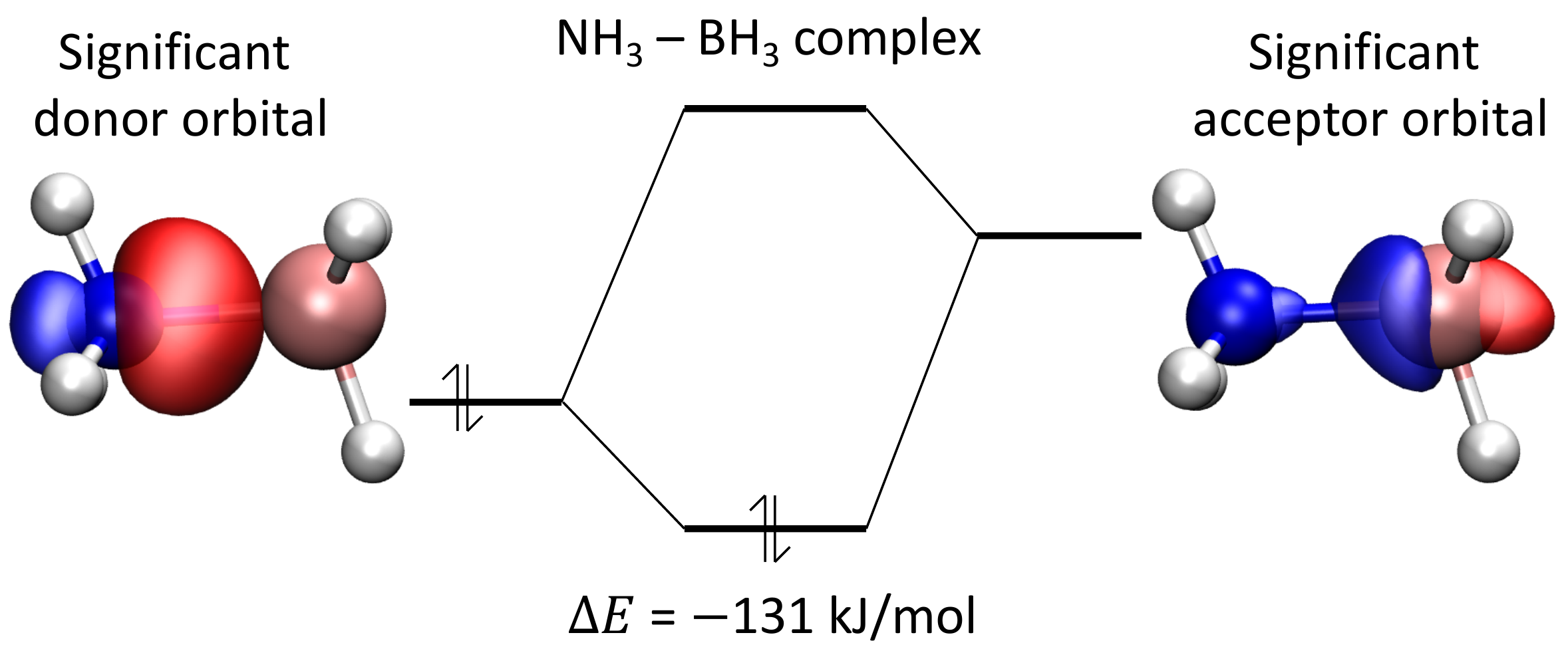}
\end{tocentry}

\begin{document}
\maketitle
\begin{abstract}
Energy decomposition analysis (EDA) based on absolutely localized molecular orbitals (ALMOs) decomposes the interaction energy between molecules into physically interpretable components like geometry distortion, frozen interactions, polarization, and charge transfer (CT, also sometimes called charge delocalization) interactions.
In this work, a numerically exact scheme to decompose the CT interaction energy into pairwise additive terms is introduced for the ALMO-EDA using density functional theory.
Unlike perturbative pairwise charge-decomposition analysis, the new approach does not break down for strongly interacting systems, or show significant exchange-correlation functional dependence in the decomposed energy components. 
%this new pairwise charge decomposition scheme decomposes 100\% of the charge transfer interaction energy into pairwise additive components irrespective of binding strength.
Both the energy lowering and the charge flow associated with CT can be decomposed. Complementary occupied-virtual orbital pairs (COVPs) that capture the dominant donor and acceptor CT orbitals are obtained for the new decomposition.
It is applied to systems with different types of interactions including DNA base-pairs, borane-ammonia adducts, and transition metal hexacarbonyls.
While consistent with most existing understanding of the nature of CT in these systems, the results also reveal some new insights into the origin of trends in donor-acceptor interactions.
%In summary, the non-perturbative charge decomposition scheme introduced in this work completes the ALMO-based EDA scheme for mean-field theories by decomposing the entire interaction energy, including the charge-transfer component, into physico-chemically meaningful parts.
\end{abstract}

\section{Introduction}
Electronic structure calculations, such as modern density functional theory (DFT),\cite{Mardirossian2017} are capable of yielding accurate results for intermolecular interactions. However no physical or chemical insight beyond the calculated observable is obtained. Many different energy decomposition analyses (EDAs) \cite{Kitaura1976new, Mitoraj2009combined, Chen1996energy, Su2009energy, Reed1988intermolecular, Jeziorski1994perturbation, Mo2000, Khaliullin2007} have attempted to address this need.
At their core, EDA attempts to partition computed interaction energies into physically and chemically motivated terms like electrostatics, dispersion, polarization, and charge transfer.
EDA thereby aims to provide insights into the fundamental nature of molecular interactions which can, in principle, help guide the design of molecules with properties of interest.
The development of force-fields based on these partitions of interaction energies is another utility of EDA schemes.\cite{Mcdaniel2016next,Das2019development}

Interaction of molecules by electron delocalization or charge transfer (CT) is one of the most fundamental drivers of complex formation.\cite{Mao2018}
CT from a filled donor orbital of one molecule to an empty acceptor orbital of another lowers the total energy of the complex, and favors binding of the molecules.
CT via forward and back donation in the Dewar-Chatt-Duncanson model\cite{Dewar1951review,Chatt1953586} plays a very important role in transition metal-ligand complex formation, and is instrumental in many transition metal catalytic transformations.
Even hydrogen bonding, which is a much weaker interaction that is ubiquitous for (bio)molecules in aqueous solution, has a significant contribution from CT.\cite{Mao2018,Van2019nature}
Another interesting aspect of charge transfer stabilization is its diversity in the strength of interaction: it can vary in strength from a few kJ/mol in hydrogen-bonding complexes to hundreds of kJ/mol in transition metal-ligand interactions.

The energy decomposition analysis scheme based on Absolutely Localized Molecular Orbitals (ALMOs) has been developed for mean-field theories (Hartree Fock and Density Functional Theory) \cite{Mo2000, Khaliullin2007, Horn2013, Horn2016c, Mao2020probing} and M\o{}ller-Plesset perturbation theory.\cite{Thirman2015energy, Thirman2017, Loipersberger2019}
The mean-field ALMO-EDA scheme partitions the interaction energy into geometric distortion, frozen, polarization, and CT components using variationally optimized intermediate wavefunctions as illustrated in Eq.~\eqref{eq::ALMO_EDA}:
\begin{align}
    \Delta E_{\text{INT}} &= \Delta E_{\text{GD}} + \Delta E_{\text{FRZ}} + \Delta E_{\text{POL}} + \Delta E_{\text{CT}} \label{eq::ALMO_EDA}
\end{align}
The energy associated with distorting the relaxed isolated fragment geometries to their complex geometry is represented by the geometric distortion term ($\Delta E_{\text{GD}}$).
The frozen term ($\Delta E_{\text{FRZ}}$) is the energy associated with bringing infinitely separated fragments to their complex geometry while retaining their infinitely-separated wavefunctions.
This term includes permanent electrostatics, Pauli repulsion, and dispersion.\cite{Horn2016}
The polarization term ($\Delta E_{\text{POL}}$) represents the energy lowering associated with relaxing the wavefunctions of each of the fragments in the presence of other fragments while not allowing CT between fragments.\cite{Khaliullin2007, Horn2015}
The energy lowering associated with CT between fragments is captured by $\Delta E_{\text{CT}}$.
CT can be further broken down into pairwise additive components for both energy lowering and its associated charge flow based on perturbation theory.\cite{Khaliullin2008}
This decomposition scheme can also be used to extract chemically relevant Complementary Occupied-Virtual orbital Pairs (COVPs) which are the most important orbitals associated with CT. \cite{Khaliullin2008}
However, this charge decomposition scheme breaks down in cases of strongly interacting species as it is based on perturbation theory.\cite{Khaliullin2007, Khaliullin2008}
In this paper, we provide an alternative non-perturbative CT decomposition scheme which works robustly for all regimes of interaction energy, while nevertheless providing an effective pairwise description of CT.

There are many other proposed measures of CT, which we can only briefly summarize here. The charge decomposition analysis method (CDA) by Dapprich and Frenking defined forward and backward donation components in intermolecular complexes.\cite{Dapprich1995investigation}
While this method is conceptually simple, some of the terms in CDA do not have a well-defined physical meaning.\cite{Khaliullin2008}
Other EDA variants define charge transfer energy as the amount of energy associated with mixing occupied orbitals of the donor with virtual orbitals of the acceptor.
Such ideas are used in the reduced variational space (RVS)\cite{Stevens1987frozen} and constrained space orbital variation (CSOV)\cite{Bagus1984new} schemes.
However, these methods only account for CT in one direction at a time and cannot provide a pairwise additive decomposition of CT such that the contributions sum up to the total CT energy. 
Methods such as EDA-NOCV (natural orbital for chemical valence) \cite{Mitoraj2009combined, Zhao2018} and localized molecular orbital (LMO)-EDA \cite{Su2009energy} report only the sum of polarization and CT, which corresponds to the ``orbital interaction'' and ``polarization'' terms in these methods, respectively. The same choice is commonly followed in symmetry-adapted perturbation theory (SAPT) calculations of intermolecular interactions,\cite{Jeziorski1994perturbation, Szalewicz2012} where polarization and CT are combined in the ``induction'' term, although several approaches have been proposed to quantify CT in SAPT. \cite{Stone2009, Misquitta2013}
%Ziegler-Rauk \cite{Mitoraj2009combined} and Bickelhaupt-Baerends\cite{Bickelhaupt2000kohn} 
%EDAs include an ``orbital interaction'' term which corresponds to the sum of polarization and charge transfer in the ALMO-EDA.  
The popular Natural Bond Orbital (NBO) method can also be used to define charge decomposition between pairs of fragments using orthogonal molecular orbitals.\cite{Reed1983natural, Reed1988intermolecular} 
However NBO and its associated Natural Energy Decomposition Analysis\cite{Glendening1994natural,Glendening2005natural} (NEDA) is known to greatly exaggerate the magnitude of CT energies.\cite{Khaliullin2009,Stone2017natural}
Constrained DFT (CDFT)\cite{Wu2005direct,Wu2006constrained, Kaduk2012constrained} has also been used to estimate the magnitude of CT energy.\cite{Wu2009,Lao2016} As it constrains charge populations, rather than preventing charge-delocalization, CDFT was shown to substantially underestimate CT in most scenarios.\cite{Mao2018}

In the ALMO-EDA, total CT is based on the energy difference between two variationally optimized wavefunctions: the fully relaxed final (FULL) wavefunction, and the constrained polarized (POL) wavefunction (which is evaluated by the self-consistent field for molecular interactions (SCF-MI) method\cite{Stoll1980, Gianinetti1996, Khaliullin2006, Horn2013}):
\begin{align}
    \Delta E_{\text{CT}} \equiv E(\Phi_{\text{FULL}}) - E(\Phi_{\text{POL}}) \label{eq::E_CT}
\end{align}
This will be a faithful description of CT if $\Phi_{\text{POL}}$ describes a ``CT-free'' state of the complex: in other words it is constrained to prohibit CT whilst allowing on-fragment polarization in response to the rest of the complex. To achieve this target, the SCF-MI wavefunction optimizes a wavefunction in which the MO coefficient matrix is block-diagonal in the molecules making up the complex. Specifically, the ALMOs of each molecule are described by the mixing of its optimal occupied orbitals in isolation (the frozen occupied orbitals) with its own set of virtual orbitals. The simplest choice is to use \textit{all} the virtual orbitals of the given isolated molecule for this purpose. However, this choice becomes ill-defined as the span of the AO basis of one molecule overlaps that of other molecules in the complex more and more. In the limit of a linearly dependent AO basis, the use of all fragment virtuals in a block-diagonal MO coefficient matrix can no longer guarantee a CT-free state. \cite{Azar2013, Horn2015, Lao2016} To overcome this formal limitation (which can also be avoided in practice by using AO basis sets without excessively diffuse functions), one can instead use ``fragment electric response functions'' (FERFs) \cite{Horn2015} as a limited set of virtuals for each molecule in the complex that describe exactly the response of the molecules to applied dipole (D) and quadrupole (Q) fields. The FERF-DQ model provides a well-defined complete basis set (CBS) limit for the resulting SCF-MI energy, and thus for $\Delta E_{\text{CT}}$.

Since the CT energy is associated with orbital mixing between occupied levels on one fragment (molecule or ligand) of a complex and virtual orbitals of another, there are observable manifestations such as red-shifting of vibrational frequencies, as well as sometimes changes in structure. To quantify such effects, the adiabatic EDA\cite{Mao2017} employs geometry optimization on each constrained surface to define a sequence of energy lowerings. The adiabatic EDA has proven useful to understand the role of charge transfer on observables\cite{Mao2017,Mao2018,Mao2019}. It has also been extended to variationally quantify the role of forward and back donation individually on observable properties, via the recently introduced variational forward-backward (VFB) analysis of the relaxations due to CT.\cite{Loipersberger2020computational} This approach uses a generalized SCF-MI scheme that permits only uni-directional CT coupled to polarization in the entire system.

At a given geometry of a complex, an existing perturbative CT analysis (CTA) within the ALMO-EDA (reviewed in detail later) can decompose the CT energy into pairwise additive components, plus a residual ``higher-order'' (HO) CT correction.\cite{Khaliullin2007}.
%\begin{align}
%    \Delta E_{\text{CT}} &= \Delta E_{\text{CT}}^{\text{RS}} + \Delta E_{\text{CT}}^{\text{HO}} \label{eq::higher_order}
%\end{align}
This scheme is general enough to also decompose the charge flow, $\Delta Q_{\text{CT}}$ associated with CT into pairwise additive components.\cite{Khaliullin2008}
%both these decompositions rely on equations with very similar mathematical structure.
%, the pairwise charge decomposition itself is based on second order perturbation theory to yield pairwise additive contributions:
%\begin{align}
%    \Delta E_{\text{CT}}^{\text{RS}} &= \sum_{i,j<i}(\Delta E^{RS}_{i \rightarrow j }+ E^{RS}_{j \rightarrow i}) \label{eq::perturbative_pairwise}
%\end{align}
%Specifically, a second-order-like noniterative Roothaan Step (RS) starting from the converged SCF-MI solution ($\Delta E_{\text{CT}}^{\text{RS}}$) is used as an approximation to the total charge transfer energy. The remaining un-decomposed CT, corresponding to fully converging the wavefunction, is categorized as the ``higher order'' (HO) term as shown Eq.~\eqref{eq::higher_order}.
The perturbative CTA relies on the HO correction being small, but unfortunately, it can vary substantially in both sign and magnitude based on the chemical system investigated (and even the choice of density functional).
%Interestingly, the higher order term also depends on the density functional used and the location of the geometry on the potential energy surface.
Furthermore, this perturbative decomposition can break down when CT becomes strong, as seen for example in the interaction between a transition metal and ligand in a transition metal complex.
%This dependence of the higher-order term, and consequently the pairwise decomposition, gives us an arbitrarily incomplete understanding of the strength of forward/backward charge donation interactions.
The recently proposed VFB approach \cite{Loipersberger2020variational} alleviates the issues associated with the perturbative CTA to some extent when it is employed to quantify forward and backward CT energies, but it does not incorporate the HO term fully. %and currently has no accompanying charge decomposition scheme. 
In addition, the current VFB formulation lacks a well-defined basis set limit.

To go beyond the perturbative CTA, in this work, we present an alternative which exactly (at least up to machine precision) decomposes the CT energy into pairwise additive terms irrespective of the strength of interaction.
This scheme is applicable to a wide range of intermolecular interactions: from weakly bound hydrogen bonded complexes to strongly bound transition metal complexes.
Improving upon its perturbative predecessor, the new CTA also provides a complete pair-wise additive decomposition of the quantity of charge transferred.

This paper is organized as follows.
After a brief summary of the ALMO-based EDA and perturbative CTA, the non-perturbative CTA is introduced.
We specify how the total CT can be exactly decomposed into pairwise additive interactions. 
%by solving for the excitation matrix and computing an effective Fock-like matrix.
Next, we illustrate the limitations of the perturbative CTA and demonstrate how the non-perturbative CTA overcomes these limitations.
We show that the non-perturbative CTA has a well-defined basis set limit.
Subsequently, we apply the new CTA to complexes ranging from hydrogen bonding in DNA base-pairs to strongly interacting transition metal hexacarbonyls.
We illustrate how this new scheme debunks certain traditional ideas about borane adducts by revealing new mechanisms of interaction.
Finally, we discuss the computational expense of the non-perturbative CTA relative to  the perturbative CTA.

\section{Theory}
%\textcolor{red}{Notation: If referring to matrix, use subscript to denote index/blocks and superscript to denote which wavefunction is used to construct that matrix (for example: POL, CT etc.,).
%If scalar, use subscript to denote POL,CT and superscript to denote RS.}

\subsection{Brief summary of the ALMO-EDA}
%Before introducing the non-perturbative pairwise CT decomposition, we briefly summarize the DFT-based ALMO-EDA.
The ALMO-EDA scheme\cite{Khaliullin2007,Horn2015} decomposes the interaction energy into chemically relevant quantities given in Eq.~\eqref{eq::ALMO_EDA} by lifting imposed constraints one by one.
The geometric distortion energy ($\Delta E_{\text{GD}}$) associated with distorting relaxed (free) monomer geometries into their corresponding supersystem geometries is the difference between the SCF energy of each monomers, $X$, at the complex geometry ($\Phi_{X}^{\text{complex}}$) and at their optimal free geometry ($\Phi_{X}^{\text{free}}$):
\begin{equation}
        \Delta E_{\text{GD}} \equiv \sum_{X}\Big(E_{\text{SCF}}(\Phi_X^{\text{complex}})-E_{\text{SCF}}(\Phi_X^{\text{free}}) \Big) \label{eq::E_GD}
\end{equation}
The frozen wavefunction of the complex, $\Phi_{\text{FRZ}}$, is a Slater determinant composed of the occupied MOs from all isolated fragments (which are non-orthogonal). The corresponding energy, $E(\Phi_{\text{FRZ}})$, accounts for Pauli repulsion by forming a valid density from the non-orthogonal frozen MOs. The frozen interaction ($\Delta E_{\text{FRZ}}$) is defined as the difference between $E(\Phi_{\text{FRZ}})$ and the isolated energies of all monomers at their complex geometry (Eq.~\eqref{eq::E_FRZ}).
\begin{equation}
    \Delta E_{\text{FRZ}} \equiv E(\Phi_{\text{FRZ}}) - \sum_X E_{\text{SCF}}(\Phi_X^{\text{complex}}) \label{eq::E_FRZ}    
\end{equation}
Physically, $\Delta E_{\text{FRZ}}$ contains contributions from permanent electrostatic interactions, dispersion, and Pauli repulsions.\cite{Horn2016}
The polarized wavefunction ($\Phi_{\text{POL}}$) is computed by relaxing the occupied MOs on each fragment in presence of the MOs of all other fragments while forbidding CT from one fragment to another by enforcing the ALMO constraint (that the MO coefficient matrix remains fragment block diagonal). This optimization of non-orthogonal orbitals with the ALMO constraint is achieved by SCF-MI,\cite{Stoll1980, Gianinetti1996, Khaliullin2006, Horn2013}
and the polarization energy is defined as the energy difference between the polarized and frozen wavefunctions.
\begin{equation}
    \Delta E_{\text{POL}} \equiv E(\Phi_{\text{POL}}) - E(\Phi_{\text{FRZ}}) \label{eq::E_POL}    
\end{equation}
The use of FERF virtual orbitals to give the polarization term a well-defined basis set limit\cite{Horn2015} has already been discussed above. So too has the definition of the energy lowering due to CT, which was shown in Eq.~\eqref{eq::E_CT}.
By construction, we can see that the EDA terms sum up to the total interaction energy, defined in Eq.~\eqref{eq::ALMO_EDA} 

\subsection{Perturbative Charge-Transfer Analysis}
To set the stage for the new non-perturbative approach to pairwise decomposing the CT energy, we first discuss the existing perturbative CTA.\cite{Khaliullin2008}
The usual second order perturbation correction to an energy may be written as $\Delta E^{(2)}=\Trace[\vb{F}^{(1)}\vb{X}^{(1)}]$, where $\vb{F}^{(1)}$ is the first order perturbed Hamiltonian and $\vb{X}^{(1)}$ is the first order perturbed wavefunction.
The polarized wavefunction, $\Phi_{\text{POL}}$, only zeros the mixing between occupied (O) and virtual (V) orbitals on fragments, but not between fragments.
Therefore, given the polarized Fock matrix, $\vb{F}^{\text{POL}}=\vb{F}(\vb{P}^{\text{POL}})$, the perturbation is the residual occupied-virtual mixing, $\vb{F}_{OV}^{\text{POL}}$, which is exclusively \textit{between} fragments.
This perturbation, or Roothaan Step (RS), in turn yields a perturbative mixing, $\vb{X}_{VO}^{\text{RS}}$, of virtuals on a given fragment into occupieds of another. 

The perturbative approximation to orbital mixing is obtained by a single diagonalization (Roothaan step) of the polarized Fock matrix, which is equivalent\cite{Liang2004,Liang2004approaching} to solving the following quadratic equations for $\vb{X}_{VO}^{\text{RS}}$:
\begin{equation}
     \label{eq::rs_inf} 
     \mathbf{F}_{\text{VO}}^{\text{POL}}+\mathbf{F}_{\text{VV}}^{\text{POL}}\mathbf{X}^{\text{RS}}_{\text{VO}}-\mathbf{X}^{\text{RS}}_{\text{VO}}\mathbf{F}_{\text{OO}}^{\text{POL}}-\mathbf{X}^{\text{RS}}_{\text{VO}}\mathbf{F}_{\text{OV}}^{\text{POL}}\mathbf{X}^{\text{RS}}_{\text{VO}} = \mathbf{0}_{\text{VO}}
\end{equation}
Equation~\eqref{eq::rs_inf} is written in the orthogonalized MO basis which is discussed in the Supplementary Information.
The resulting $\mathbf{X}^{\text{RS}}$ is still a perturbative solution, because it gives the energy lowering of a single diagonalization rather than iterating to self-consistency.
\begin{equation}
     \Delta E_{\text{CT}}^{\text{RS}} = 2 \Trace[\vb{F}_{\text{VO}}\vb{X}^{\text{RS}}_{\text{OV}}] \label{eq::excitation_mat}    
\end{equation}
 Note that solving Eqs.~\eqref{eq::rs_inf} and \eqref{eq::excitation_mat} for $\Delta E_{\text{CT}}^{\text{RS}}$ is equivalent to infinite-order single excitation perturbation theory with a fixed Fock matrix,\cite{Liang2004,Liang2004approaching} and thus is preferable to $\Delta E^{(2)}_{\text{CT}}$. However the form still couples occupied MOs on one fragment with virtuals on another through $\vb{X}_{VO}^{\text{RS}}$, and therefore $\Delta E_{\text{CT}}^{\text{RS}}$ is pairwise decomposable, just like $\Delta E^{(2)}_{\text{CT}}$. The correction to $\Delta E^{\text{RS}}_{\text{CT}}$ for its lack of self-consistency is a non-pairwise decomposable, higher order (HO) term, $\Delta E_{\text{CT}}^{\text{HO}} = \Delta E_{\text{CT}} - \Delta E^{\text{RS}}_{\text{CT}}$, which means the CT energy is represented as the sum of the RS contribution, and the residual HO term:
\begin{align}
    \Delta E_{\text{CT}} = \Delta E_{\text{CT}}^{\text{RS}}+\Delta E_{\text{CT}}^{\text{HO}} \label{eq::ct_decomp}
\end{align}

%($\Delta E_{\text{CT}}^{\text{HO}}$, the ``higher-order'' term) as shown in Eq.~\eqref{eq::ct_decomp}.
%The pairwise decomposable charge transfer energy lowering can further be written in terms of an excitation matrix ($\mathbf{X}^p$, where is superscript $p$ denotes that the matrix was obtained using perturbative treatment) and Fock matrix ($\mathbf{F}$) as shown in Eq.~\eqref{eq::excitation_mat}.
%The Fock matrix here is constructed using the density of the polarized wavefunction ($\psi_{\text{POL}}$).
%The subscripts $O$ and $V$ represent the occupied and virtual blocks of the corresponding matrices.

%This excitation matrix represents the excitations from occupied to virtual orbitals and has been shown that it can be written as the Cayley generator of unitary transformation ($\mathbf{U}$).\cite{Liang2004}
%\begin{align}
%    \mathbf{U}^p &= (\mathbf{I+X}^p-\mathbf{X}^{p\dagger})(\mathbf{I+X}^{p\dagger}\mathbf{X}^p+\mathbf{X}^p\mathbf{X}^{p\dagger})^{1/2}
%\end{align}
%Here, the unitary matrix $\mathbf{U}^p$ connects the molecular orbitals of the polarized wavefunction to the ones obtained after a single Roothaan step transformation.

Equation~\eqref{eq::excitation_mat} can be rewritten in terms of projectors onto the occupied space ($\mathbf{P}^{\text{POL}}$) and virtual space ($\mathbf{Q}^{\text{POL}}$) of the polarized wavefunction as follows.
\begin{align}
    \Delta E_{\text{CT}}^{\text{RS}} 
    = \Tr[\vb{F}^{\text{POL}}\vb{P}^{\text{POL}}\vb{X}^{\text{RS}}\vb{Q}^{\text{POL}}] \label{eq::tr_qfpxq}
    %&= \Tr[\mathbf{(QFP)(PX}^p\mathbf{Q)}] \\
\end{align}
$\vb{P}^{\text{POL}}$ and $\vb{Q}^{\text{POL}}$ are sums of projectors onto the polarized occupied orbitals and virtual orbitals on all fragments respectively.
\begin{align}
\begin{split}
    \vb{P}^{\text{POL}} = \sum_{X,i}\vb{P}_{Xi}^{\text{POL}} \\
    \vb{Q}^{\text{POL}} = \sum_{X,a}\vb{Q}_{Xa}^{\text{POL}}  \label{eq::projector_sum}
\end{split}
\end{align}
Eqs.~\eqref{eq::projector_sum} can be inserted into Eq.~\eqref{eq::tr_qfpxq} to obtain the corresponding energy lowering due to transfer of charge from occupied orbital $i$ on fragment $X$ to virtual orbital $a$ on fragment $Y$ as follows.
\begin{align}
    \Delta E_{\text{CT}}^{\text{RS}} &= \sum_{X,Y}\sum_{i,a} \Delta E_{Xi\rightarrow Ya} \nonumber \\ 
    \Delta E_{Xi\rightarrow Ya} &= \Tr{\vb{F}^{\text{POL}} \vb{P}_{Xi}^{\text{POL}} \vb{X}^{\text{RS}} \vb{Q}_{Ya}^{\text{POL}}} \label{eq::pairwise_energy}
\end{align}
Similarly, it has been shown that the total charge transferred can also be decomposed into pairwise additive components:\cite{Khaliullin2008}
\begin{align}
    \Delta Q_{\text{CT}}^{\text{RS}} &= \sum_{X,Y}\sum_{i,a} \Delta Q_{Xi\rightarrow Ya} \nonumber \\ 
    \Delta Q_{Xi\rightarrow Ya} &= \Tr{\vb{P}^{\text{RS}} \vb{P}_{Xi}^{\text{POL}} \vb{X}^{\text{RS}} \vb{Q}_{Ya}^{\text{POL}}} \label{eq::pairwise_charge}
    %\mathbf{P'} &= \mathbf{UPU}^\dagger 
\end{align}
Here, $\mathbf{P}^{\text{RS}}$ is the density matrix of the RS occupied orbitals. 
Using $\mathbf{P}^{\text{RS}}$ to replace the Fock operator in Eq.~\eqref{eq::pairwise_energy} yields the charge transfer decomposition of Eq.~\eqref{eq::pairwise_charge}.

While this approach is appealing and useful, its perturbative nature (i.e.~its lack of self-consistency) is a disadvantage, because $\Delta E_{\text{CT}}^{\text{RS}}$ can never be exact. In particular, $\vb{X}^{\text{RS}}$ generates the energy lowering, $\Delta E_{\text{CT}}^{\text{RS}}$, of a single diagonalization rather than the proper self-consistent energy lowering, $\Delta E_{\text{CT}}$, that results from lifting the SCF-MI constraint of no charge delocalization between fragments. %Eq.~\eqref{eq::excitation_mat} is linear in $\mathbf{X}$, leading to an error of the order of $\mathcal{O}(\mathbf{X}^2)$.
In complexes containing large charge transfer, this can lead to the higher order term becoming large, not because CT cannot be partitioned in a pairwise additive fashion, but because the approximation of a single diagonalization becomes inadequate. We therefore take up the challenge of lifting this approximation.
%$\mathbf{X}$ is no longer small and the corresponding error $\mathcal{O}(\mathbf{X}^2)$ can be large.

\subsection{Exactly Pairwise-Additive Charge-Transfer Analysis}
In this section, we will show that it is possible to generate an exactly pairwise-additive charge transfer analysis (at least to as much precision as we wish). The working form of the result for the charge-transfer energy is:
\begin{align}
    \Delta E_{\text{CT}} = E(\Phi_{\text{FULL}}) - E(\Phi_{\text{POL}}) =  2\Tr[\vb{F}_{VO}^{\text{CT}}\vb{X}^{\text{CT}}_{OV}] \label{eq::vct_Feff}
\end{align}
This has the same form as the approximate expression, Eq.~\eqref{eq::excitation_mat} discussed above, but with new definitions of the Fock matrix and the occupied-virtual mixings.

To begin, since we know the polarized state (i.e.~$E(\Phi_{\text{POL}})$) and the final, fully relaxed state (i.e. $E(\Phi_{\text{FULL}})$), we also, at least implicitly, know the occupied-virtual mixings, $\vb{X}^{\text{CT}}_{OV}$, necessary to connect them. We shall discuss how we explictly obtain them after establishing the exactly pairwise addition decomposition. Let us define $\lambda=0$ as the state $\Phi_{\text{POL}}$, and $\lambda=1$ as the state $\Phi_{\text{FULL}}$ which are connected along the straight-line path specified by $\lambda \vb{X}^{\text{CT}}$. Applying the fundamental theorem of line integrals along this path, it is true by definition that:
\begin{align}
    \Delta E_{\text{CT}} = E(1) - E(0) =  \sum_{i,a} \int_0^1 \frac{\partial E(\lambda)}{\partial \lambda X_{ia}^{\text{CT}}} \cdot X_{ia}^{\text{CT}} d\lambda  
\end{align}
Since:
\begin{equation}
    {F}_{ai}(\vb{X}^{\text{CT}}) \equiv  \frac12\frac{\partial E(\vb{X}^{\text{CT}})}{\partial X_{ia}^{\text{CT}}}
\end{equation}
We are led directly to the desired result, Eq.~\eqref{eq::vct_Feff}, where evidently the appropriate Fock matrix arises from integrating along the line:
\begin{align}
    \vb{F}_{VO}^{\text{CT}} =  \int_0^1 {\vb{F}_{VO}[\lambda \vb{X}^{\text{CT}}]} d\lambda 
    \label{eq::Feff}
\end{align}

Let us turn next to the charge reorganization that occurs as the electron density is rearranged. 
To connect directly to the energy changes discussed above, we choose to write it in an isomorphic form (other more obvious possibilities\cite{Mitoraj2009combined} also exist, of course, and are independently useful). 
The charge that is rearranged, $\Delta Q_{\text{CT}}$, as the system evolves from the POL state to the final FULL state from $\lambda=0$ to $\lambda=1$ along the straight-line path given by $\lambda \vb{X}$ can likewise be represented by a line integral:
\begin{align}
    \Delta Q_{\text{CT}} = Q(1) - Q(0)  &= \int_{\lambda = 0}^{\lambda =1} \sum_{i,a} \frac{\partial Q[\lambda \vb{X}^{\text{CT}}]}{\partial \lambda X_{ia}^{\text{CT}} } X_{ia}^{\text{CT}} d\lambda \label{eq::delQ_CT}
\end{align}
Here $Q[\lambda \vb{X}^{\text{CT}}]$ is the charge that is promoted from occupied levels in the polarized state to orbitals that are virtual in the polarized state:
\begin{align}
    Q[\lambda \vb{X}^{\text{CT}}] = \Tr{\vb{Q}_0 \vb{P}[\lambda \vb{X}^{\text{CT}}]\vb{Q}_0} 
                      = \Tr{         \vb{P}[\lambda \vb{X}^{\text{CT}}]\vb{Q}_0}\end{align}
The variation of $Q[\lambda \vb{X}^{\text{CT}}]$ with respect to elements, $X_{ia}^{\text{CT}}$ of $X^{\text{CT}}$ is:
\begin{align}
    \frac{\partial \Tr{\vb{P}[\lambda\vb{X}^{\text{CT}}]\vb{Q}^{\text{POL}}}}{\partial \lambda X_{ia}^{\text{CT}}} &= -\lambda (\vb{P}^{\text{POL}}\vb{X}\vb{Q}^{\text{POL}})_{ai}
    -\lambda(\vb{Q}^{\text{POL}}\vb{X}\vb{P}^{\text{POL}})_{ai} \nonumber \\
    &= -\lambda (\vb{X}_{VO}^{\text{CT}})_{ai} \nonumber \\
    &= P_{ai}(\lambda \vb{X}^{\text{CT}}) \label{eq::del_trace}
\end{align}
Substituting Eq.~\eqref{eq::del_trace} into Eq.~\eqref{eq::delQ_CT}, we get
\begin{align}
     \Delta Q_{\text{CT}} &= \int_{\lambda=0}^{\lambda=1} \sum_{i,a} P_{ia}(\lambda \vb{X}^{\text{CT}}) X_{ai}^{\text{CT}} d\lambda -  \int_{\lambda=0}^{\lambda=1} \sum_{i,a} P_{ai}(\lambda \vb{X}^{\text{CT}}) X_{ia}^{\text{CT}} d\lambda \\
     &= 2 \Tr{\vb{P}_{VO}^{\text{CT}}\vb{X}_{OV}^{\text{CT}}} 
\end{align}
where 
\begin{align}
    \vb{P}^{\text{CT}}_{VO} &= \int_{\lambda=0}^{\lambda=1} \vb{P}_{VO}[\lambda \vb{X}^{\text{CT}}]d\lambda \label{eq::Peff}
\end{align}

Two issues must be addressed to employ this approach in practice: (i) we must find the orbital mixings associated with CT, $\vb{X}^{\text{CT}}_{OV}$, and (ii) we must develop a suitable quadrature to efficiently and accurately numerically evaluate Eq.~\eqref{eq::Feff} and Eq.~\eqref{eq::Peff}. 
We address these points in turn below.

%It is useful here to note that the excitation matrix ($\mathbf{X}$) is the key to partitioning the charge transfer energy and charge into pairwise additive terms.
%It also determines the total amount of charge transfer that is decomposed into pairwise additive terms.
%Now, it is important to ask if we can find an excitation matrix which can completely recover the total amount of non-perturbative charge transfer given by Eq.~\eqref{eq::E_CT}.
%In the perturbative treatment, $\mathbf{X}$ is the generator of the unitary transformation that rotates the SCF-MI molecule orbitals to the RS transformed molecular orbitals, and consequently can only recover the energy difference between these two wavefunctions.

%In order recover the total amount of charge transfer, we should find a unitary transformation (and subsequently its generator) that would transform the polarized wavefunction molecular orbitals ($\mathbf{C}_{\text{POL}}$) to the molecular orbitals of the CT wavefunction ($\mathbf{C}_{\text{CT}}$) as shown in equation Eq.~\eqref{eq::C_connect}.
%\begin{align}
%    \mathbf{C}_{\text{CT}} &= \mathbf{C}_{\text{POL}}\mathbf{U}  \\  
%    \mathbf{U} &= e^{\mathbf{X}^{\text{CT}}} \label{eq::C_connect}
%\end{align}

Density matrices are independent of the redundant occupied-occupied and virtual-virtual mixings that are needed to fully specify molecular orbitals. Working in an orthonormal basis, we are seeking the unitary transformation, $\vb{U}^{\text{CT}}$ connecting the polarized density matrix, $\vb{P}^{\text{POL}}$ and the fully relaxed density matrix, $\vb{P}^{\text{FULL}}$:
\begin{equation}
\vb{P}^{\text{FULL}} = \left(\vb{U}^{\text{CT}}\right)^\dagger \vb{P}^{\text{POL}} \vb{U}^{\text{CT}}
\label{eq::P_connect}
\end{equation}
$\vb{U}^{\text{CT}}$ can be written just in terms of occupied-virtual mixings, $\vb{X}_{OV}^{\text{CT}}$, as the matrix exponential:
\begin{equation}
\vb{U}^{\text{CT}} = 
    \begin{bmatrix}
        \mathbf{U}_{OO}^{\text{CT}} & \mathbf{U}_{OV}^{\text{CT}} \\
        \mathbf{U}_{VO}^{\text{CT}} & \mathbf{U}_{VV}^{\text{CT}} 
    \end{bmatrix}
    = \exp{
    \begin{bmatrix}
        \mathbf{0} & \mathbf{X}_{OV}^{\text{CT}} \\
        -(\mathbf{X}_{OV}^{\text{CT}})^\dagger & \mathbf{0}
    \end{bmatrix}}
%    \begin{bmatrix}
%        \mathbf{0} & -\mathbf{X}_{OV}^{\text{CT}} \\
%        (\mathbf{X}_{OV}^{\text{CT}})^\dagger & \mathbf{0} 
%    \end{bmatrix}}
\label{eq::U_CT}
\end{equation}
%$\vb{U}=\exp(\vb{X}_{OV})$. Specifically, the unitary transformation, $\vb{U}^{\text{CT}}$ connecting the polarized density matrix, , and the fully relaxed density matrix, $\vb{P}^{\text{FULL}}$ can thus be expressed in terms of $\vb{X}^{\text{CT}}_{OV}$, according to: %Rewriting Eq.~\eqref{eq::C_connect} in terms of POL and CT densities ($\mathbf{P}^{\text{POL}}$ in the POL MO basis and $\mathbf{P}^{\text{CT}}$) and along with the further simplification that O-O and V-V rotations do not change the energy expressed, we have
%\begin{align}
%\vb{P}^\text{FULL} = 
%    \begin{bmatrix}
%        \mathbf{P}_{OO}^{\text{FULL}} & \mathbf{P}_{OV}^{\text{FULL}} \\
%        \mathbf{P}_{VO}^{\text{FULL}} & \mathbf{P}_{VV}^{\text{FULL}} 
%    \end{bmatrix}
%    = \mathrm{exp}
%    \begin{bmatrix}
%        \mathbf{0} & \mathbf{X}_{OV}^{\text{CT}} \\
%        -(\mathbf{X}_{OV}^{\text{CT}})^\dagger & \mathbf{0} 
%    \end{bmatrix}
%    \begin{bmatrix}
%        \mathbf{P}_{OO}^{\text{POL}} & \mathbf{0} \\
%        \mathbf{0} & \mathbf{0}
%    \end{bmatrix}
%    \mathrm{exp}
%    \begin{bmatrix}
%        \mathbf{0} & -\mathbf{X}_{OV}^{\text{CT}} \\
%        (\mathbf{X}_{OV}^{\text{CT}})^\dagger & \mathbf{0} 
%    \end{bmatrix} \label{eq::P_connect}
%\end{align}
To solve for $\vb{X}_{OV}^{\text{CT}}$ we define a cost function that vanishes when  Eq.~\eqref{eq::P_connect} is satisfied:
\begin{align}
    C &= || \vb{P}^{\text{FULL}}-\vb{U}^{\text{CT}}\vb{P}^{\text{POL}}\left(\vb{U}^{\text{CT}}\right)^{\dagger} || _F^2 \label{eq::Pconnect_cost} 
\end{align}
As shown in the Supplementary Information, the analytical gradient of the cost function, $C$, with respect to the $ia^{\text{th}}$ element of $\vb{X}_{OV}^{\text{CT}}$ is given by:
\begin{align}
    \frac{\partial C}{\partial \left(\vb{X}_{OV}^{\text{CT}}\right)_{ia}}\Big|_{\vb{X}_{OV}^{\text{CT}}=\boldsymbol{0}} &= 4 \left[ \left(\vb{U}^{\text{CT}}_{\text{curr}}\right)^{\dagger}\vb{P}^{\text{FULL}} \vb{U}^{\text{CT}}_{\text{curr}} \right]_{ia}
\end{align}
Here $\vb{U}^{\text{CT}}_{\text{curr}}$ is the current approximation to the unitary transformation. We choose a working orthonormal basis that symmetrically orthogonalizes the ALMO occupieds, and the ALMO virtuals (after projecting out their occupied components), and canonically orthogonalizes the projected virtual void orbitals (which were excluded from the evaluation of $\vb{P}^{\text{POL}}$). Minimizing $C$ using the analytical gradient is performed via standard iterative techniques for unconstrained non-linear equations like quasi-Newton methods and DIIS in order to obtain $\vb{X}_{OV}^{\text{CT}}$.
In principle, one can solve for $\mathbf{X}^{\text{CT}}$ analytically by taking the logarithm of $\mathbf{U}^{\text{CT}}$.
However, we do not use this method as this leads to non-zero $OO$ and $VV$ blocks of $\mathbf{X}^{\text{CT}}$ and hence cannot be used for pairwise decomposition directly.

%The working basis and gradient for solving Eq.~\eqref{eq::P_connect} is included in the Supplementary Information.

Now, we will address how to develop a suitable numerical quadrature scheme to evaluate $\vb{F}_{VO}^{\text{CT}}$ and $\vb{P}_{VO}^{\text{CT}}$.
We can numerically evaluate the integral in Eq.~\eqref{eq::Feff} and Eq.~\eqref{eq::Peff} using Gauss quadrature rules in the interval $[0, \mathbf{X}^{\text{CT}}]$.
Specifically, we use the 5-point Gauss-Lobatto quadrature rules,\cite{Abramowitz1974} as they include the end points of the interval which have already been evaluated for the purpose of EDA.
The expressions for $\vb{F}_{VO}^{\text{CT}}$ and $\vb{P}_{VO}^{\text{CT}}$ are shown in Eq.~\eqref{eq::F_quadrature} and Eq.~\eqref{eq::P_quadrature}.
\begin{align}
    \mathbf{F}^{\text{CT}}_{VO} &= \frac{1}{20}\mathbf{F}_{VO}[0]+\frac{49}{180}\mathbf{F}_{VO}\Bigg[\frac{1}{2}\Big(1-\sqrt{\frac{3}{7}}\Big)\mathbf{X}^{\text{CT}}\Bigg] \nonumber \\
    &+\frac{16}{45}\mathbf{F}_{VO}\Big[\frac{1}{2}\mathbf{X}^{\text{CT}}\Big]+\frac{49}{180}\mathbf{F}_{VO}\Bigg[\frac{1}{2}\Big(1+\sqrt{\frac{3}{7}}\Big)\mathbf{X}^{\text{CT}}\Bigg] + \frac{1}{20}\mathbf{F}_{VO}[\mathbf{X}^{\text{CT}}] \label{eq::F_quadrature} \\
    \mathbf{P}^{\text{CT}}_{VO} &= \frac{1}{20}\mathbf{P}_{VO}[0]+\frac{49}{180}\mathbf{P}_{VO}\Bigg[\frac{1}{2}\Big(1-\sqrt{\frac{3}{7}}\Big)\mathbf{X}^{\text{CT}}\Bigg] \nonumber \\
    &+\frac{16}{45}\mathbf{P}_{VO}\Big[\frac{1}{2}\mathbf{X}^{\text{CT}}\Big]+\frac{49}{180}\mathbf{P}_{VO}\Bigg[\frac{1}{2}\Big(1+\sqrt{\frac{3}{7}}\Big)\mathbf{X}^{\text{CT}}\Bigg] + \frac{1}{20}\mathbf{P}_{VO}[\mathbf{X}^{\text{CT}}] \label{eq::P_quadrature} 
\end{align}
As $\vb{F}_{VO}[0]$ and $\vb{P}_{VO}[0]$ are the Fock and density matrices of the polarized wavefunction $(\Phi^{\text{POL}})$ and $\vb{F}_{VO}[\vb{X}^{\text{CT}}]$ and $\vb{P}_{VO}[\vb{X}^{\text{CT}}]$ are the Fock and density matrices of the fully relaxed wavefunction $(\Phi^{\text{FULL}})$, we would have to evaluate only three new Fock and density matrices at points specified by Eqs.~\eqref{eq::F_quadrature} and \eqref{eq::P_quadrature}.
Empirically, for the systems studied in this work, using the 5-point quadrature formula is sufficient to recover the variational charge transfer energy with sub-Joule per mole accuracy (See Table S6). 
%Empirically, using the 5-point quadrature formula gives us an error of $\mathcal{O}((\vb{X}^{\text{CT}})^9)$ which is small enough to recover the charge transfer energy to an accuracy of better than a Joule per mole. 
%{\color{blue} We need to mention that we are assuming that the spectral norm of X is small here. How did you get a Joule per mole estimation?}

%\subsection{Orbital projectors and linear dependence}

To partition CT we need an appropriate set of projectors onto the occupied and unoccupied subspaces of each of the interacting fragments.
In the second-generation ALMO-EDA,\cite{Horn2016} the Hilbert space of each polarized fragment $X$ is spanned by its occupied frozen orbitals and the virtual Fragment Electric Response Functions (FERFs)\cite{Horn2015} that are basis-independent and provide a  well-defined basis set limit for $\Delta E_{\text{POL}}$. After polarization, let us denote these spaces as $\mathbb{P}_X$ and $\mathbb{V}_X$. 
The unoccupied AO space on each fragment not spanned by FERF virtuals (hereinafter virtuals) is termed the ``void space'' (denoted by $\mathbb{R}_X$) and the entire Hilbert space on fragment $X$ (denoted by $\mathbb{H}_X$) can be written as shown in Eq.~\eqref{eq::frgm_hilbert_space_partition}
\begin{align}
    \mathbb{H}_X &= \mathbb{P}_X \oplus \mathbb{V}_X \oplus \mathbb{R}_X \label{eq::frgm_hilbert_space_partition} \\
    &= \mathbb{P}_X \oplus \mathbb{Q}_X
\end{align}
where $\mathbb{Q}_X = \mathbb{V}_X \oplus \mathbb{R}_X$ denotes the total unoccupied space formed by combining virtuals and voids.

Following Ref.~\citenum{Khaliullin2008}, the projector onto the  $i^{th}$ occupied orbital on fragment $X$, denoted by $\hat{P}_{Xi}$, can be formed as shown below:
\begin{align}
    \hat{P}_{Xi} &= \ket{\phi_{Xi}}\bra{\phi^{Xi}} \\ \label{eq::occ_projector}
    &= \sum_{Y,j} \ket{\phi_{Xi}}(\boldsymbol\sigma_{OO}^{-1})^{Xi, Yj}\bra{\phi_{Yj}} \\
    (\mathbf{P}_{Xi})^{\mu, \nu} &= \sum_{Y,j}(\mathbf{C}_{\text{POL}})^{X\mu\bullet}_{\bullet Xi} (\boldsymbol\sigma_{OO}^{-1})^{Xi,Yj}((\mathbf{C_\text{POL}})^T)^{\bullet Y\nu}_{Yj \bullet}
\end{align} 
In the equations above, the biorthogonal covariant and contravariant notation was used for dealing with non-orthogonal molecular orbitals.\cite{Head-Gordon1997}
One further consideration regarding the virtual space projector is the treatment of near and exact linear dependence in large AO basis sets, which is part of ensuring a decomposition with well-defined basis set limits.
Linear dependence between basis functions within the same fragment can be resolved by simply discarding one of the linearly dependent molecular orbital, for example by canonical orthogonalization.
However, this is not possible with linear dependence between fragments without losing the association of the retained functions to fragments. Therefore inter-fragment linear dependence in the virtual space is treated by using the Moore-Penrose generalized inverse of the overlap matrix ($\boldsymbol\sigma_{VV}^+$) by discarding the null space.\cite{Head-Gordon1997}
%This allows for an exact treatment of interfragment linear dependencies in $\mathbb{Q}$ space without double counting charge transfer.
For CT decomposition, we do not make any distinction between FERF virtuals and the void space.
\begin{align}
    \hat{Q}_{Xa} &= \ket{\phi_{Xa}}\bra{\phi^{Xa}} \\
    &= \sum_{X,a}\ket{\phi_{Xa}}(\boldsymbol\sigma_{VV}^+)^{Xa,Yb}\bra{\phi_{Yb}} \\ \label{eq::virt_projector}
    (\mathbf{Q}_{Xa})^{\mu,\nu} &= \sum_{Y,b}(\mathbf{C}_{\text{POL}})^{X\mu \bullet}_{\bullet Xa}(\boldsymbol\sigma_{VV}^+)^{Xa,Yb}((\mathbf{C}_{\text{POL}})^T)^{\bullet Y\nu}_{Yb \bullet}
\end{align}

With the aid of these projectors, we can now exactly (up to arbitrary precision) decompose the CT energy, Eq.~\eqref{eq::vct_Feff} into contributions from each pair of orbitals, on each pair of interacting fragments in the complex:
\begin{align}
    \Delta E_{\text{CT}} &= \sum_{X,Y}\sum_{i,a} \Delta E_{Xi\rightarrow Ya} \nonumber \\ 
    \Delta E_{Xi\rightarrow Ya} &= \Tr{\vb{F}^{\text{CT}} \vb{P}_{Xi}^{\text{POL}} \vb{X}^{\text{CT}} \vb{Q}_{Ya}^{\text{POL}}} \label{eq::exact_pairwise_energy}
\end{align}
Similarly, the total charge transferred can be exactly decomposed into pairwise additive orbital components:
\begin{align}
    \Delta Q_{\text{CT}} &= \sum_{X,Y}\sum_{i,a} \Delta Q_{Xi\rightarrow Ya} \nonumber \\ 
    \Delta Q_{Xi\rightarrow Ya} &= \Tr{\vb{P}^{\text{CT}} \vb{P}_{Xi}^{\text{POL}} \vb{X}^{\text{CT}} \vb{Q}_{Ya}^{\text{POL}}} \label{eq::exact_pairwise_charge}
\end{align}
In practice, the results are compacted into the CT energy and flow between each pair of fragments. For the occupied orbitals on one fragment, $X$, coupling to virtual orbitals on another fragment, $Y$, it is convenient to singular value decompose their couplings,  $\left(\vb{X}^{\text{CT}}_{OV}\right)^{XY}$.
\begin{equation}
    \left(\vb{X}^{\text{CT}}_{OV}\right)^{XY} = \vb{L}_O^X \vb{s} \left(\vb{R}_V^Y\right)^\dagger
\end{equation}
Often there is only 1 significant singular value, in which case there is a single complementary donor-acceptor orbital pair (COVP) controlling CT, where the donor orbital is represented by the first column of $\vb{L}_O^X$ and the acceptor is given by the first column of $\vb{R}_V^Y$. 

Finally we note that $\vb{X}^{\text{CT}}_{OV}$ describes CT couplings from all occupied orbitals to all unoccupied orbitals irrespective of the fragment that the virtuals belong to. This applies not just to the case where the fragments $X$ and $Y$ are different, but also to the case where they are the same. 
%To put it in other words, $\mathbf{X}^{\text{CT}}$ can describe excitations from occupied orbital on one fragment to unoccupied orbitals of all other fragments and also unoccupied orbitals belonging to the same fragment.
As a consequence of these non-zero couplings on the same fragment, we will have non-zero $\Delta E_{X\rightarrow X}$ terms, because the on-fragment blocks of $\mathbf{F}^{\text{CT}}_{VO}$ are also non-zero.
This is in contrast to the perturbative mixing, $\mathbf{X}^{\text{RS}}_{OV}$, that has zero on-fragment blocks of $\mathbf{F}^{\text{POL}}_{VO}$ (a result of the SCF-MI iterations) leading to zero $E_{X\rightarrow X}$ terms.  
The interpretation of the non-zero pairwise on-fragment terms for the exactly pairwise additive scheme is that this a re-polarization of the fragments in response to charge transfer.

\section{Computational Details}
The first consideration in decomposing charge transfer is computing the densities of the polarized ($\mathbf{P}^{\text{POL}}$) and FULL ($\mathbf{P}^{\text{FULL}}$) wavefunctions.
After computing these quantities, we can solve for the non-perturbative mixing matrix ($\mathbf{X}^{\text{CT}}$) by solving Eq.~\eqref{eq::P_connect}.

The perturbative and non-perturbative pairwise charge decomposition schemes were implemented in a developmental version of Q-Chem 5.0.\cite{Shao2015} In particular, the new CTA method was implemented within the \texttt{libgscf} and \texttt{libloco} libraries which are, respectively, new SCF and SCFMI modules in Q-Chem.\cite{Lee2018}
$\omega$B97X-D\cite{Chai2008} with def2-TZVPD\cite{Weigend2005,Rappoport2010a} basis set was used for geometry optimization and energy decomposition analyses unless stated otherwise.
$\omega$B97X-D is a range-separated hybrid density functional and has shown to give superior performance on a large number of systems for a wide range of chemical properties including non-covalent interaction energies.\cite{Mardirossian2017}
All geometries were confirmed to be a minimum on the potential energy surface by confirming that the Hessian has no negative eigenvalues.
All interaction energies computed adiabatically by including the relaxation energies of the individual fragments when they are infinitely separated from each other.
All Complementary Occupied-Virtual Pairs (COVPs) are plotted with an isosurface value of $\pm 0.07$ au.
All plots were created using \texttt{Matplotlib}\cite{Hunter2007} and molecule figures were generated using VMD.\cite{HUMP96}

\section{Results and Discussion}
\subsection{Perturbative CTA vs non-perturbative CTA}

\begin{figure}
    \centering
     \begin{subfigure}[t]{0.95\textwidth}
         \centering
        %  \vskip 10pt
         \includegraphics[width=\textwidth]{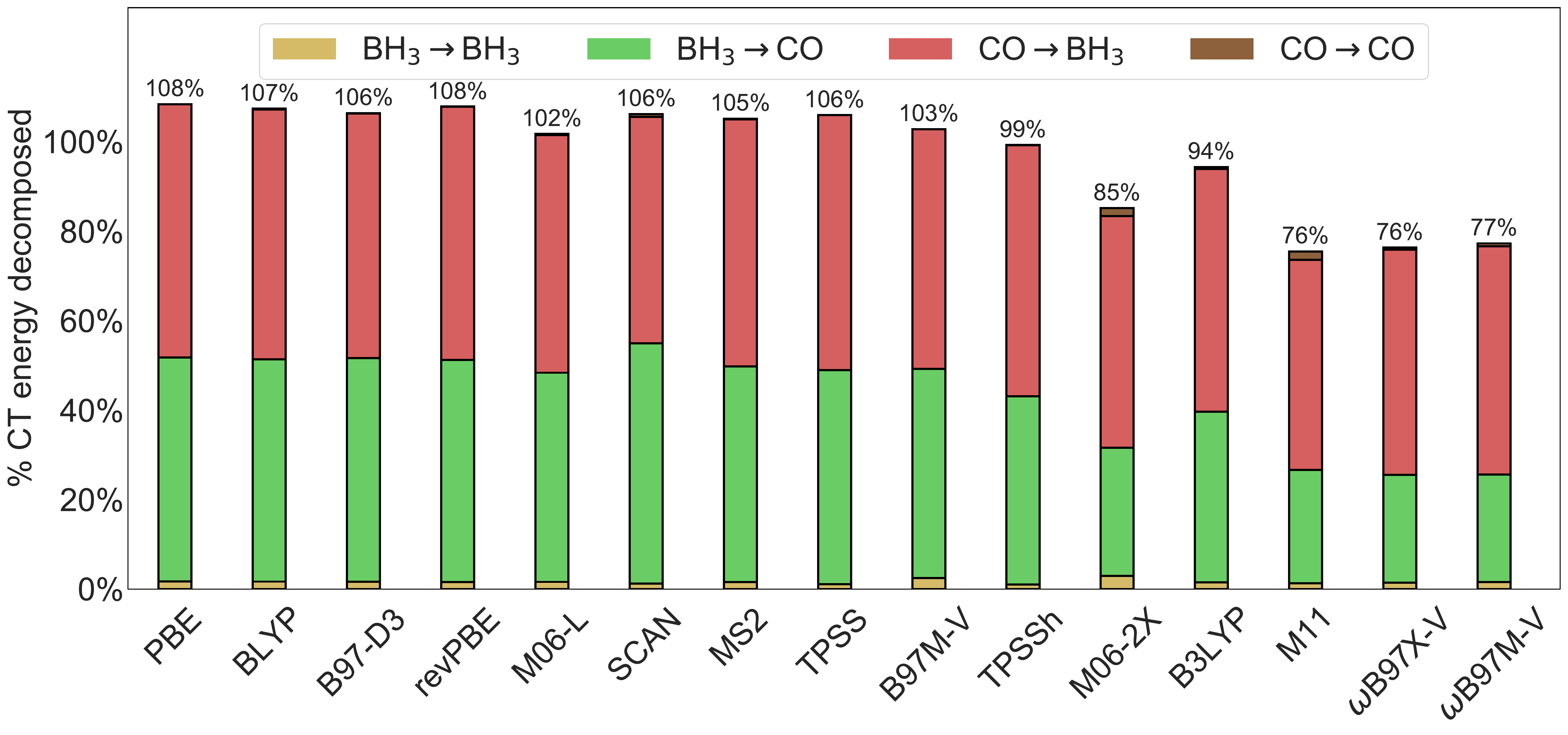}
         \caption{}
         \label{fig:func_dep_pert}
     \end{subfigure}
     \begin{subfigure}[b]{0.95\textwidth}
         \centering
         \includegraphics[width=\textwidth]{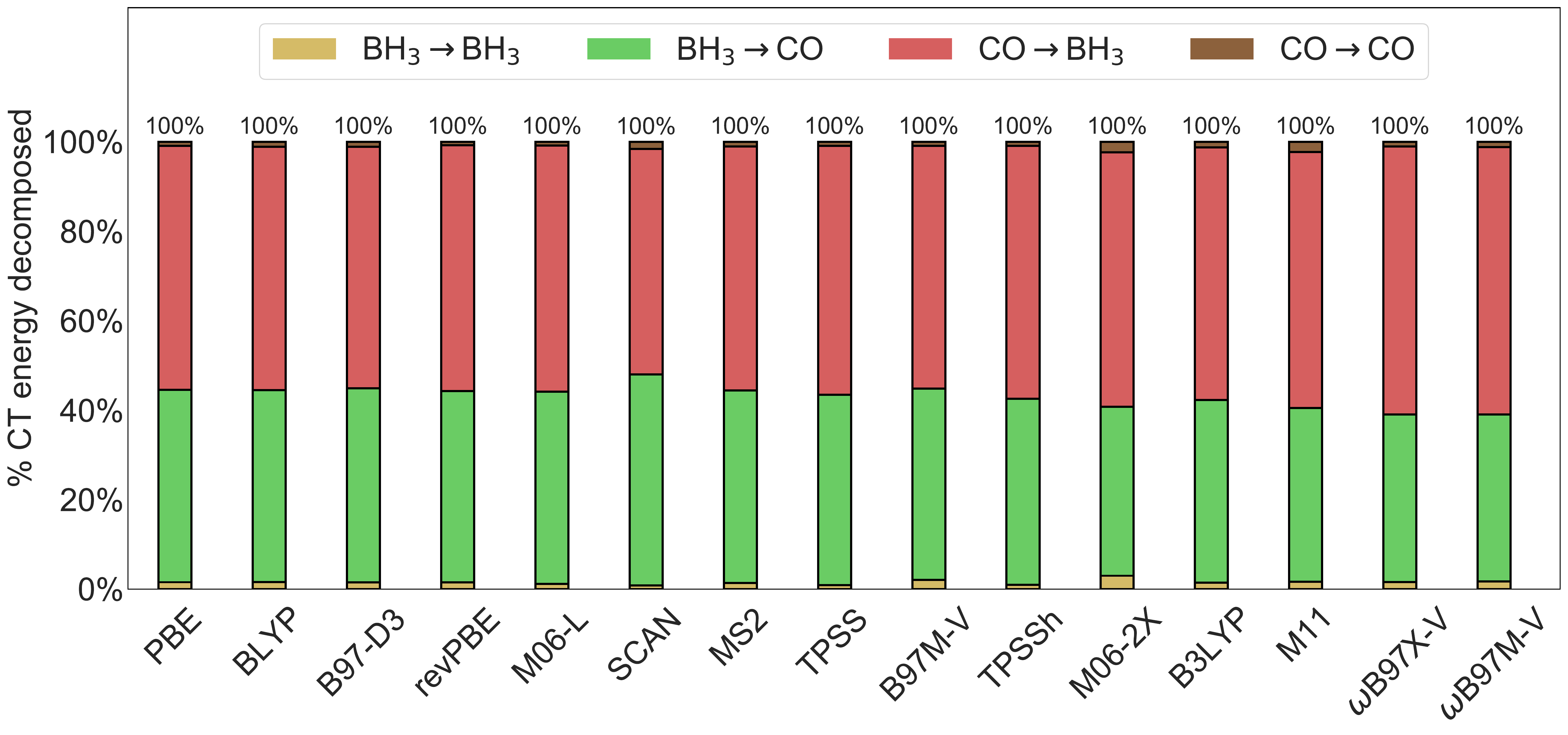}
         \caption{}
         \label{fig:func_dep_nonperturb}
     \end{subfigure}
    \caption{Dependence of pairwise decomposition of charge transfer energy on different density functionals in the perturbative (upper panel) and non-perturbative (lower panel) treatment of charge transfer energy decomposition for the \ce{BH3-CO} complex in the def2-TZVPD basis set. The on-fragment CT terms (\ce{BH3}$\rightarrow$\ce{BH3} and \ce{CO}$\rightarrow$\ce{CO}) are very small in most cases and cannot be seen in this figure for multiple density functionals. The different components of CT are shown as bars stacked on each other and appear in this order from bottom to top: \ce{BH3}$\rightarrow$\ce{BH3}, \ce{BH3}$\rightarrow$\ce{CO}, \ce{CO}$\rightarrow$\ce{BH3}, and \ce{CO}$\rightarrow$\ce{CO}. }
    \label{fig:func_dep}
\end{figure}

The perturbative CTA has some well-known limitations, especially its inability to pairwise decompose the entire CT energy. 
%In this section, we illustrate how the non-perturbative charge decomposition analysis scheme introduced in this paper can overcome all the limitations of the perturbative analysis.
A manifestation of this problem is that the fraction of CT energy decomposed into pairwise additive terms depends on the density functional used, as illustrated in Fig.~\eqref{fig:func_dep} for the 
case of the borane-carbonyl complex (\ce{BH3-CO}). 
\ce{BH3-CO} is bound by $-123.6$ kJ/mol at the $\omega$B97X-D/def2-TZVPD level of theory as a result of strong CT character, with active forward ($-171.0$ kJ/mol) and backward donation ($-113.8$ kJ/mol).
As shown in Fig.~\eqref{fig:func_dep}, the perturbative Roothaan step treatment overestimates CT for semi-local functionals by amounts ranging from 8\% for PBE to about 2\% for M06-L.
Concurrently, the Roothaan step underestimates CT for hybrid and range-separated hybrid functionals by amounts ranging from 1\% for TPSSh to rather severe values of 24\% for M11 and $\omega$B97X-V.
Among the pairwise components, the perturbative \ce{BH3}$\rightarrow$\ce{CO} back-donation sees the most variation, fluctuating by more than 90 kJ/mol across the density functionals tested.
%The magnitude of underestimation is also more severe than the magnitude of overestimation.
This functional-dependent performance of the Roothaan step in the perturbative CTA %underestimation of overestimation of charge transfer for semi-local density functionals and overestimation of charge transfer for hybrid functionals in the perturbative treatment 
is reminiscent of density functional minimal adaptive basis calculations based on the same perturbative corrections.\cite{Mao2016approaching}

On the other hand, the non-perturbative pairwise CTA consistently decomposes 100\% of the charge transfer energy irrespective of the density functional employed. It is particularly encouraging to see from Fig.~\eqref{fig:func_dep} that the functional-dependence of pairwise contributions in the non-perturbative approach is also significantly lower than their corresponding perturbative counterparts.

\begin{figure}
     \centering
     \begin{subfigure}[t]{0.25\textwidth}
         \centering
        %  \vskip 10pt
         \includegraphics[width=\textwidth]{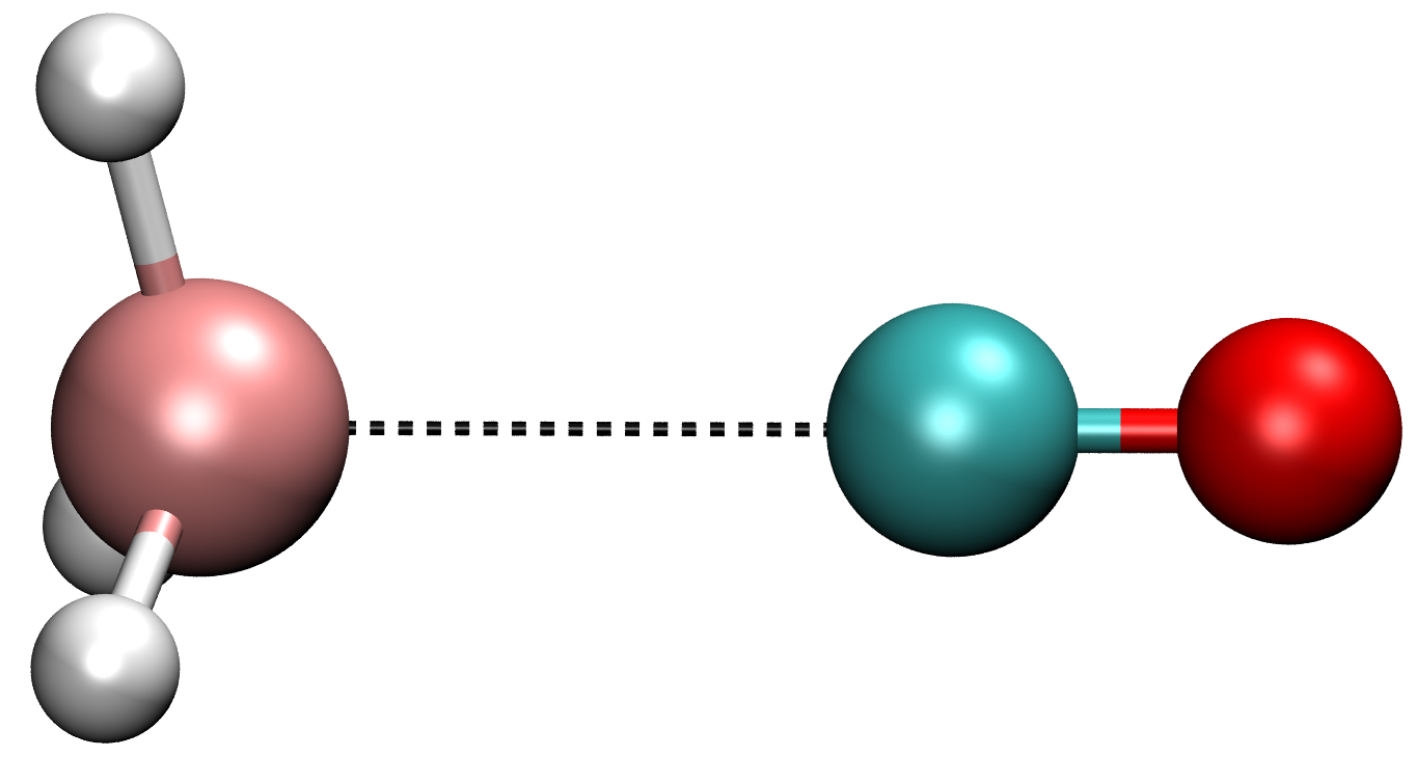}
         \caption{}
         \label{fig:BH_CO_PES}
     \end{subfigure}
     \begin{subfigure}[b]{0.74\textwidth}
         \centering
         \includegraphics[width=\textwidth]{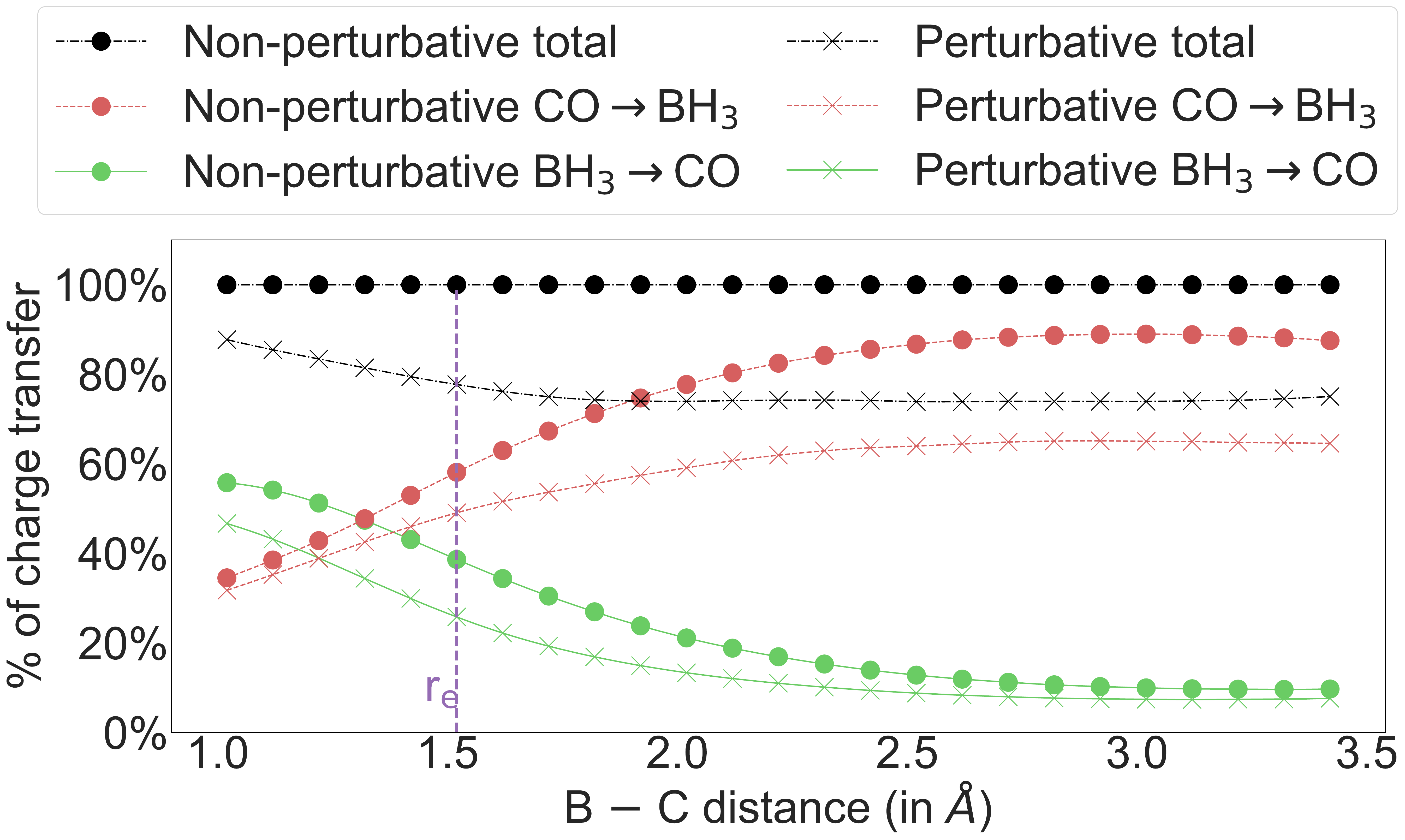}
         \caption{}
         \label{fig:PES_dep}
     \end{subfigure}
        \caption{(a) \ce{BH3}--\ce{CO} complex stretched along the dashed line (b) \ce{BH3}$\rightarrow$\ce{CO}, \ce{CO}$\rightarrow$\ce{BH3}, and total charge decomposed is shown as a percentage of the total charge transfer energy for the perturbative and non-perturbative CTAs for different B -- C bond distances. The equilibrium bond distance ($r$\textsubscript{e}) is shown in purple.}
        \label{fig:PES_dependence}
\end{figure}

Consistent performance of EDA and CTA across the potential energy surface (PES) is necessary for training EDA-based force-fields which can benefit from these decompositions for parametrizing force-field terms.
The limitation of perturbative CTA manifests in another form when considering different points on a PES.
Consider a simple one-dimensional rigid PES formed by stretching the \ce{BH3}--\ce{CO} complex along the \ce{B-C} bond with the geometry of \ce{BH3} and \ce{CO} fixed to be the same as their equilibrium geometries.
Fig.~\eqref{fig:PES_dep} shows the total percentage of CT decomposed into pairwise additive terms for both perturbative and non-perturbative schemes.
The total percentage of CT decomposed by the perturbative scheme varies along the potential energy surface.
When the \ce{B-C} bond is compressed to 1 \AA, 88\% of total CT is decomposed, while 
at equilibrium, this number falls to 77\% and saturates at 74\% as the bond is stretched.
By contrast, and by virtue of its design, the non-perturbative CTA does not have this dependence at all. 
%and consistently decomposes all of the change transfer energy into pairwise additive components at all points on the PES.
%The non-perturbative treatment is robust and can handle all magnitudes of charge transfer.
The charge transfer energy in Fig.~\eqref{fig:PES_dep} varies over two orders of magnitude from $-736.8$ kJ/mol at 1 \AA\ separation to $-1.5$ kJ/mol at 3.4 \AA.
%Non-perturbative charge decomposition scheme, by construction, can decompose the total charge transfer completely regardless of the magnitude of the total charge transfer.
From Fig.~\eqref{fig:PES_dep}, it is also interesting to note that the perturbative CTA yields components that deviate from the exact non-perturbative values by significantly different fractions across the PES: the perturbative \ce{CO}$\rightarrow$\ce{BH3} component deviates significantly from the exact fraction at long distances, while it agrees much better at the shortest distances.
%converges to around 65\% while its non-perturbative counterpart converges to 88\%.
On the other hand, the perturbative \ce{BH3}$\rightarrow$\ce{CO} fraction agrees well with the exact value at long-range, and deviates significantly at short-range.
%These examples stand testimony to the fact that the limitations of the perturbative treatment are not very systematic and a simple-minded scaling is insufficient to correct its deficiencies.
%The variation scheme introduced in this work does overcome all the limitations of the perturbative treatment and treats the pairwise decomposition in a more even-handed way.
The advantage of the new non-perturbative CTA for pairwise decomposition is quite clear.

\subsection{Basis set dependence}

\begin{figure}
    \centering
    \includegraphics[width=\textwidth]{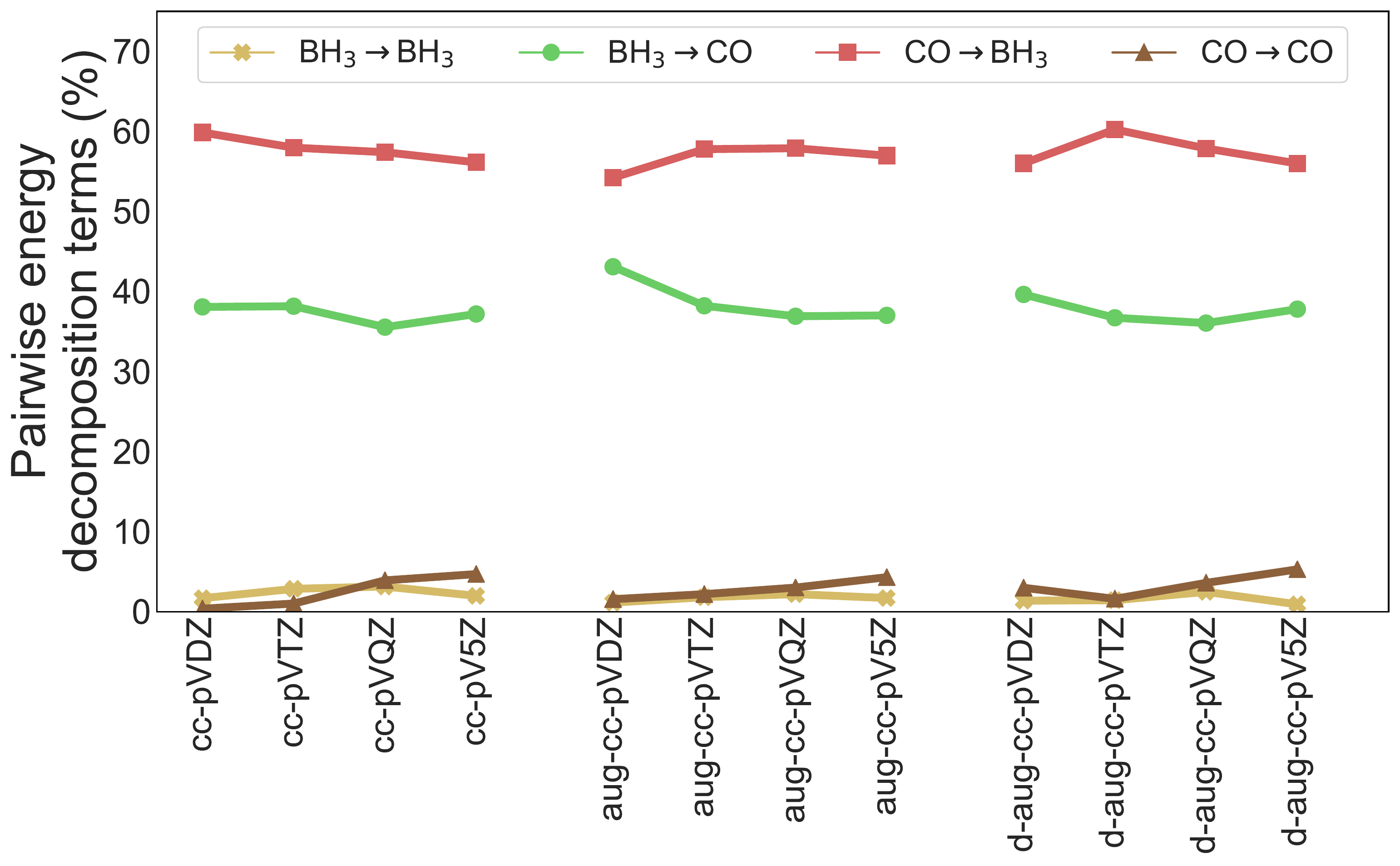}
    \caption{Convergence properties of the non-perturbative pairwise decomposition components of the total CT energy with respect to increasing the highest angular momentum of the basis set for the Dunning basis set sequence: cc-pVXZ, aug-cc-pVXZ, and d-aug-cc-pVXZ (X=D, T, Q, and 5) for the \ce{BH3}--\ce{CO} system at its equilibrium geometry using $\omega$B97X-D.}
    \label{fig:basis_lim_energy}
\end{figure}

\begin{figure}
    \centering
    \includegraphics[width=\textwidth]{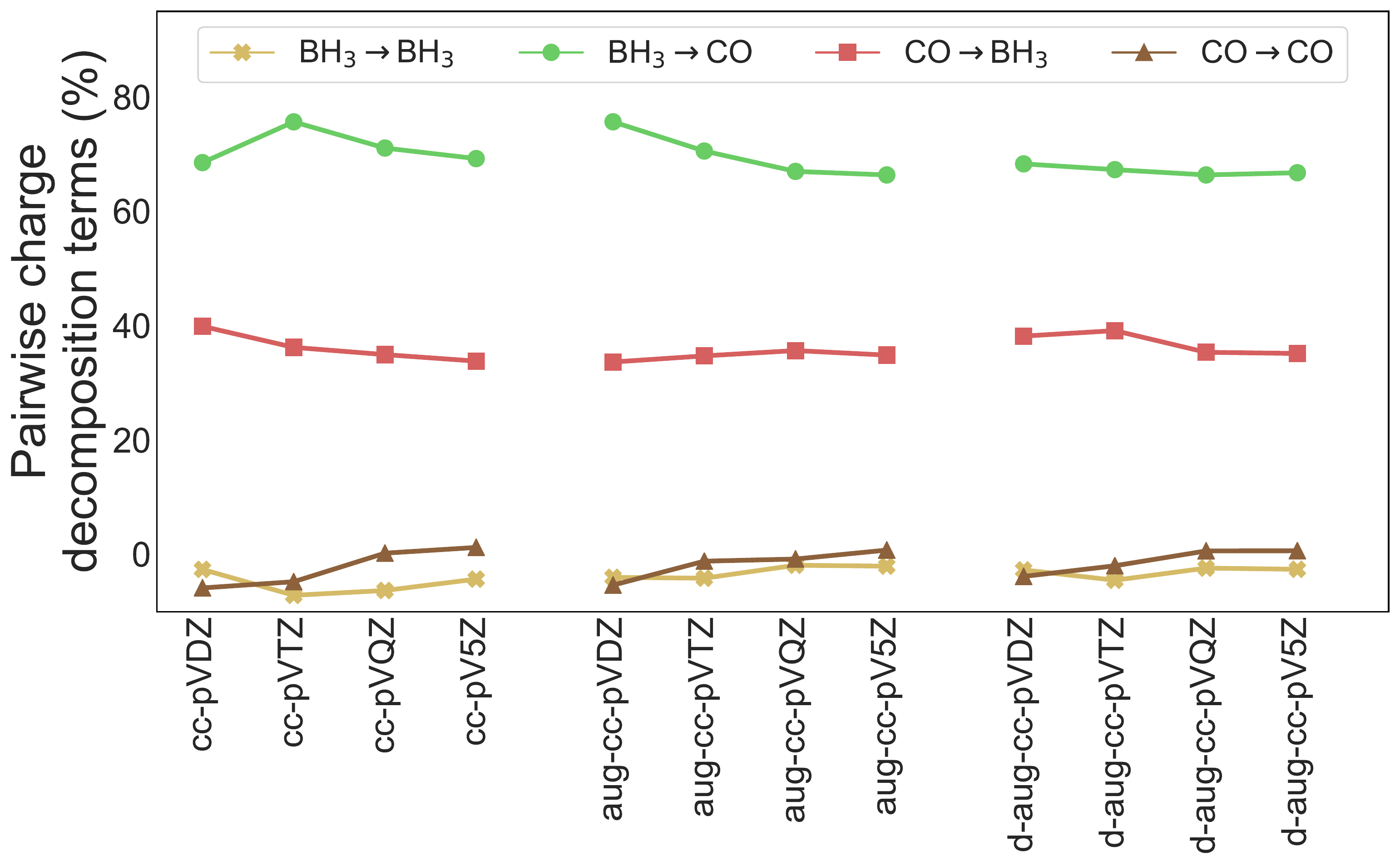}
    \caption{Convergence properties of the non-perturbative pairwise decomposition components of  total charge transfer with respect to increasing the highest angular momentum of the basis set for the Dunning basis set sequence: cc-pVXZ, aug-cc-pVXZ, and d-aug-cc-pVXZ (X=D, T, Q, and 5) for the \ce{BH3}--\ce{CO} system at equilibrium geometry using $\omega$B97X-D.}
    \label{fig:basis_lim_charge}
\end{figure}

As modern density functionals show their best performance when approaching the basis set limit, it is advisable to run all calculations with as large a basis set as computationally feasible~\cite{Mardirossian2017}.
Hence, one of the essential properties an energy decomposition scheme should have is a reasonably stable and physically meaningful basis set limit for all the contributing energy terms.
The Fragment Electric Response Functions (FERFs),\cite{Horn2015} which construct a polarization subspace for each fragment based on its response to electric fields, are used in the ALMO-EDA to give the polarization energy and CT energy well-defined basis set limits.
In this section, we assess the basis set convergence properties of each of the pairwise charge and energy transfer components by again considering the example of \ce{BH3}--\ce{CO} adduct at its equilibrium distance.
Fig.~\eqref{fig:basis_lim_energy} shows the partition of the CT energy into its 4 pairwise contributing terms as a function of increasing size of the AO basis.
Similarly, Fig.~\eqref{fig:basis_lim_charge} shows the corresponding partition of the pairwise decomposed charge flow terms.
The size of the basis set can be increased in two complementary ways: (1) Increasing the highest angular momentum (cardinal number, X) of the one-particle basis set (denoted by the DZ, TZ, QZ, and 5Z sequence) (2) Increasing the level of augmentation of diffuse basis functions (denoted by the augmented (aug-)and doubly-augmented (d-aug-) prefixes).
The latter effect is important for treating ions, excited states and strongly polarized systems with comparable accuracy to relatively non-polar ground states, but leads to near linear-dependence of the basis, which presents a challenge for methods such as CTA that use Hilbert space partitioning.
%Both of these techniques for increasing the size of one-particle basis set are important as they dictate the shape and size of the molecular orbitals that can be formed, consequently deciding the level of precision with which electron density can be represented.


Figures~\eqref{fig:basis_lim_energy} and \eqref{fig:basis_lim_charge} show that the CT partition into  four components is well-behaved with respect to increasing X and augmentation.
Considering the two major components of the CT energy, we can see that the \ce{BH3}$\rightarrow$\ce{CO} component converges to 37\% and the \ce{CO}$\rightarrow$\ce{BH3} converges to 56\%.
Similarly, the \ce{BH3}$\rightarrow$\ce{CO} charge flow component converges to 67\% and the \ce{CO}$\rightarrow$\ce{BH3} component converges to 35\%.
The double-$\zeta$ basis sets, contaminated with basis set superposition error (BSSE) as well as incompleteness error, are too small to reliably capture either the CT energy or charge flow.
If the DZ basis sets are excluded, the two major components are almost converged at the triple-$\zeta$ basis set level, while the repolarization (on-fragment) components contribute little to CT regardless of  basis set.

An interesting aspect of CT analysis for \ce{BH3}$-$\ce{CO} is the fact that the pairwise energy and charge components are not correlated.
The \ce{CO}$\rightarrow$\ce{BH3} energy component is larger than the \ce{BH3}$\rightarrow$\ce{CO} CT energy component, while the opposite is true for the pairwise charge flow components.
This result is consistent with perturbative analysis presented earlier.\cite{Khaliullin2008} The underlying reason has its origin in the different quantities being decomposed: the CT energy versus the charge flow associated with CT. Let us denote $V_{\text{DA}}$ as the matrix element coupling donor (D) and acceptor (A) orbitals, with orbital energies $\epsilon_{\text{D}}$ and $\epsilon_{\text{A}}$. From dimensionality (or perturbation theory) arguments, the CT energy is $\sim V_{\text{DA}}^2/(\epsilon_{\text{D}} - \epsilon_{\text{A}})$. On the other hand, the charge flow behaves as $\sim V_{\text{DA}}/(\epsilon_{\text{D}} - \epsilon_{\text{A}})$.
%We infer that a smaller gap between energy levels controls the dominance of \ce{BH3}$\rightarrow$\ce{CO} back-donation in charge flow, while a larger coupling between levels controls the dominance of \ce{CO}$\rightarrow$\ce{BH3} forward donation in the CT energy.
Figures showing the magnitude of these pairwise decomposed components in kJ/mol and me\textsuperscript{-} are included in the Supporting Information.

\subsection{DNA base-pairs}

\begin{figure}
    \centering
    \includegraphics[width=\textwidth]{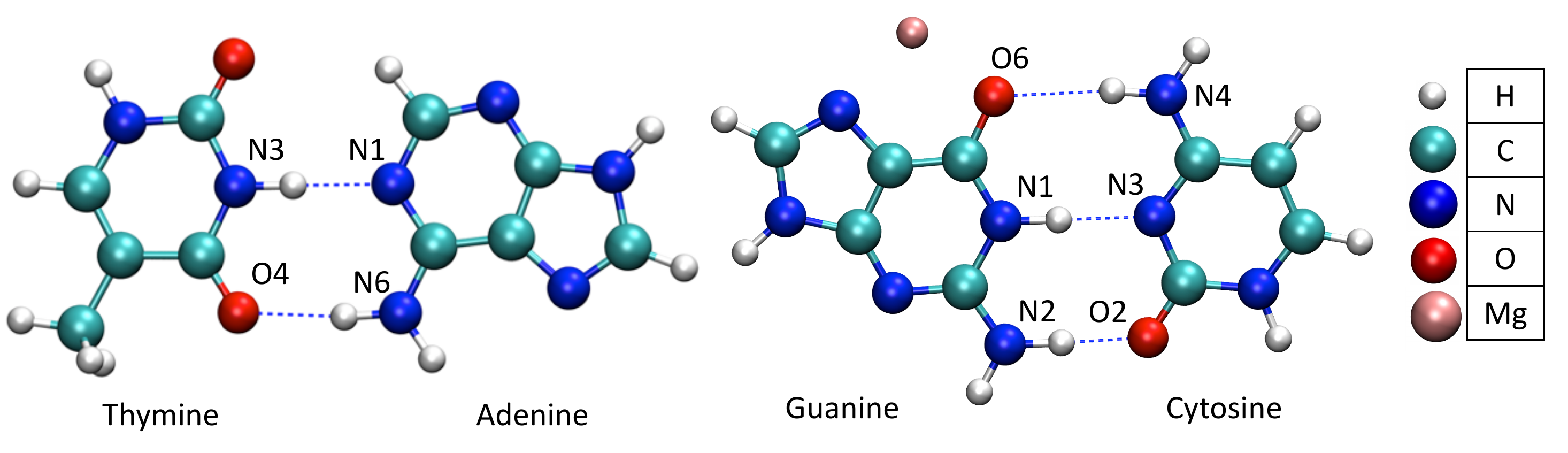}
    \caption{DNA base pairs adenine-thymine (A-T; left panel) and guanine-cytosine (G-C) with \ce{Mg^2+} (right panel)}
    \label{fig:DNA}
\end{figure}

Hydrogen bonding between DNA bases is one of the most important non-covalent interactions as it modulates a myriad of biological phenomena, such as the melting temperature of oligonucleotide sequences which is a critical parameter in molecular biology experiments.\cite{Owczarzy2008predicting}
The Watson-Crick base pairs, adenine-thymine (A-T) and guanine-cytosine (G-C), shown in Fig.~\eqref{fig:DNA}, interact by characteristic hydrogen bonding.
The A-T base pair is bound by two hydrogen bonds: one from N3 of thymine to N1 of adenine, and the other from N6 of adenine to O4 of thymine.
The G-C binding energy is $-120.8$ kJ/mol which is much larger than the A-T binding energy of $-69.4$ kJ/mol as G-C has three hydrogen bonds (two H-bond donors on G and one donor on C) while A-T has only two. 

\begin{figure}
     \centering
     \begin{subfigure}[b]{0.49\textwidth}
         \centering
         \includegraphics[width=\textwidth]{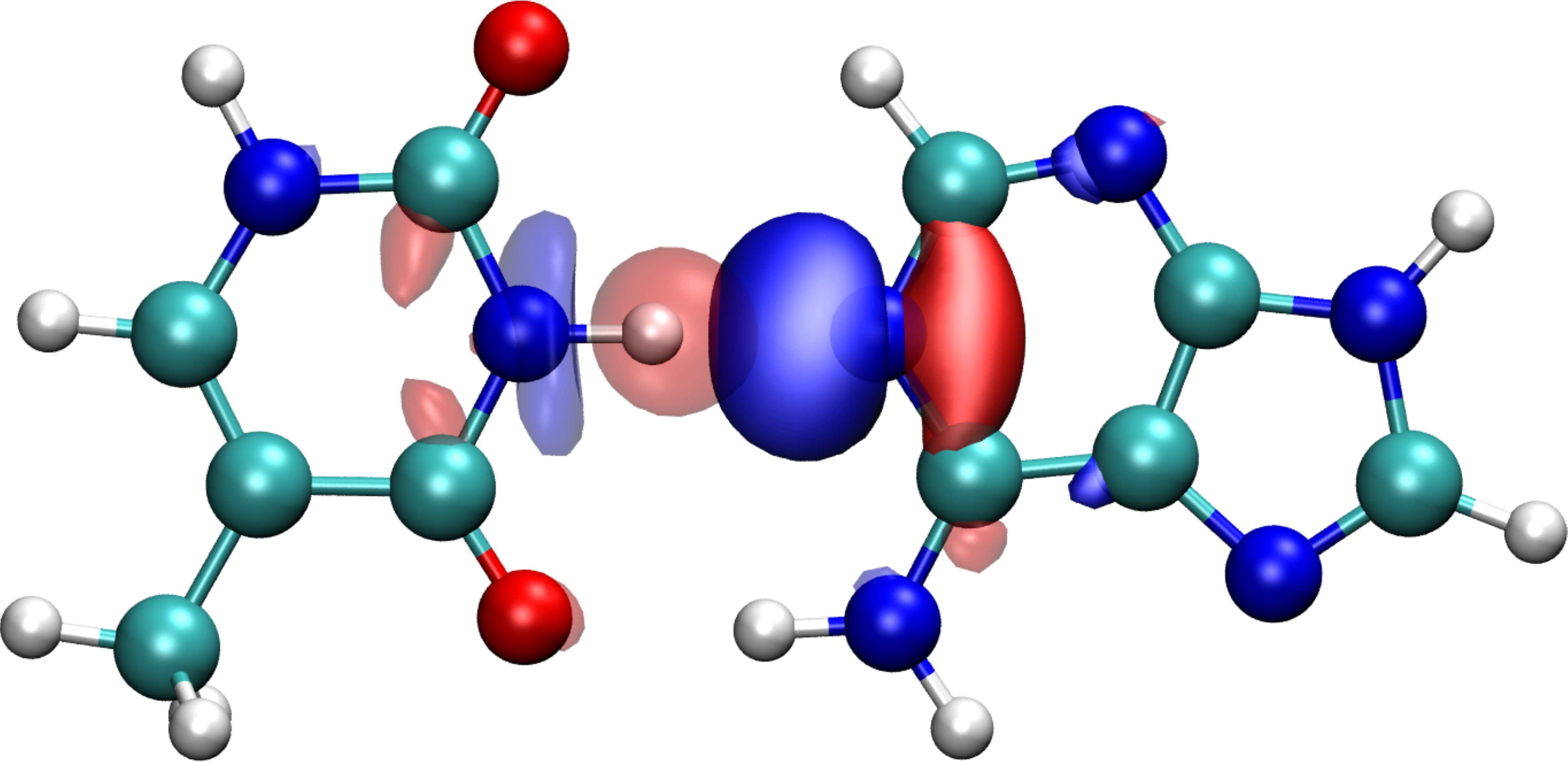}
         \caption{$\Delta E_{\text{CT}}^{\text{COVP1}}=-23.9$ kJ/mol}
         \label{fig:DNA_AT_covp1}
     \end{subfigure}
     \begin{subfigure}[b]{0.49\textwidth}
         \centering
         \includegraphics[width=\textwidth]{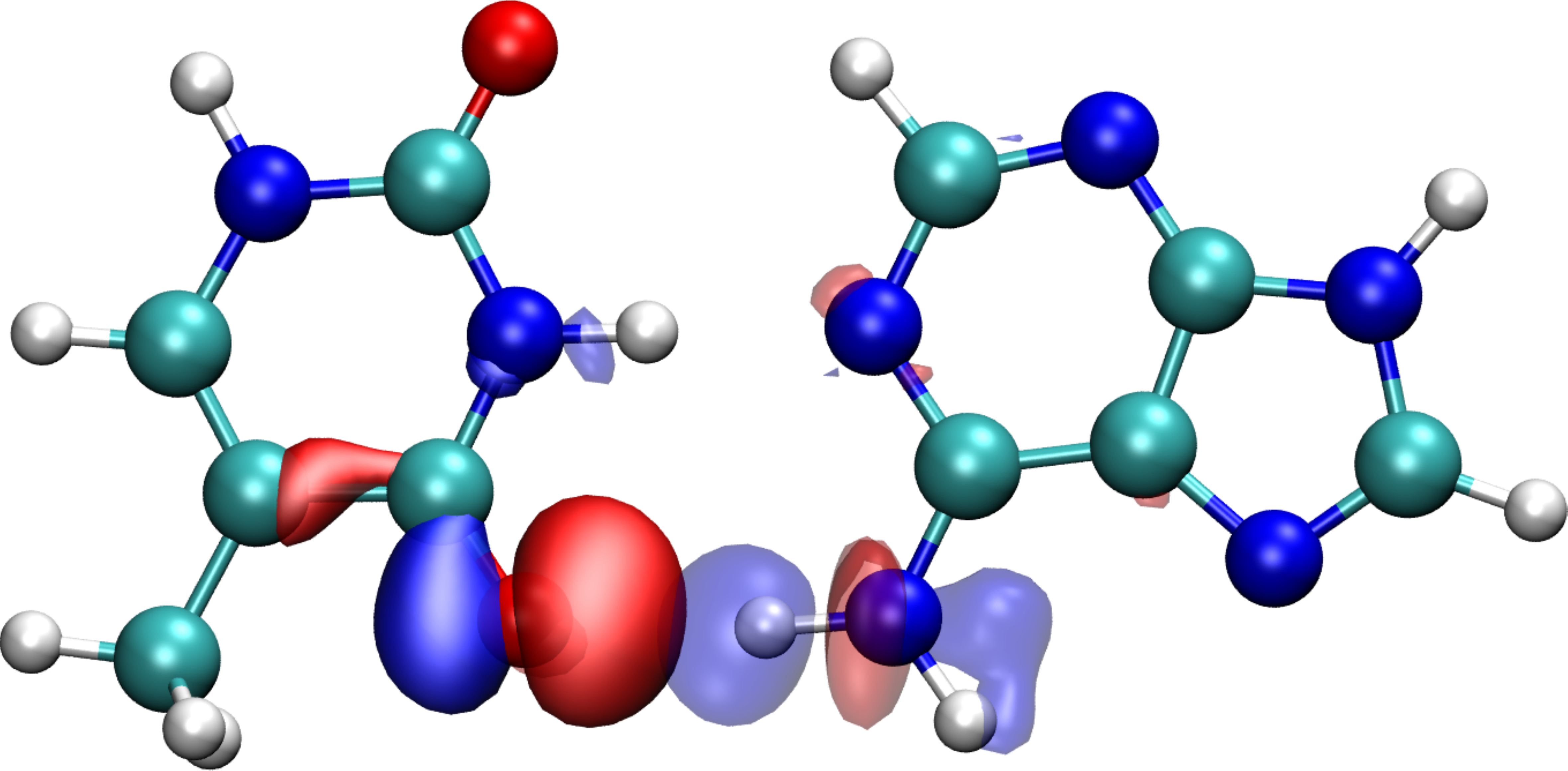}
         \caption{$\Delta E_{\text{CT}}^{\text{COVP2}}=-8.7$ kJ/mol}
         \label{fig:DNA_AT_covp2}
     \end{subfigure}
        \caption{(a) The most significant Complementary Occupied-Virtual Pair (COVP) for the A-T DNA base pair (b) Second most significant COVP for the A-T DNA base pair. Atom color codes are shown in Fig.~\eqref{fig:DNA}.}
        \label{fig:DNA_AT_covp}
\end{figure}

The pairwise CTA for these base pairs is shown in Table~\eqref{tab:DNA}. From a CTA perspective, the G-C pair really has two N lone pair donors on cytosine, coupled to two charge accepting \ce{H-N} groups on guanine, rather than the opposite proton donor perspective. This is reflected in the larger C$\rightarrow$G term versus the smaller G$\rightarrow$C contribution.
The main Complementary Occupied-Virtual Pairs (COVPs) shown in Fig.~\eqref{fig:DNA_AT_covp} for the A-T pair are a very convenient way to understand these hydrogen bonds.
The most significant COVP of the A-T base pair contributes a majority (about 67\%) of the CT interactions.
The COVP donor is located on N1 of adenine and the COVP acceptor is located on N3\ce{-H} of thymine.
The second hydrogen bond has much weaker CT, which accounts for the remaining CT stabilization.

\begin{table}[]
\caption{Energy decomposition analysis and non-perturbative charge decomposition analysis (in kJ/mol) for the DNA base pairs thymine(T)-adenine(A), guanine(G)-cytosine(C), and their corresponding metallated versions.}
\label{tab:DNA}
\resizebox{\textwidth}{!}{%
\begin{tabular}{c|p{1.5cm}p{1.5cm}p{1.5cm}p{1.5cm}p{1.5cm}|p{2.0cm}p{2.0cm}p{2.0cm}p{2.0cm}}
\hline
                            & \multicolumn{5}{c}{Energy decomposition   analysis} & \multicolumn{4}{|c}{Non-perturbative decomposition of CT energy} \\ \hline
\multicolumn{1}{c}{} & \multicolumn{1}{|c}{$\Delta E_{\text{INT}}$} & \multicolumn{1}{c}{$\Delta E_{\text{GD}}$} & \multicolumn{1}{c}{$\Delta E_{\text{FRZ}}$} & \multicolumn{1}{c}{$\Delta E_{\text{POL}}$} & \multicolumn{1}{c|}{$\Delta E_{\text{CT}}$} & 1$\rightarrow$1        & 1$\rightarrow$2         & 2$\rightarrow$1         & 2$\rightarrow$2       \\ \hline
T(1):A(2)       & -63.8      & 5.6     & -7.4     & -26.2    & -35.8    & 0.0         & -10.7        & -25.4        & 0.3        \\
G(1):C(2)      & -120.8      & 13.5     & -35.6    & -52.2    & -46.5    & 0.0         & -19.0        & -27.6        & 0.2        \\
\ce{Na+} G(1):C(2)   & -138.2      & 11.0     & -45.4    & -55.8    & -48.1    & -0.2        & -10.2        & -38.0        & 0.4        \\
\ce{Mg^2+} G(1):C(2) & -195.7      & 18.5     & -56.3    & -90.9    & -67.1    & -0.3        & -5.3         & -62.2        & 0.7        \\
\ce{Ca^2+} G(1):C(2) & -178.9      & 16.3    & -54.6    & -80.9    & -59.7    & -0.3        & -5.4         & -54.6        & 0.6   \\ \hline    
\end{tabular}%
}
\end{table}

\begin{figure}
     \centering
     \begin{subfigure}[b]{0.49\textwidth}
         \centering
         \includegraphics[width=\textwidth]{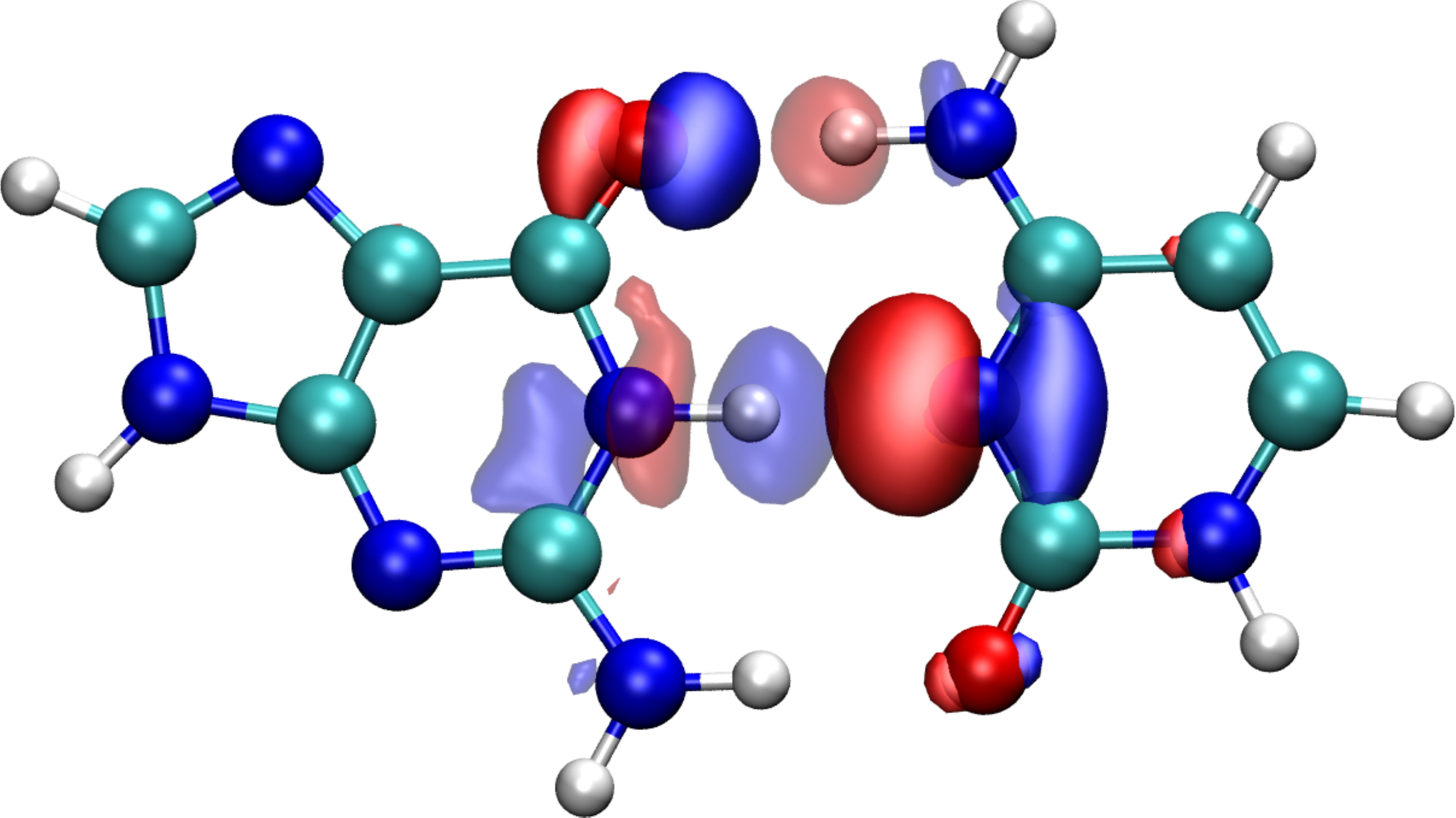}
         \caption{$\Delta E_{\text{CT}}^{\text{COVP1}}=-16.7$ kJ/mol}
         \label{fig:DNA_GC_covp1}
     \end{subfigure}
     \begin{subfigure}[b]{0.49\textwidth}
         \centering
         \includegraphics[width=\textwidth]{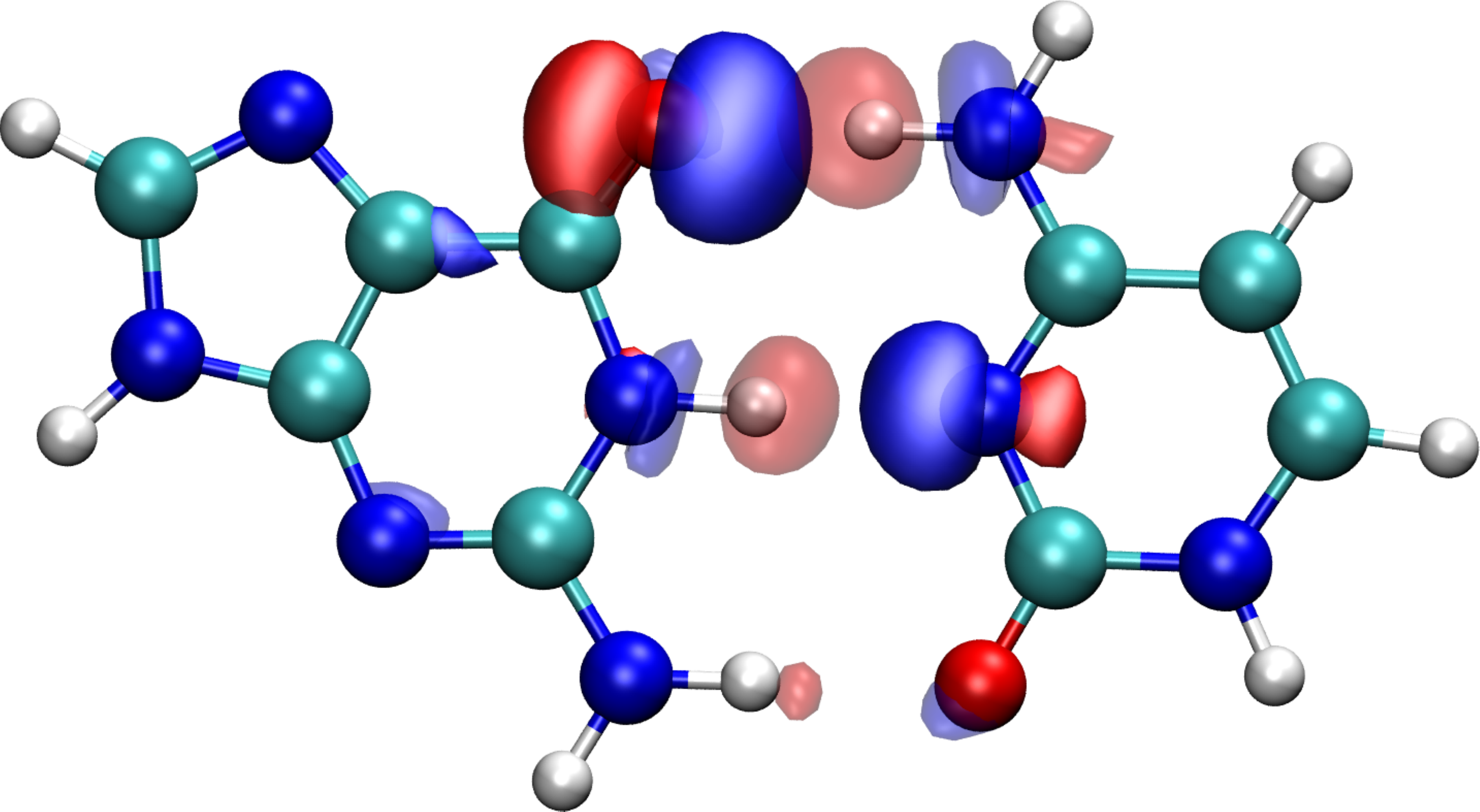}
         \caption{$\Delta E_{\text{CT}}^{\text{COVP2}}=-16.6$ kJ/mol}
         \label{fig:DNA_GC_covp2}
     \end{subfigure}
     \begin{subfigure}[b]{0.49\textwidth}
         \centering
         \includegraphics[width=\textwidth]{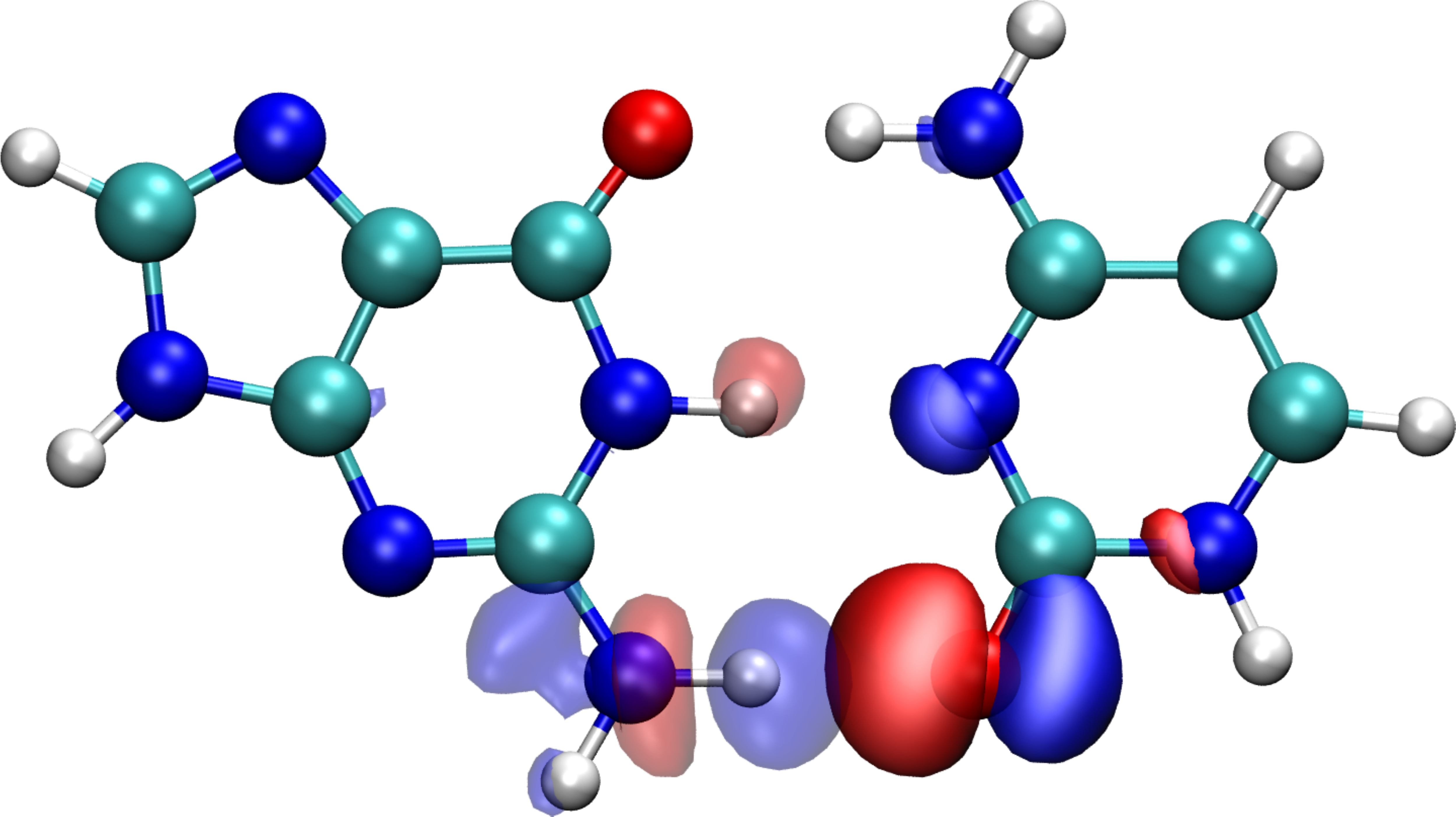}
         \caption{$\Delta E_{\text{CT}}^{\text{COVP3}}=-9.3$ kJ/mol}
         \label{fig:DNA_GC_covp3}
     \end{subfigure}
        \caption{(a) The most significant Complementary Occupied-Virtual Pair (COVP) for the G-C DNA base pair  (b) Second most significant COVP for the G-C DNA base pair (c) Third most significant COVP for the G-C DNA base pair. Atom color codes are shown in Fig.~\eqref{fig:DNA}.}
        \label{fig:DNA_GC_covp}
\end{figure}

By contrast, as shown in Fig.~\eqref{fig:DNA_GC_covp}, the G-C base pair with its three hydrogen bonds contains two equally significant COVPs, each of which contributes $-16.6$ kJ/mol to the CT energy.
Unlike the A-T COVPs, these G-C COVPs do not completely localize on any one particular hydrogen bond although the first pair (shown in Fig.~\eqref{fig:DNA_GC_covp1}) is mainly on the O6-N4 hydrogen bond, while the second pair (shown in Fig.~\eqref{fig:DNA_GC_covp2}) dominates the N3\ce{-N}1 interaction. For two almost degenerate COVPs such as this, it is possible to localize the two occupied and two virtual orbitals if a more localized picture is desired.

The presence of metal cations has a significant effect on biophysical processes such as stabilization of DNA triple and quadruple helices.
At low concentrations, metal cations produce a stabilizing effect by neutralizing the negatively charged phosphate backbone.
However, at high concentrations they affect the structural integrity of DNA by disrupting the hydrogen bonding interactions.\cite{Lippard1994principles, Kaim2013bioinorganic}
Metal cations interact with the N7 position of guanine as shown in the right panel of Fig.~\eqref{fig:DNA}.\cite{Sigel1993interactions, Lippert2000multiplicity, Valls2004cubic}
An understanding of the effect of metal cation coordination on the binding energy of DNA base pairs will help illuminate whether this 3-body interaction affects the stability of the DNA duplex.
The effect of metal cation coordination of the strength of hydrogen bonds in DNA base pairs has been studied using Natural Bond Orbital Analysis.\cite{Stasyuk2020effect}

As a primitive model for this effect, we consider guanine of a G-C dimer binding in a bidentate fashion via its N7 and O6 sites to three metal cations (\ce{Na+}, \ce{Mg^2+}, and \ce{Ca^2+}), as shown for \ce{Mg^2+} in the right panel of  Fig.~\eqref{fig:DNA}. 
The EDA and non-perturbative CTA of the three metallated G-C complexes are shown in Table~\eqref{tab:DNA}.
All 3 metallations of guanine increase the binding energy of the complex.
This increment is small ($17.4$ kJ/mol) when the metal cation is \ce{Na+}, but quite large when the metal cation is \ce{Mg^2+} or \ce{Ca^2+}.
The EDA reveals that this increment comes from increases in all major mechanisms of interaction: frozen interactions, polarization, and charge transfer.
The adiabatic EDA\cite{Mao2017} (see Table S1) confirms that the increase in frozen and polarization interactions is larger than the increase in CT.
The predominance of permanent and induced electrostatics agrees with chemical understanding of this toy model of metallated DNA base pairs, which is controlled by interactions with the unscreened charge.
%The adiabatic EDA calculations are included in the \textcolor{red}{supplementary information}.

Considering the case of \ce{Mg^2+}-G-C versus G-C, frozen interactions, polarization, and CT contribute an additional 20.6 kJ/mol, $38.7$ kJ/mol, and $20.6$ kJ/mol respectively to the interaction energy.
Applying the non-perturbative CTA to the charge transfer contribution reveals an interesting pattern.
Metallation of guanine increases the C$\rightarrow$G CT energy while it decreases the G$\rightarrow$C CT.
This serves to decrease the strength of the O6(G)$\cdots$\ce{H}-N4(C) hydrogen bond and increase the strength of the N3(C)$\cdots$\ce{H}-N1(G) and O2(C)$\cdots$\ce{H}-N2(G) hydrogen bonds.
The increase in the strength of the latter CT interactions is larger than the decrease of the former, thereby leading to an overall increase in CT upon metallation.
%Another aspect to note is that divalent metal cations (\ce{Mg^2+} and \ce{Ca^2+})  increase the strength of the interactions to a much larger extent than the monovalent \ce{Na+}.
COVPs for the metallated G-C base pairs are included in the Supporting Information (See Figures S3, S4, and S5). 
The significant COVPs are localized on the N3-N1 and O2-N2 hydrogen bonds, while
the COVP localized on the O6-N4 hydrogen bond contributes little to the total CT energy.

These aspects of hydrogen bonding in nucleobases have been previously studied using delocalization indices from Atom-In-Molecule (AIM) theory.\cite{Poater2005hydrogen}
While it is difficult to compare our method with AIM theory, both analysis methods agree on the selective strengthening and weakening of hydrogen bonds upon metallation in the G-C base pair.
While AIM theory measures this change in terms of change in delocalization index, our analysis can directly compute an energy value associated with these interactions, thus enabling a richer and more direct comparison to the interaction energy.

\subsection{Borane-amine adducts}
Borane-amine adducts are textbook examples of Lewis acid-base pairs.
The ammonia-borane complex has been studied particularly in detail as it has been considered a promising hydrogen storage material that contains 19.6 wt\% of hydrogen.\cite{Marder2007will,Peng2008ammonia}
This adduct consists of an electron-deficient group 13 center and an electron-rich group 15 center.
Traditionally, bonding in this adduct has been understood as a dative bond as a result of donation of an electron pair from the nitrogen to the boron center. This is supported by the fact that borane, which is a planar molecule, pyramidalizes upon complexation with ammonia.

% Please add the following required packages to your document preamble:
% \usepackage{graphicx}
\begin{table}[]
\caption{EDA and non-perturbative CT energy decomposition analysis (in kJ/mol) for the series of adducts \ce{BX3}--\ce{NH3} where X~=~F, Cl, or Br.}
\label{tab:BX3_NH3}
\resizebox{\textwidth}{!}{%
\begin{tabular}{cccccc|cccc}
\hline
\multicolumn{1}{c}{} & \multicolumn{5}{c}{Energy Decomposition   Analysis}                                                                                     & \multicolumn{4}{|c}{Non-perturbative   decomposition of CT energy}                                                                                                                                                           \\ \hline
\multicolumn{1}{c}{} & \multicolumn{1}{c}{$\Delta E_{\text{INT}}$} & \multicolumn{1}{c}{$\Delta E_{\text{GD}}$} & \multicolumn{1}{c}{$\Delta E_{\text{FRZ}}$} & \multicolumn{1}{c}{$\Delta E_{\text{POL}}$} & \multicolumn{1}{c|}{$\Delta E_{\text{CT}}$} & \multicolumn{1}{c}{\ce{BX3}$\rightarrow$\ce{BX3}} & \multicolumn{1}{c}{\ce{BX3}$\rightarrow$\ce{NH3}} & \multicolumn{1}{c}{\ce{NH3}$\rightarrow$\ce{BX3}} & \multicolumn{1}{c}{\ce{NH3}$\rightarrow$\ce{NH3}} \\ \hline
\ce{BF3}--\ce{NH3}    & -90.8                          & 99.7                    & 122.8                    & -157.4                   & -156.0                  & 0.0                                                & -5.1                                               & -151.2                                             & 0.2                                                \\
\ce{BCl3}--\ce{NH3}   & -107.7                         & 101.2                   & 259.7                    & -282.2                   & -186.4                  & 0.2                                                & -9.8                                               & -179.3                                             & 2.6                                                \\
\ce{BBr3}--\ce{NH3}   & -119.2                         & 94.3                    & 295.5                    & -322.6                   & -186.5                  & 0.2                                                & -11.6                                              & -177.8                                             & 2.7                                            \\ \hline   
\end{tabular}%
}
\end{table}

Understanding the nature and strength of the dative bond in these adducts is key to tuning the strength of such dative interactions, and can potentially be used to design ligands on a catalyst or engineer protein-drug interactions.
One common example is the series of halogenated boranes binding ammonia: \ce{BX3}--\ce{NH3} where X~=~\ce{F}, \ce{Cl}, or \ce{Br}.
The order of stability of these adducts is \ce{BBr3}--\ce{NH3} $>$ \ce{BCl3}--\ce{NH3} $>$ \ce{BF3}--\ce{NH3} as shown in Table~\eqref{tab:BX3_NH3}.
This ordering is rather counter-intuitive as one would naively expect the reverse ordering consistent with ordering of electronegativity of the halogens (F $>$ Cl $>$ Br).
%Multiple theories have been proposed to explain this ordering by invoking different chemical phenomenon.
One explanation proposed for this ordering is that electron donation from the halogen to the empty p-orbital of boron reduces the Lewis acidity of boron.\cite{Hirao1999lewis}
As shape and sizes of B and F match better than B and Cl, it was suggested that donation from F to B was stronger.
While this explanation suggests that \ce{BF3} has a stronger $\pi$-bond character than \ce{BCl3}, and consequently should have a higher energy of pyramidalization, the opposite has been found to be true (See Ref.~\citenum{Hirao1999lewis} and $\Delta E_{\text{GD}}$ in Table~\eqref{tab:BX3_NH3}).
It was also found that the Natural Bond Orbital $\pi$-overlap between F and B in \ce{BF3} and Cl and B in \ce{BCl3} are more or less identical.\cite{Plumley2009periodic}
Another explanation correlating the Lewis acidity of halogenated boranes with its LUMO level was proposed\cite{Branchadell1991lewis, Bessac2003bcl3} and then contradicted.\cite{Plumley2009periodic}

In this section, we attempt to explain the trend in the strength of interaction of trihalo-borane-ammonia adducts.
The interaction energy increases upon going from \ce{BF3}--\ce{NH3} to \ce{BBr3}--\ce{NH3} from $-90.8$ to $-119.2$ kJ/mol.
Upon going down the periodic table from \ce{BF3}--\ce{NH3} to \ce{BCl3}--\ce{NH3}, the polarization and charge transfer components of the interaction energy increase (Table~\eqref{tab:BX3_NH3}).
While the increase in polarization is cancelled out by the increase in the repulsive frozen interactions, charge transfer causes a true increase in the strength of interaction.
Upon decomposition of CT into pairwise additive terms, it is clear that the increase in the \ce{NH3}$\rightarrow$\ce{BX3} component is the most significant and leads to enhanced binding energy.
On the other hand, upon going from \ce{BCl3}--\ce{NH3} to \ce{BBr3}--\ce{NH3}, total CT and the \ce{NH3}$\rightarrow$\ce{BX3} component remain unchanged.
Most of the increase in binding energy arises from a decrease in the geometry distortion term.
While most of the increase in the polarization energy is offset by the increase in repulsive frozen interactions, the increase in the polarization term still contributes a little to the overall increase in the binding energy.
Energy decomposition in an adiabatic picture (see Table S3) further emphasizes the importance of CT relative to polarization and frozen interactions.
The adiabatic energy decomposition also shows an increase in CT upon going from \ce{BF3}--\ce{NH3} to \ce{BBr3}--\ce{NH3}.

\begin{table}[]
\caption{Energy decomposition analysis and non-perturbative charge decomposition analysis (in kJ/mol) for the adduct \ce{BH3}--\ce{NMe_pH_q} ($p+q=3$)} 
\label{tab:BH3_NMepHq}
\resizebox{\textwidth}{!}{%
\begin{tabular}{cccccc|cccc}
\hline
                     & \multicolumn{5}{c|}{Energy Decomposition   Analysis}       & \multicolumn{4}{c}{Non-perturbative Decomposition of CT energy}                                                                                    \\ \hline
                     & $\Delta E_{\text{INT}}$  & $\Delta E_{\text{GD}}$         & $\Delta E_{\text{FRZ}}$      & $\Delta E_{\text{POL}}$       & $\Delta E_{\text{CT}}$       & \ce{BH3}$\rightarrow$\ce{BH3} & \ce{BH3}$\rightarrow$\ce{NMe_pH_q} & \ce{NMe_pH_q}$\rightarrow$\ce{BH3} & \ce{NMe_pH_q}$\rightarrow$\ce{NMe_pH_q} \\ \hline
\ce{BH3}--\ce{NH3}   & -133.1     & 56.3       & 116.5    & -148.1    & -157.7   & -2.7                             & -16.4                            & -138.1                           & -0.5                             \\
\ce{BH3}--\ce{NMeH2} & -153.2     & 59.7       & 114.0    & -172.9    & -154.0   & -3.4                             & -17.4                            & -131.4                           & -1.9                             \\
\ce{BH3}--\ce{NMe2H} & -163.1     & 62.8       & 110.7    & -185.9    & -150.6   & -3.7                             & -17.8                            & -127.7                           & -1.4                             \\
\ce{BH3}--\ce{NMe3}  & -163.8 & 65.8 & 107.7 & -178.4 & -159.0 & -3.6                             & -18.1                            & -136.5                           & -0.8     \\ \hline                       
\end{tabular}%
}
\end{table}
Another interesting observation is that placing electron donating groups on B decreases the binding energy of the adduct, while placing the same group on N increases the binding strength of the adduct.
This phenomenon can be classically understood as the enhancement of Lewis basicity when electron donating groups are placed on nitrogen.
In order to gain a further understanding of this phenomenon, we consider a series of boranes bound to ammonia substituted with increasing numbers of methyl (\ce{-Me}) groups.
Methyl is an electron donating group, and with addition of each methyl group on the nitrogen center the interaction energy increases as shown in Table~\eqref{tab:BH3_NMepHq}.
Experimentally, gas phase dissociation enthalpies also increase.\cite{Staubitz2010amine}
Performing EDA on the interaction energies of this series, we can see that the total CT energy shows no trend with increasing the number of methyl groups on ammonia. 
EDA in the adiabatic framework is also in agreement with this although the relative magnitude of CT is much larger than the polarization (See Table S4).
This shows that understanding the enhancement in binding energy of these adducts in terms of increased Lewis basicity of substituted ammonia is incorrect.
The CTA further supports this claim by not showing a clear trend in the CT energy associated with donation from \ce{NMe_pH_q} to \ce{BH3}, which lies in the range of $-138.1$ to $-127.7$ kJ/mol, and does not become stronger with increasing number of methyl substituents at N.
By contrast, the polarization energy increases with increasing number of methyl substituents and is thus the main origin of enhanced binding. The reason is likely because \ce{-Me} is more polarizable than \ce{-H}, and consequently more \ce{-Me} groups cause larger polarization interactions.
%Our ALMO-based EDA method combined with the non-perturbative charge decomposition scheme reveals that polarization causes enhanced binding upon increasing the number of methyl groups on ammonia in borane-ammonia adducts.

\begin{figure}
     \centering
     \begin{subfigure}[b]{0.49\textwidth}
         \centering
         \includegraphics[width=\textwidth]{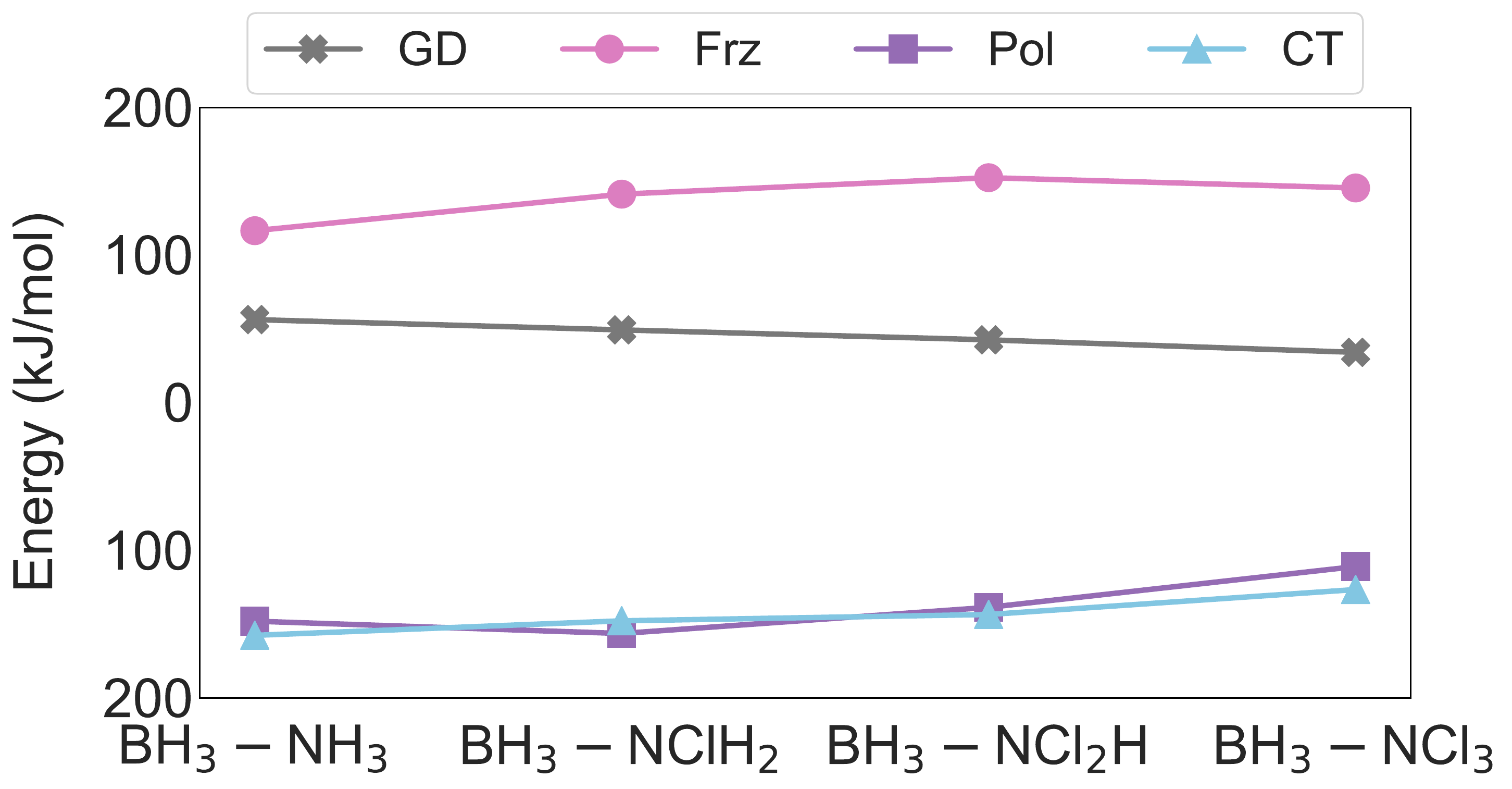}
         \caption{}
         \label{fig:BH3_NFpHq_EDA}
     \end{subfigure}
     \begin{subfigure}[b]{0.49\textwidth}
         \centering
         \includegraphics[width=\textwidth]{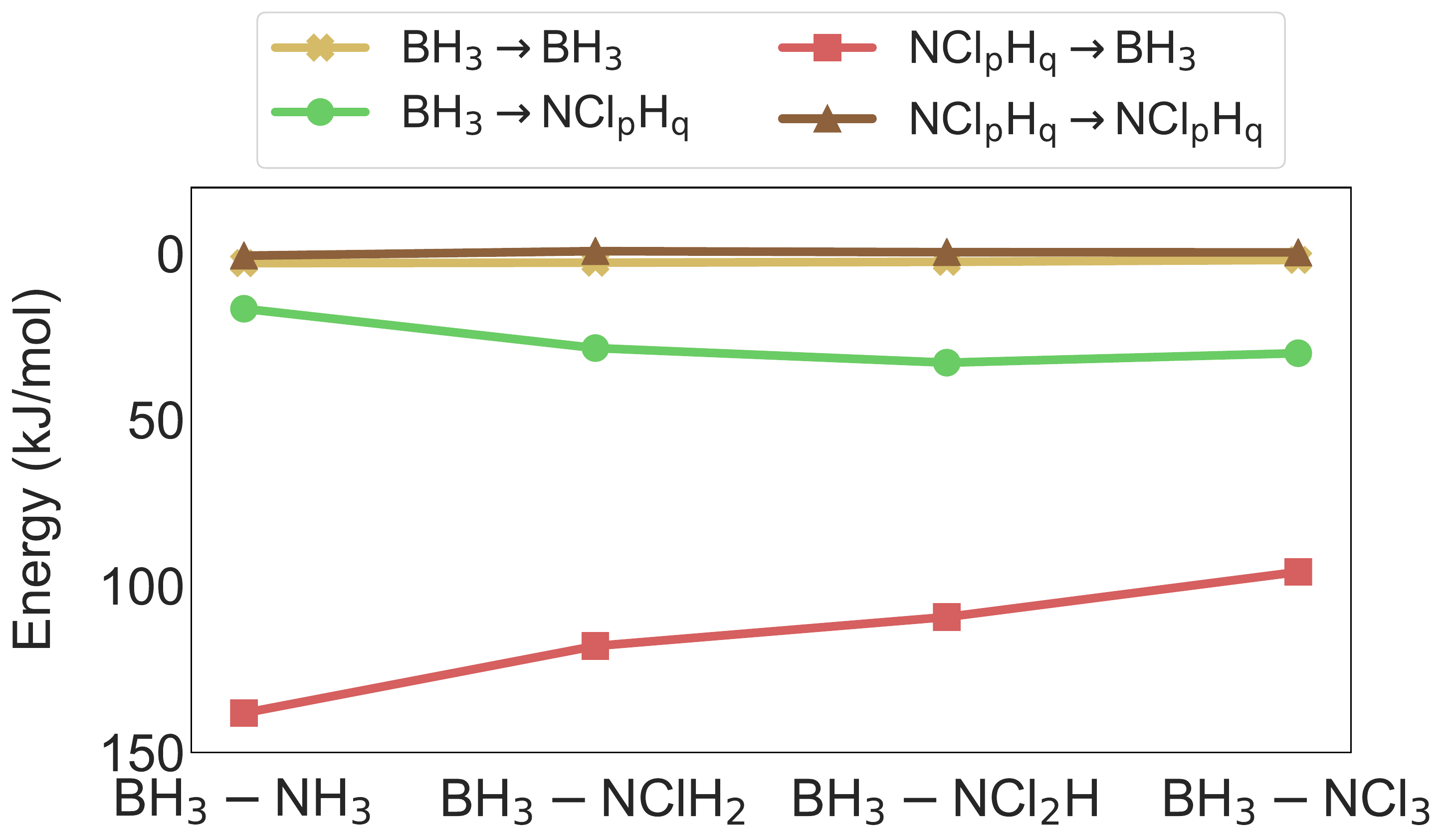}
         \caption{}
         \label{fig:BH3_NFpHq_VCT}
     \end{subfigure}
        \caption{(a) Energy decomposition of \ce{BH3}--\ce{NCl_pH_q} binding energy into geometry distortion (GD), frozen (Frz), polarization (Pol), and charge transfer (CT) terms. (b) Non-perturbative decomposition of charge transfer into pairwise additive terms. A table containing the energetics shown in this figure is included in the Supporting Information (Table S2).}
        \label{fig:BH3_NFpHq}
\end{figure}

Now, let us consider a chloride (\ce{-Cl}) substitution at N instead of \ce{-Me} substitution.
The chloride, being an electron withdrawing group, reduces the electron density available for donation in forming an adduct.
As the number of chloride groups on N increases, the binding energy of the adduct decreases from $-133.1$ kJ/mol in the case of \ce{BH3}--\ce{NH3} adduct to just $-58.2$ kJ/mol for \ce{BH3}--\ce{NCl3}.
EDA shown in Fig.~\eqref{fig:BH3_NFpHq_EDA} shows that this is a result of the decrease in both polarization and charge transfer.
The non-perturbative CTA reveals two important trends upon chloride substitution: First, as expected, \ce{NCl_pH_q}$\rightarrow$\ce{BH3} CT decreases significantly from $-138.1$ kJ/mol to $-95.6$ kJ/mol on going from \ce{NH3} to \ce{NCl3}.
Second, \ce{BH3}$\rightarrow$\ce{NCl_pH_q} CT energy increases upon chloride substitution.
However, the latter increase is rather small and is eclipsed by the former leading to an overall decrease in the CT energy lowering.
Thus, the non-perturbative CTA can elucidate the interplay of different pairwise CT energy components in order to explain the observed binding energies of borane-amine adducts. 

\subsection{Carbonyl complexes}

Transition metal -- carbonyl interactions are ubiquitous in transition metal chemistry and catalysis,\cite{Xu2000transition, Sunley2000high}
such as in important intermediates in \ce{CO2} reduction. \cite{Loipersberger2020computational}
These are cases of synergic bonding with significant binding arising from both forward electron donation (carbonyl to metal complex) and backward (metal complex to carbonyl) donation.
The Dewar-Chatt-Duncanson\cite{Dewar1951review,Chatt1953586} model has been invoked to explain bonding in metal carbonyls with differing degrees of success.\cite{Rossomme2020electronic}
According to this model, the red shift in the carbonyl stretching frequency is understood as a consequence of back-donation from the metal to the anti-bonding 2$\pi^*$ orbitals of \ce{CO}, decreasing its bond strength, lengthening the \ce{C-O} bond, and consequently decreasing its vibrational stretching frequency. 

\begin{figure}
    \centering
    \includegraphics[width=\textwidth]{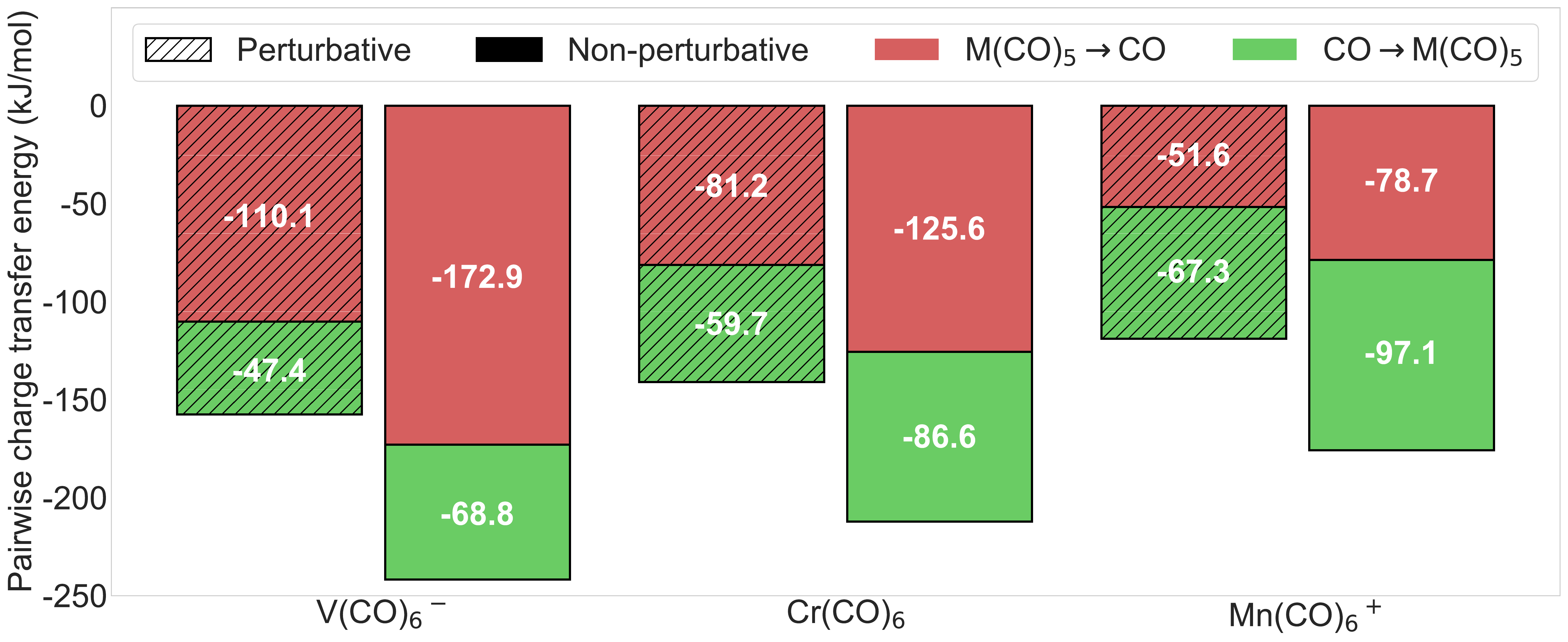}
    \caption{Stabilization energy associated with forward (\ce{CO}$\rightarrow$\ce{M(CO)5}) and backward (\ce{M(CO)5}$\rightarrow$\ce{CO}) CT interactions in the perturbative and non-perturbative schemes for 3 isoelectronic transition metal hexacarbonyls. The \ce{M(CO)5}$\rightarrow$\ce{CO} component is shown on the top and the \ce{CO}$\rightarrow$\ce{M(CO)5} is shown below it.}
    \label{fig:MCO}
\end{figure}

Free CO is IR active and its stretching frequency appears at 2143 cm\textsuperscript{-1}.\cite{Huber2013molecular}
Consider the sequence of isoelectronic 3d transition metal hexacarbonyls: \ce{V(CO)6-}, \ce{Cr(CO)6}, and \ce{Mn(CO)6+}.\cite{Abel1969vibrational, Reed2010infrared}
The electron density on the metal decreases as we go from anionic \ce{V(CO)6-} to cationic \ce{Mn(CO)6+}.
As a consequence, the metal has less electrons to donate to the 2$\pi^*$ orbitals of \ce{CO}.
Figure~\eqref{fig:MCO} shows that the non-perturbative \ce{M(CO)5}$\rightarrow$\ce{CO} CT
decreases from $-172.9$ kJ/mol in \ce{V(CO)6-} to $-78.7$ kJ/mol in \ce{Mn(CO)6+}.
The experimental red shift also decreases in the same order: 285 cm\textsuperscript{-1} for \ce{V(CO)6-}, 140 cm\textsuperscript{-1} for \ce{Cr(CO)6}, and 37 cm\textsuperscript{-1} for \ce{Mn(CO)6+}.
The computed red shifts show a similar trend: 262 cm\textsuperscript{-1} for \ce{V(CO)6-}, 127 cm\textsuperscript{-1} for \ce{Cr(CO)6}, and 11 cm\textsuperscript{-1} for \ce{Mn(CO)6+}.
As back-donation is a major component of the CT interaction, total CT energy also decreases as one moves to the right in the 3d transition metal sequence.
The non-perturbative CTA also reveals the contrasting mechanisms dominating the nature of CT in \ce{V(CO)6-} and \ce{Mn(CO)6+}: back donation is the major component of CT contributing about 73\% of the CT energy in \ce{V(CO)6-}, whereas forward donation is the major driving force accounting for 55\% of the charge transfer energy in \ce{Mn(CO)6+}.
Thus, the CTA reveals the intricate interplay between different mechanisms driving CT in 3d metal hexacarbonyls.
The variational forward-backward CTA\cite{Loipersberger2020computational} also shows that the adiabatic back-donation (\ce{M(CO)5}$\rightarrow$\ce{CO}) decreases upon moving from \ce{V(CO)6-} to \ce{Mn(CO)6+} while the forward (\ce{CO}$\rightarrow$\ce{M(CO)5}) component decreases (See Table S5).

This example also brings out the main limitation of the perturbative CTA, which is capable of capturing only $\sim$67\% of the total CT energy.
In the \ce{V(CO)6-} case, for example, the perturbative treatment underestimates the forward component by only 21.5 kJ/mol compared to the non-perturbative counterpart, but makes a much larger error in the backward component underestimating it by 62.8 kJ/mol.
This inconsistency in the magnitude of underestimation in the perturbative scheme underscores the advantage of the new non-perturbative CTA.

\subsection{Computational cost}
As established above, the non-perturbative CTA is a superior alternative to perturbative analysis for studying charge decomposition.
%The non-perturbative scheme, by construction, can decompose 100\% of the charge transfer energy into pairwise additive terms.
To address the computational cost of obtaining these improved results, we will briefly compare the cost of the non-perturbative scheme against its perturbative predecessor.
The perturbative scheme computes $\vb{X}_{VO}^{\text{RS}}$ by solving for the Roothaan step, Eq.~\eqref{eq::rs_inf}, which involves only cubic-scaling matrix multiplications, where the dimension of these matrices involves the number of occupied ($O$) and virtual ($V$) orbitals of the complex.
Given $\vb{X}_{VO}^{\text{RS}}$, the pairwise energy decomposition follows by taking its trace with the Fock matrix constructed from the polarized ALMOs ($\Phi_{\text{POL}}$). This Fock matrix is already constructed in the order to converge those orbitals and compute the energy of the polarized wavefunction ($E(\Phi_{\text{POL}})$).

By contrast, the non-perturbative CTA involves solving Eq.~\eqref{eq::P_connect} in order to compute the non-perturbative CT matrix, $\vb{X}_{VO}^{\text{CT}}$.
The process of solving this equation involves cubic scaling matrix diagonalization and multiplications of matrices whose dimension is the number of MOs, $N=O+V$ of the complex.
Given $\vb{X}_{VO}^{\text{CT}}$, the non-perturbative pairwise energy decomposition requires computation of the $\mathbf{F}^{\text{CT}}$ using Eq.~\eqref{eq::F_quadrature}.
This requires five different Fock matrices at different interpolating density points between $\Phi_{\text{POL}}$ and $\Phi_{\text{CT}}$.
Two of these five Fock matrices ($\mathbf{F}(0)$ and $\mathbf{F}(\mathbf{X})$ in Eq.~\eqref{eq::F_quadrature}) are available from computing the energy of the polarized wavefunction and energy of the fully-relaxed complex, leaving three additional Fock matrix constructions. 
Given that the complete ALMO-EDA scheme typically requires tens of Fock matrix constructions for determining the polarized and fully-relaxed wavefunctions, the addition of three more Fock matrix constructions only adds very little to the total computational cost.
%This additional computational expense is well-justified for the advantages the non-perturbative scheme provides over the perturbative scheme, particularly given its robustness for a variety of interacting complexes.

\section{Conclusions}
We have introduced a non-perturbative approach that (numerically) exactly decomposes the variational energy lowering due to charge transfer (CT) in molecular complexes into pairwise additive terms in the context of ALMO-based energy decomposition analysis (ALMO-EDA) for density functional theory.\cite{Khaliullin2007,Horn2016}
The non-perturbative CT analysis (CTA) is a superior replacement for the existing perturbative CTA,\cite{Khaliullin2008} which relied on a perturbative approximation to incompletely and inexactly extract the pairwise contributions. As demonstrated here, the perturbative CTA decomposes different percentages of the total charge transfer energy at different points on a potential energy surface, and also has a dependence on the density functional used, typically underestimating CT for hybrid and range-separated density functionals while overestimating it for pure functionals.

This new method introduced in this work finds the generator, $\vb{X}_{VO}^{\text{CT}}$, of the unitary transformation that transforms the polarized wavefunction into that of the fully-relaxed wavefunction.
Next, an effective Fock-like matrix, $\mathbf{F}^{\text{CT}}$, is constructed from Fock matrices computed at different density points connecting the polarized wavefunction and fully-relaxed wavefunction along the path given by $\vb{X}_{VO}^{\text{CT}}$. 
Taking the trace of the product of the $\vb{F}_{OV}^{\text{CT}}$ and $\mathbf{X}_{VO}^{\text{CT}}$ along with appropriate orbital projectors sandwiched in the middle gives the non-perturbative charge decomposition in terms of donor and acceptor orbital pairs of the 
fragments comprising the complex. This completes the (numerically) exact pairwise energy decomposition scheme for CT. We also extended the CTA to decompose the charge flow in a pairwise fashion as well. Finally, singular value decomposition (SVD) of $\vb{X}_{VO}^{\text{CT}}$ yields complementary occupied-virtual orbital pairs (COVPs), which are the most important orbitals involved in CT.

%The perturbative pairwise charge decomposition scheme, which is based on perturbation theory, breaks down drastically for strongly interacting complexes severely limiting its applicability to a large class of systems.
%The main limitation of the perturbative charge decomposition analysis manifests in multiple ways when investigating real chemical systems.
%The perturbative treatment decomposes different percentages of the total charge transfer energy at different points on the PES.
%The non-perturbative charge decomposition, on the other hand, performs consistently at all points on the PES decomposing 100\% of the charge transfer energy into pairwise additive terms as illustrated using the \ce{BH3}$-$\ce{CO} PES.
%The percentage of charge transfer energy decomposed into pairwise additive components in the perturbative treatment ($ \Delta E_{\text{CT}}^{\text{RS}} / \Delta E_{\text{CT}} \times 100 $ ) also has a severe dependence on the density functional used, underestimating the charge transfer for some density functionals and while overestimating it for others.
%In contrast, the non-perturbative scheme completely decomposes the total charge transfer energy into pairwise additive terms for all density functionals.
%Thus, the non-perturbative scheme presented in this work not only overcomes this major limitation but also retains other advantageous features like pairwise decomposition of charge delocalization and the formalism of Complementary Occupied-Virtual Pairs (COVPs).

We have demonstrated the usefulness of the new CTA by applying it to various chemical systems with varying strengths of charge transfer interaction.
Application to DNA base-pairs reveals the nature of hydrogen bonding in the thymine:adenine and guanine:cytosine complexes.
Additionally, CTA reveals the effect of metallation on the hydrogen bonding patterns of the guanine:cytosine base-pair.
Investigating the interaction energy of a series of borane adducts of the form \ce{BX3}$-$\ce{NH3} (X~=~F, Cl, or Br) reveals that an increase in polarization and CT energy from \ce{NH3} to \ce{BX3} enhances the binding energy of the \ce{BX3}$-$\ce{NH3} complex as we go down the halogen group.
%This understanding advances some of the traditional hypothesis, which invokes orbital overlap between the Boron and halogen, to explain the trend in the binding energy of the \ce{BX3}$-$\ce{NH3} series of adducts.
The CTA also revealed interesting aspects of the effect of methyl and halogen substitution on the nitrogen center.
Additionally, the CTA provided insight into the delicate interplay of forward and backward donation in a series of isoelectronic transitional metal hexacarbonyls.
It is likely to be useful for a wide variety of other interpretive problems in intermolecular interactions, as well as other  applications such as for training force-fields to account for pairwise decomposition of charge transfer.\cite{Mcdaniel2016next,Das2019development}

\begin{acknowledgement}
S.P.V thanks Matthias Loipersberger for helpful discussions.
This work was supported by the U.S. National Science Foundation through Grant No. CHE-1955643. We also acknowledge additional support from the Hydrogen Materials - Advanced Research Consortium (HyMARC), established as part of the Energy Materials Network under the U.S. Department of Energy, Office of Energy Efficiency and Renewable Energy, under Contract No. DE-AC02-05CH11231. The following author declares a competing financial interest. M. H. G. is a part owner of Q-Chem, Inc.
\end{acknowledgement}

\begin{suppinfo}
Additional information supporting convergence to basis set limit, COVPs for metallated guanine:cytosine base pairs, and raw data for \ce{BH3}$-$\ce{NCl_pH_q} complexes is included in the Supporting Information.
In addition to this, supporting adiabatic EDA calculations and variational forward-backward charge transfer analysis calculations are also included.
All geometries used for calculations are provided in the Supporting Information. 
\end{suppinfo}

\bibliography{references}

\providecommand{\latin}[1]{#1}
\makeatletter
\providecommand{\doi}
  {\begingroup\let\do\@makeother\dospecials
  \catcode`\{=1 \catcode`\}=2 \doi@aux}
\providecommand{\doi@aux}[1]{\endgroup\texttt{#1}}
\makeatother
\providecommand*\mcitethebibliography{\thebibliography}
\csname @ifundefined\endcsname{endmcitethebibliography}
  {\let\endmcitethebibliography\endthebibliography}{}
\begin{mcitethebibliography}{83}
\providecommand*\natexlab[1]{#1}
\providecommand*\mciteSetBstSublistMode[1]{}
\providecommand*\mciteSetBstMaxWidthForm[2]{}
\providecommand*\mciteBstWouldAddEndPuncttrue
  {\def\EndOfBibitem{\unskip.}}
\providecommand*\mciteBstWouldAddEndPunctfalse
  {\let\EndOfBibitem\relax}
\providecommand*\mciteSetBstMidEndSepPunct[3]{}
\providecommand*\mciteSetBstSublistLabelBeginEnd[3]{}
\providecommand*\EndOfBibitem{}
\mciteSetBstSublistMode{f}
\mciteSetBstMaxWidthForm{subitem}{(\alph{mcitesubitemcount})}
\mciteSetBstSublistLabelBeginEnd
  {\mcitemaxwidthsubitemform\space}
  {\relax}
  {\relax}

\bibitem[Mardirossian and Head-Gordon(2017)Mardirossian, and
  Head-Gordon]{Mardirossian2017}
Mardirossian,~N.; Head-Gordon,~M. {Thirty years of density functional theory in
  computational chemistry: an overview and extensive assessment of 200 density
  functionals}. \emph{Mol. Phys.} \textbf{2017}, \emph{8976}, 1--58\relax
\mciteBstWouldAddEndPuncttrue
\mciteSetBstMidEndSepPunct{\mcitedefaultmidpunct}
{\mcitedefaultendpunct}{\mcitedefaultseppunct}\relax
\EndOfBibitem
\bibitem[Kitaura and Morokuma(1976)Kitaura, and Morokuma]{Kitaura1976new}
Kitaura,~K.; Morokuma,~K. A new energy decomposition scheme for molecular
  interactions within the Hartree-Fock approximation. \emph{Int. J. Quantum
  Chem.} \textbf{1976}, \emph{10}, 325--340\relax
\mciteBstWouldAddEndPuncttrue
\mciteSetBstMidEndSepPunct{\mcitedefaultmidpunct}
{\mcitedefaultendpunct}{\mcitedefaultseppunct}\relax
\EndOfBibitem
\bibitem[Mitoraj \latin{et~al.}(2009)Mitoraj, Michalak, and
  Ziegler]{Mitoraj2009combined}
Mitoraj,~M.~P.; Michalak,~A.; Ziegler,~T. A combined charge and energy
  decomposition scheme for bond analysis. \emph{J. Chem. Theory Comput.}
  \textbf{2009}, \emph{5}, 962--975\relax
\mciteBstWouldAddEndPuncttrue
\mciteSetBstMidEndSepPunct{\mcitedefaultmidpunct}
{\mcitedefaultendpunct}{\mcitedefaultseppunct}\relax
\EndOfBibitem
\bibitem[Chen and Gordon(1996)Chen, and Gordon]{Chen1996energy}
Chen,~W.; Gordon,~M.~S. Energy decomposition analyses for many-body interaction
  and applications to water complexes. \emph{J. Phys. Chem.} \textbf{1996},
  \emph{100}, 14316--14328\relax
\mciteBstWouldAddEndPuncttrue
\mciteSetBstMidEndSepPunct{\mcitedefaultmidpunct}
{\mcitedefaultendpunct}{\mcitedefaultseppunct}\relax
\EndOfBibitem
\bibitem[Su and Li(2009)Su, and Li]{Su2009energy}
Su,~P.; Li,~H. Energy decomposition analysis of covalent bonds and
  intermolecular interactions. \emph{J. Chem. Phys.} \textbf{2009}, \emph{131},
  014102\relax
\mciteBstWouldAddEndPuncttrue
\mciteSetBstMidEndSepPunct{\mcitedefaultmidpunct}
{\mcitedefaultendpunct}{\mcitedefaultseppunct}\relax
\EndOfBibitem
\bibitem[Reed \latin{et~al.}(1988)Reed, Curtiss, and
  Weinhold]{Reed1988intermolecular}
Reed,~A.~E.; Curtiss,~L.~A.; Weinhold,~F. Intermolecular interactions from a
  natural bond orbital, donor-acceptor viewpoint. \emph{Chem. Rev.}
  \textbf{1988}, \emph{88}, 899--926\relax
\mciteBstWouldAddEndPuncttrue
\mciteSetBstMidEndSepPunct{\mcitedefaultmidpunct}
{\mcitedefaultendpunct}{\mcitedefaultseppunct}\relax
\EndOfBibitem
\bibitem[Jeziorski \latin{et~al.}(1994)Jeziorski, Moszynski, and
  Szalewicz]{Jeziorski1994perturbation}
Jeziorski,~B.; Moszynski,~R.; Szalewicz,~K. Perturbation theory approach to
  intermolecular potential energy surfaces of van der Waals complexes.
  \emph{Chem. Rev.} \textbf{1994}, \emph{94}, 1887--1930\relax
\mciteBstWouldAddEndPuncttrue
\mciteSetBstMidEndSepPunct{\mcitedefaultmidpunct}
{\mcitedefaultendpunct}{\mcitedefaultseppunct}\relax
\EndOfBibitem
\bibitem[Mo \latin{et~al.}(2000)Mo, Gao, and Peyerimhoff]{Mo2000}
Mo,~Y.; Gao,~J.; Peyerimhoff,~S.~D. {Energy decomposition analysis of
  intermolecular interactions using a block-localized wave function approach}.
  \emph{J. Chem. Phys.} \textbf{2000}, \emph{112}, 5530--5538\relax
\mciteBstWouldAddEndPuncttrue
\mciteSetBstMidEndSepPunct{\mcitedefaultmidpunct}
{\mcitedefaultendpunct}{\mcitedefaultseppunct}\relax
\EndOfBibitem
\bibitem[Khaliullin \latin{et~al.}(2007)Khaliullin, Cobar, Lochan, Bell, and
  Head-Gordon]{Khaliullin2007}
Khaliullin,~R.~Z.; Cobar,~E.~A.; Lochan,~R.~C.; Bell,~A.~T.; Head-Gordon,~M.
  {Unravelling the origin of intermolecular interactions using absolutely
  localized molecular orbitals}. \emph{J. Phys. Chem. A} \textbf{2007},
  \emph{111}, 8753--8765\relax
\mciteBstWouldAddEndPuncttrue
\mciteSetBstMidEndSepPunct{\mcitedefaultmidpunct}
{\mcitedefaultendpunct}{\mcitedefaultseppunct}\relax
\EndOfBibitem
\bibitem[McDaniel and Schmidt(2016)McDaniel, and Schmidt]{Mcdaniel2016next}
McDaniel,~J.~G.; Schmidt,~J. Next-generation force fields from symmetry-adapted
  perturbation theory. \emph{Annu. Rev. Phys. Chem.} \textbf{2016}, \emph{67},
  467--488\relax
\mciteBstWouldAddEndPuncttrue
\mciteSetBstMidEndSepPunct{\mcitedefaultmidpunct}
{\mcitedefaultendpunct}{\mcitedefaultseppunct}\relax
\EndOfBibitem
\bibitem[Das \latin{et~al.}(2019)Das, Urban, Leven, Loipersberger, Aldossary,
  Head-Gordon, and Head-Gordon]{Das2019development}
Das,~A.~K.; Urban,~L.; Leven,~I.; Loipersberger,~M.; Aldossary,~A.;
  Head-Gordon,~M.; Head-Gordon,~T. Development of an advanced force field for
  water using variational energy decomposition analysis. \emph{J. Chem. Theory
  Comput.} \textbf{2019}, \emph{15}, 5001--5013\relax
\mciteBstWouldAddEndPuncttrue
\mciteSetBstMidEndSepPunct{\mcitedefaultmidpunct}
{\mcitedefaultendpunct}{\mcitedefaultseppunct}\relax
\EndOfBibitem
\bibitem[Mao \latin{et~al.}(2018)Mao, Ge, Horn, and Head-Gordon]{Mao2018}
Mao,~Y.; Ge,~Q.; Horn,~P.~R.; Head-Gordon,~M. {On the Computational
  Characterization of Charge-Transfer Effects in Noncovalently Bound Molecular
  Complexes}. \emph{J. Chem. Theory Comput.} \textbf{2018}, \emph{14},
  2401--2417\relax
\mciteBstWouldAddEndPuncttrue
\mciteSetBstMidEndSepPunct{\mcitedefaultmidpunct}
{\mcitedefaultendpunct}{\mcitedefaultseppunct}\relax
\EndOfBibitem
\bibitem[Dewar(1951)]{Dewar1951review}
Dewar,~J. A review of the pi-complex theory. \emph{Bull. Soc. Chim. Fr.}
  \textbf{1951}, \emph{18}, C71--C79\relax
\mciteBstWouldAddEndPuncttrue
\mciteSetBstMidEndSepPunct{\mcitedefaultmidpunct}
{\mcitedefaultendpunct}{\mcitedefaultseppunct}\relax
\EndOfBibitem
\bibitem[Chatt and Duncanson(1953)Chatt, and Duncanson]{Chatt1953586}
Chatt,~J.; Duncanson,~L. Olefin co-ordination compounds. Part III. Infra-red
  spectra and structure: attempted preparation of acetylene complexes. \emph{J.
  Chem. Soc.} \textbf{1953}, 2939--2947\relax
\mciteBstWouldAddEndPuncttrue
\mciteSetBstMidEndSepPunct{\mcitedefaultmidpunct}
{\mcitedefaultendpunct}{\mcitedefaultseppunct}\relax
\EndOfBibitem
\bibitem[van~der Lubbe and Fonseca~Guerra(2019)van~der Lubbe, and
  Fonseca~Guerra]{Van2019nature}
van~der Lubbe,~S.~C.; Fonseca~Guerra,~C. The nature of hydrogen bonds: A
  delineation of the role of different energy components on hydrogen bond
  strengths and lengths. \emph{Chem. - Asian J.} \textbf{2019}, \emph{14},
  2760--2769\relax
\mciteBstWouldAddEndPuncttrue
\mciteSetBstMidEndSepPunct{\mcitedefaultmidpunct}
{\mcitedefaultendpunct}{\mcitedefaultseppunct}\relax
\EndOfBibitem
\bibitem[Horn \latin{et~al.}(2013)Horn, Sundstrom, Baker, and
  Head-Gordon]{Horn2013}
Horn,~P.~R.; Sundstrom,~E.~J.; Baker,~T.~A.; Head-Gordon,~M. {Unrestricted
  absolutely localized molecular orbitals for energy decomposition analysis:
  Theory and applications to intermolecular interactions involving radicals}.
  \emph{J. Chem. Phys.} \textbf{2013}, \emph{138}, 30--32\relax
\mciteBstWouldAddEndPuncttrue
\mciteSetBstMidEndSepPunct{\mcitedefaultmidpunct}
{\mcitedefaultendpunct}{\mcitedefaultseppunct}\relax
\EndOfBibitem
\bibitem[Horn \latin{et~al.}(2016)Horn, Mao, and Head-Gordon]{Horn2016c}
Horn,~P.~R.; Mao,~Y.; Head-Gordon,~M. Probing non-covalent interactions with a
  second generation energy decomposition analysis using absolutely localized
  molecular orbitals. \emph{Phys. Chem. Chem. Phys.} \textbf{2016}, \emph{18},
  23067--23079\relax
\mciteBstWouldAddEndPuncttrue
\mciteSetBstMidEndSepPunct{\mcitedefaultmidpunct}
{\mcitedefaultendpunct}{\mcitedefaultseppunct}\relax
\EndOfBibitem
\bibitem[Mao \latin{et~al.}(2020)Mao, Levine, Loipersberger, Horn, and
  Head-Gordon]{Mao2020probing}
Mao,~Y.; Levine,~D.~S.; Loipersberger,~M.; Horn,~P.~R.; Head-Gordon,~M. Probing
  radical–molecule interactions with a second generation energy decomposition
  analysis of DFT calculations using absolutely localized molecular orbitals.
  \emph{Phys. Chem. Chem. Phys.} \textbf{2020}, \emph{22}, 12867--12885\relax
\mciteBstWouldAddEndPuncttrue
\mciteSetBstMidEndSepPunct{\mcitedefaultmidpunct}
{\mcitedefaultendpunct}{\mcitedefaultseppunct}\relax
\EndOfBibitem
\bibitem[Thirman and Head-Gordon(2015)Thirman, and
  Head-Gordon]{Thirman2015energy}
Thirman,~J.; Head-Gordon,~M. An energy decomposition analysis for second-order
  M{\o}ller--Plesset perturbation theory based on absolutely localized
  molecular orbitals. \emph{J. Chem. Phys.} \textbf{2015}, \emph{143},
  084124\relax
\mciteBstWouldAddEndPuncttrue
\mciteSetBstMidEndSepPunct{\mcitedefaultmidpunct}
{\mcitedefaultendpunct}{\mcitedefaultseppunct}\relax
\EndOfBibitem
\bibitem[Thirman and Head-Gordon(2017)Thirman, and Head-Gordon]{Thirman2017}
Thirman,~J.; Head-Gordon,~M. Efficient Implementation of Energy Decomposition
  Analysis for Second-Order M{\o}ller--Plesset Perturbation Theory and
  Application to Anion--$\pi$ Interactions. \emph{J. Phys. Chem. A}
  \textbf{2017}, \emph{121}, 717--728\relax
\mciteBstWouldAddEndPuncttrue
\mciteSetBstMidEndSepPunct{\mcitedefaultmidpunct}
{\mcitedefaultendpunct}{\mcitedefaultseppunct}\relax
\EndOfBibitem
\bibitem[Loipersberger \latin{et~al.}(2019)Loipersberger, Lee, Mao, Das, Ikeda,
  Thirman, Head-Gordon, and Head-Gordon]{Loipersberger2019}
Loipersberger,~M.; Lee,~J.; Mao,~Y.; Das,~A.~K.; Ikeda,~K.; Thirman,~J.;
  Head-Gordon,~T.; Head-Gordon,~M. Energy Decomposition Analysis for
  Interactions of Radicals: Theory and Implementation at the MP2 Level with
  Application to Hydration of Halogenated Benzene Cations and Complexes between
  CO$_{2}^{-\bullet}$ and Pyridine and Imidazole. \emph{J. Phys. Chem. A}
  \textbf{2019}, \emph{123}, 9621--9633\relax
\mciteBstWouldAddEndPuncttrue
\mciteSetBstMidEndSepPunct{\mcitedefaultmidpunct}
{\mcitedefaultendpunct}{\mcitedefaultseppunct}\relax
\EndOfBibitem
\bibitem[Horn \latin{et~al.}(2016)Horn, Mao, and Head-Gordon]{Horn2016}
Horn,~P.~R.; Mao,~Y.; Head-Gordon,~M. {Defining the contributions of permanent
  electrostatics, Pauli repulsion, and dispersion in density functional theory
  calculations of intermolecular interaction energies}. \emph{J. Chem. Phys.}
  \textbf{2016}, \emph{144}\relax
\mciteBstWouldAddEndPuncttrue
\mciteSetBstMidEndSepPunct{\mcitedefaultmidpunct}
{\mcitedefaultendpunct}{\mcitedefaultseppunct}\relax
\EndOfBibitem
\bibitem[Horn and Head-Gordon(2015)Horn, and Head-Gordon]{Horn2015}
Horn,~P.~R.; Head-Gordon,~M. {Polarization contributions to intermolecular
  interactions revisited with fragment electric-field response functions}.
  \emph{J. Chem. Phys.} \textbf{2015}, \emph{143}\relax
\mciteBstWouldAddEndPuncttrue
\mciteSetBstMidEndSepPunct{\mcitedefaultmidpunct}
{\mcitedefaultendpunct}{\mcitedefaultseppunct}\relax
\EndOfBibitem
\bibitem[Khaliullin \latin{et~al.}(2008)Khaliullin, Bell, and
  Head-Gordon]{Khaliullin2008}
Khaliullin,~R.~Z.; Bell,~A.~T.; Head-Gordon,~M. {Analysis of charge transfer
  effects in molecular complexes based on absolutely localized molecular
  orbitals.} \emph{J. Chem. Phys.} \textbf{2008}, \emph{128}, 184112\relax
\mciteBstWouldAddEndPuncttrue
\mciteSetBstMidEndSepPunct{\mcitedefaultmidpunct}
{\mcitedefaultendpunct}{\mcitedefaultseppunct}\relax
\EndOfBibitem
\bibitem[Dapprich and Frenking(1995)Dapprich, and
  Frenking]{Dapprich1995investigation}
Dapprich,~S.; Frenking,~G. Investigation of donor-acceptor interactions: a
  charge decomposition analysis using fragment molecular orbitals. \emph{J.
  Phys. Chem.} \textbf{1995}, \emph{99}, 9352--9362\relax
\mciteBstWouldAddEndPuncttrue
\mciteSetBstMidEndSepPunct{\mcitedefaultmidpunct}
{\mcitedefaultendpunct}{\mcitedefaultseppunct}\relax
\EndOfBibitem
\bibitem[Stevens and Fink(1987)Stevens, and Fink]{Stevens1987frozen}
Stevens,~W.~J.; Fink,~W.~H. Frozen fragment reduced variational space analysis
  of hydrogen bonding interactions. Application to the water dimer. \emph{Chem.
  Phys. Lett.} \textbf{1987}, \emph{139}, 15--22\relax
\mciteBstWouldAddEndPuncttrue
\mciteSetBstMidEndSepPunct{\mcitedefaultmidpunct}
{\mcitedefaultendpunct}{\mcitedefaultseppunct}\relax
\EndOfBibitem
\bibitem[Bagus \latin{et~al.}(1984)Bagus, Hermann, and
  Bauschlicher~Jr]{Bagus1984new}
Bagus,~P.~S.; Hermann,~K.; Bauschlicher~Jr,~C.~W. A new analysis of charge
  transfer and polarization for ligand--metal bonding: Model studies of Al4CO
  and Al4NH3. \emph{J. Chem. Phys.} \textbf{1984}, \emph{80}, 4378--4386\relax
\mciteBstWouldAddEndPuncttrue
\mciteSetBstMidEndSepPunct{\mcitedefaultmidpunct}
{\mcitedefaultendpunct}{\mcitedefaultseppunct}\relax
\EndOfBibitem
\bibitem[Zhao \latin{et~al.}(2018)Zhao, von Hopffgarten, Andrada, and
  Frenking]{Zhao2018}
Zhao,~L.; von Hopffgarten,~M.; Andrada,~D.~M.; Frenking,~G. Energy
  decomposition analysis. \emph{WIREs: Comput. Mol. Sci.} \textbf{2018},
  \emph{8}, e1345\relax
\mciteBstWouldAddEndPuncttrue
\mciteSetBstMidEndSepPunct{\mcitedefaultmidpunct}
{\mcitedefaultendpunct}{\mcitedefaultseppunct}\relax
\EndOfBibitem
\bibitem[Szalewicz(2012)]{Szalewicz2012}
Szalewicz,~K. Symmetry-adapted perturbation theory of intermolecular forces.
  \emph{Wiley Interdiscip. Rev.: Comput. Mol. Sci.} \textbf{2012}, \emph{2},
  254--272\relax
\mciteBstWouldAddEndPuncttrue
\mciteSetBstMidEndSepPunct{\mcitedefaultmidpunct}
{\mcitedefaultendpunct}{\mcitedefaultseppunct}\relax
\EndOfBibitem
\bibitem[Stone and Misquitta(2009)Stone, and Misquitta]{Stone2009}
Stone,~A.~J.; Misquitta,~A.~J. Charge-transfer in symmetry-adapted perturbation
  theory. \emph{Chem. Phys. Lett.} \textbf{2009}, \emph{473}, 201--205\relax
\mciteBstWouldAddEndPuncttrue
\mciteSetBstMidEndSepPunct{\mcitedefaultmidpunct}
{\mcitedefaultendpunct}{\mcitedefaultseppunct}\relax
\EndOfBibitem
\bibitem[Misquitta(2013)]{Misquitta2013}
Misquitta,~A.~J. Charge transfer from regularized symmetry-adapted perturbation
  theory. \emph{J. Chem. Theory Comput.} \textbf{2013}, \emph{9},
  5313--5326\relax
\mciteBstWouldAddEndPuncttrue
\mciteSetBstMidEndSepPunct{\mcitedefaultmidpunct}
{\mcitedefaultendpunct}{\mcitedefaultseppunct}\relax
\EndOfBibitem
\bibitem[Reed and Weinhold(1983)Reed, and Weinhold]{Reed1983natural}
Reed,~A.~E.; Weinhold,~F. Natural bond orbital analysis of near-Hartree--Fock
  water dimer. \emph{J. Chem. Phys.} \textbf{1983}, \emph{78}, 4066--4073\relax
\mciteBstWouldAddEndPuncttrue
\mciteSetBstMidEndSepPunct{\mcitedefaultmidpunct}
{\mcitedefaultendpunct}{\mcitedefaultseppunct}\relax
\EndOfBibitem
\bibitem[Glendening and Streitwieser(1994)Glendening, and
  Streitwieser]{Glendening1994natural}
Glendening,~E.~D.; Streitwieser,~A. Natural energy decomposition analysis: An
  energy partitioning procedure for molecular interactions with application to
  weak hydrogen bonding, strong ionic, and moderate donor--acceptor
  interactions. \emph{J. Chem. Phys.} \textbf{1994}, \emph{100},
  2900--2909\relax
\mciteBstWouldAddEndPuncttrue
\mciteSetBstMidEndSepPunct{\mcitedefaultmidpunct}
{\mcitedefaultendpunct}{\mcitedefaultseppunct}\relax
\EndOfBibitem
\bibitem[Glendening(2005)]{Glendening2005natural}
Glendening,~E.~D. Natural energy decomposition analysis: Extension to density
  functional methods and analysis of cooperative effects in water clusters.
  \emph{J. Phys. Chem. A} \textbf{2005}, \emph{109}, 11936--11940\relax
\mciteBstWouldAddEndPuncttrue
\mciteSetBstMidEndSepPunct{\mcitedefaultmidpunct}
{\mcitedefaultendpunct}{\mcitedefaultseppunct}\relax
\EndOfBibitem
\bibitem[Khaliullin \latin{et~al.}(2009)Khaliullin, Bell, and
  Head-Gordon]{Khaliullin2009}
Khaliullin,~R.~Z.; Bell,~A.~T.; Head-Gordon,~M. {Electron donation in the
  water-water hydrogen bond}. \emph{Chem. - Eur. J.} \textbf{2009}, \emph{15},
  851--855\relax
\mciteBstWouldAddEndPuncttrue
\mciteSetBstMidEndSepPunct{\mcitedefaultmidpunct}
{\mcitedefaultendpunct}{\mcitedefaultseppunct}\relax
\EndOfBibitem
\bibitem[Stone(2017)]{Stone2017natural}
Stone,~A.~J. Natural bond orbitals and the nature of the hydrogen bond.
  \emph{J. Phys. Chem. A} \textbf{2017}, \emph{121}, 1531--1534\relax
\mciteBstWouldAddEndPuncttrue
\mciteSetBstMidEndSepPunct{\mcitedefaultmidpunct}
{\mcitedefaultendpunct}{\mcitedefaultseppunct}\relax
\EndOfBibitem
\bibitem[Wu and Van~Voorhis(2005)Wu, and Van~Voorhis]{Wu2005direct}
Wu,~Q.; Van~Voorhis,~T. Direct optimization method to study constrained systems
  within density-functional theory. \emph{Phys. Rev. A} \textbf{2005},
  \emph{72}, 024502\relax
\mciteBstWouldAddEndPuncttrue
\mciteSetBstMidEndSepPunct{\mcitedefaultmidpunct}
{\mcitedefaultendpunct}{\mcitedefaultseppunct}\relax
\EndOfBibitem
\bibitem[Wu and Van~Voorhis(2006)Wu, and Van~Voorhis]{Wu2006constrained}
Wu,~Q.; Van~Voorhis,~T. Constrained density functional theory and its
  application in long-range electron transfer. \emph{J. Chem. Theory Comput.}
  \textbf{2006}, \emph{2}, 765--774\relax
\mciteBstWouldAddEndPuncttrue
\mciteSetBstMidEndSepPunct{\mcitedefaultmidpunct}
{\mcitedefaultendpunct}{\mcitedefaultseppunct}\relax
\EndOfBibitem
\bibitem[Kaduk \latin{et~al.}(2012)Kaduk, Kowalczyk, and
  Van~Voorhis]{Kaduk2012constrained}
Kaduk,~B.; Kowalczyk,~T.; Van~Voorhis,~T. Constrained density functional
  theory. \emph{Chem. Rev.} \textbf{2012}, \emph{112}, 321--370\relax
\mciteBstWouldAddEndPuncttrue
\mciteSetBstMidEndSepPunct{\mcitedefaultmidpunct}
{\mcitedefaultendpunct}{\mcitedefaultseppunct}\relax
\EndOfBibitem
\bibitem[Wu \latin{et~al.}(2009)Wu, Ayers, and Zhang]{Wu2009}
Wu,~Q.; Ayers,~P.~W.; Zhang,~Y. Density-based energy decomposition analysis for
  intermolecular interactions with variationally determined intermediate state
  energies. \emph{J. Chem. Phys.} \textbf{2009}, \emph{131}, 164112\relax
\mciteBstWouldAddEndPuncttrue
\mciteSetBstMidEndSepPunct{\mcitedefaultmidpunct}
{\mcitedefaultendpunct}{\mcitedefaultseppunct}\relax
\EndOfBibitem
\bibitem[Lao and Herbert(2016)Lao, and Herbert]{Lao2016}
Lao,~K.~U.; Herbert,~J.~M. {Energy Decomposition Analysis with a Stable
  Charge-Transfer Term for Interpreting Intermolecular Interactions}. \emph{J.
  Chem. Theory Comput.} \textbf{2016}, \emph{12}, 2569--2582\relax
\mciteBstWouldAddEndPuncttrue
\mciteSetBstMidEndSepPunct{\mcitedefaultmidpunct}
{\mcitedefaultendpunct}{\mcitedefaultseppunct}\relax
\EndOfBibitem
\bibitem[Stoll \latin{et~al.}(1980)Stoll, Wagenblast, and Preuss]{Stoll1980}
Stoll,~H.; Wagenblast,~G.; Preuss,~H. {On the Use of Local Basis-Sets for
  Localized Molecular-Orbitals}. \emph{Theor. Chim. Acta} \textbf{1980},
  \emph{57}, 169--178\relax
\mciteBstWouldAddEndPuncttrue
\mciteSetBstMidEndSepPunct{\mcitedefaultmidpunct}
{\mcitedefaultendpunct}{\mcitedefaultseppunct}\relax
\EndOfBibitem
\bibitem[Gianinetti \latin{et~al.}(1996)Gianinetti, Raimondi, and
  Tornaghi]{Gianinetti1996}
Gianinetti,~E.; Raimondi,~M.; Tornaghi,~E. {Modification of the Roothaan
  equations to exclude BSSE from molecular interaction calculations}.
  \emph{Int. J. Quantum Chem.} \textbf{1996}, \emph{60}, 157--166\relax
\mciteBstWouldAddEndPuncttrue
\mciteSetBstMidEndSepPunct{\mcitedefaultmidpunct}
{\mcitedefaultendpunct}{\mcitedefaultseppunct}\relax
\EndOfBibitem
\bibitem[Khaliullin \latin{et~al.}(2006)Khaliullin, Head-Gordon, and
  Bell]{Khaliullin2006}
Khaliullin,~R.~Z.; Head-Gordon,~M.; Bell,~A.~T. An efficient self-consistent
  field method for large systems of weakly interacting components. \emph{J.
  Chem. Phys.} \textbf{2006}, \emph{124}, 204105\relax
\mciteBstWouldAddEndPuncttrue
\mciteSetBstMidEndSepPunct{\mcitedefaultmidpunct}
{\mcitedefaultendpunct}{\mcitedefaultseppunct}\relax
\EndOfBibitem
\bibitem[Azar \latin{et~al.}(2013)Azar, Horn, Sundstrom, and
  Head-Gordon]{Azar2013}
Azar,~R.~J.; Horn,~P.~R.; Sundstrom,~E.~J.; Head-Gordon,~M. Useful lower limits
  to polarization contributions to intermolecular interactions using a minimal
  basis of localized orthogonal orbitals: Theory and analysis of the water
  dimer. \emph{J. Chem. Phys.} \textbf{2013}, \emph{138}, 084102\relax
\mciteBstWouldAddEndPuncttrue
\mciteSetBstMidEndSepPunct{\mcitedefaultmidpunct}
{\mcitedefaultendpunct}{\mcitedefaultseppunct}\relax
\EndOfBibitem
\bibitem[Mao \latin{et~al.}(2017)Mao, Horn, and Head-Gordon]{Mao2017}
Mao,~Y.; Horn,~P.~R.; Head-Gordon,~M. {Energy decomposition analysis in an
  adiabatic picture}. \emph{Phys. Chem. Chem. Phys.} \textbf{2017}, \emph{19},
  5944--5958\relax
\mciteBstWouldAddEndPuncttrue
\mciteSetBstMidEndSepPunct{\mcitedefaultmidpunct}
{\mcitedefaultendpunct}{\mcitedefaultseppunct}\relax
\EndOfBibitem
\bibitem[Mao and Head-Gordon(2019)Mao, and Head-Gordon]{Mao2019}
Mao,~Y.; Head-Gordon,~M. Probing Blue-Shifting Hydrogen Bonds with Adiabatic
  Energy Decomposition Analysis. \emph{J. Phys. Chem. Lett.} \textbf{2019},
  \emph{10}, 3899--3905\relax
\mciteBstWouldAddEndPuncttrue
\mciteSetBstMidEndSepPunct{\mcitedefaultmidpunct}
{\mcitedefaultendpunct}{\mcitedefaultseppunct}\relax
\EndOfBibitem
\bibitem[Loipersberger \latin{et~al.}(2020)Loipersberger, Zee, Panetier, Chang,
  Long, and Head-Gordon]{Loipersberger2020computational}
Loipersberger,~M.; Zee,~D.~Z.; Panetier,~J.~A.; Chang,~C.~J.; Long,~J.~R.;
  Head-Gordon,~M. Computational Study of an Iron (II) Polypyridine
  Electrocatalyst for CO2 Reduction: Key Roles for Intramolecular Interactions
  in CO2 Binding and Proton Transfer. \emph{Inorg. Chem.} \textbf{2020}, \relax
\mciteBstWouldAddEndPunctfalse
\mciteSetBstMidEndSepPunct{\mcitedefaultmidpunct}
{}{\mcitedefaultseppunct}\relax
\EndOfBibitem
\bibitem[Loipersberger \latin{et~al.}(2020)Loipersberger, Mao, and
  Head-Gordon]{Loipersberger2020variational}
Loipersberger,~M.; Mao,~Y.; Head-Gordon,~M. Variational Forward--Backward
  Charge Transfer Analysis Based on Absolutely Localized Molecular Orbitals:
  Energetics and Molecular Properties. \emph{J. Chem. Theory Comput.}
  \textbf{2020}, \emph{16}, 1073--1089\relax
\mciteBstWouldAddEndPuncttrue
\mciteSetBstMidEndSepPunct{\mcitedefaultmidpunct}
{\mcitedefaultendpunct}{\mcitedefaultseppunct}\relax
\EndOfBibitem
\bibitem[Liang and Head-Gordon(2004)Liang, and Head-Gordon]{Liang2004}
Liang,~W.; Head-Gordon,~M. {An exact reformulation of the diagonalization step
  in electronic structure calculations as a set of second order nonlinear
  equations}. \emph{J. Chem. Phys.} \textbf{2004}, \emph{120},
  10379--10384\relax
\mciteBstWouldAddEndPuncttrue
\mciteSetBstMidEndSepPunct{\mcitedefaultmidpunct}
{\mcitedefaultendpunct}{\mcitedefaultseppunct}\relax
\EndOfBibitem
\bibitem[Liang and Head-Gordon(2004)Liang, and
  Head-Gordon]{Liang2004approaching}
Liang,~W.; Head-Gordon,~M. Approaching the basis set limit in density
  functional theory calculations using dual basis sets without diagonalization.
  \emph{J. Phys. Chem. A} \textbf{2004}, \emph{108}, 3206--3210\relax
\mciteBstWouldAddEndPuncttrue
\mciteSetBstMidEndSepPunct{\mcitedefaultmidpunct}
{\mcitedefaultendpunct}{\mcitedefaultseppunct}\relax
\EndOfBibitem
\bibitem[Abramowitz(1974)]{Abramowitz1974}
Abramowitz,~M. \emph{Handbook of Mathematical Functions, With Formulas, Graphs,
  and Mathematical Tables,}; Dover Publications, Inc.: USA, 1974\relax
\mciteBstWouldAddEndPuncttrue
\mciteSetBstMidEndSepPunct{\mcitedefaultmidpunct}
{\mcitedefaultendpunct}{\mcitedefaultseppunct}\relax
\EndOfBibitem
\bibitem[Head-Gordon \latin{et~al.}(1998)Head-Gordon, Maslen, and
  White]{Head-Gordon1997}
Head-Gordon,~M.; Maslen,~P.~E.; White,~C.~A. {A tensor formulation of
  many-electron theory in a nonorthogonal single-particle basis}. \emph{J.
  Chem. Phys.} \textbf{1998}, \emph{108}, 616--625\relax
\mciteBstWouldAddEndPuncttrue
\mciteSetBstMidEndSepPunct{\mcitedefaultmidpunct}
{\mcitedefaultendpunct}{\mcitedefaultseppunct}\relax
\EndOfBibitem
\bibitem[Shao \latin{et~al.}(2015)Shao, Gan, Epifanovsky, Gilbert, Wormit,
  Kussmann, Lange, Behn, Deng, Feng, Ghosh, Goldey, Horn, Jacobson, Kaliman,
  Khaliullin, Ku\'{s}, Landau, Liu, Proynov, Rhee, Richard, Rohrdanz, Steele,
  Sundstrom, Woodcock, Zimmerman, Zuev, Albrecht, Alguire, Austin, Beran,
  Bernard, Berquist, Brandhorst, Bravaya, Brown, Casanova, Chang, Chen, Chien,
  Closser, Crittenden, Diedenhofen, Distasio, Do, Dutoi, Edgar, Fatehi,
  Fusti-Molnar, Ghysels, Golubeva-Zadorozhnaya, Gomes, Hanson-Heine, Harbach,
  Hauser, Hohenstein, Holden, Jagau, Ji, Kaduk, Khistyaev, Kim, Kim, King,
  Klunzinger, Kosenkov, Kowalczyk, Krauter, Lao, Laurent, Lawler, Levchenko,
  Lin, Liu, Livshits, Lochan, Luenser, Manohar, Manzer, Mao, Mardirossian,
  Marenich, Maurer, Mayhall, Neuscamman, Oana, Olivares-Amaya, Oneill,
  Parkhill, Perrine, Peverati, Prociuk, Rehn, Rosta, Russ, Sharada, Sharma,
  Small, Sodt, Stein, St{\"{u}}ck, Su, Thom, Tsuchimochi, Vanovschi, Vogt,
  Vydrov, Wang, Watson, Wenzel, White, Williams, Yang, Yeganeh, Yost, You,
  Zhang, Zhang, Zhao, Brooks, Chan, Chipman, Cramer, Goddard, Gordon, Hehre,
  Klamt, Schaefer, Schmidt, Sherrill, Truhlar, Warshel, Xu, Aspuru-Guzik, Baer,
  Bell, Besley, Chai, Dreuw, Dunietz, Furlani, Gwaltney, Hsu, Jung, Kong,
  Lambrecht, Liang, Ochsenfeld, Rassolov, Slipchenko, Subotnik, Van~Voorhis,
  Herbert, Krylov, Gill, and Head-Gordon]{Shao2015}
Shao,~Y.; Gan,~Z.; Epifanovsky,~E.; Gilbert,~A.~T.; Wormit,~M.; Kussmann,~J.;
  Lange,~A.~W.; Behn,~A.; Deng,~J.; Feng,~X.; Ghosh,~D.; Goldey,~M.;
  Horn,~P.~R.; Jacobson,~L.~D.; Kaliman,~I.; Khaliullin,~R.~Z.; Ku\'{s},~T.;
  Landau,~A.; Liu,~J.; Proynov,~E.~I.; Rhee,~Y.~M.; Richard,~R.~M.;
  Rohrdanz,~M.~A.; Steele,~R.~P.; Sundstrom,~E.~J.; Woodcock,~H.~L.;
  Zimmerman,~P.~M.; Zuev,~D.; Albrecht,~B.; Alguire,~E.; Austin,~B.;
  Beran,~G.~J.; Bernard,~Y.~A.; Berquist,~E.; Brandhorst,~K.; Bravaya,~K.~B.;
  Brown,~S.~T.; Casanova,~D.; Chang,~C.~M.; Chen,~Y.; Chien,~S.~H.;
  Closser,~K.~D.; Crittenden,~D.~L.; Diedenhofen,~M.; Distasio,~R.~A.; Do,~H.;
  Dutoi,~A.~D.; Edgar,~R.~G.; Fatehi,~S.; Fusti-Molnar,~L.; Ghysels,~A.;
  Golubeva-Zadorozhnaya,~A.; Gomes,~J.; Hanson-Heine,~M.~W.; Harbach,~P.~H.;
  Hauser,~A.~W.; Hohenstein,~E.~G.; Holden,~Z.~C.; Jagau,~T.~C.; Ji,~H.;
  Kaduk,~B.; Khistyaev,~K.; Kim,~J.; Kim,~J.; King,~R.~A.; Klunzinger,~P.;
  Kosenkov,~D.; Kowalczyk,~T.; Krauter,~C.~M.; Lao,~K.~U.; Laurent,~A.~D.;
  Lawler,~K.~V.; Levchenko,~S.~V.; Lin,~C.~Y.; Liu,~F.; Livshits,~E.;
  Lochan,~R.~C.; Luenser,~A.; Manohar,~P.; Manzer,~S.~F.; Mao,~S.~P.;
  Mardirossian,~N.; Marenich,~A.~V.; Maurer,~S.~A.; Mayhall,~N.~J.;
  Neuscamman,~E.; Oana,~C.~M.; Olivares-Amaya,~R.; Oneill,~D.~P.;
  Parkhill,~J.~A.; Perrine,~T.~M.; Peverati,~R.; Prociuk,~A.; Rehn,~D.~R.;
  Rosta,~E.; Russ,~N.~J.; Sharada,~S.~M.; Sharma,~S.; Small,~D.~W.; Sodt,~A.;
  Stein,~T.; St{\"{u}}ck,~D.; Su,~Y.~C.; Thom,~A.~J.; Tsuchimochi,~T.;
  Vanovschi,~V.; Vogt,~L.; Vydrov,~O.; Wang,~T.; Watson,~M.~A.; Wenzel,~J.;
  White,~A.; Williams,~C.~F.; Yang,~J.; Yeganeh,~S.; Yost,~S.~R.; You,~Z.~Q.;
  Zhang,~I.~Y.; Zhang,~X.; Zhao,~Y.; Brooks,~B.~R.; Chan,~G.~K.;
  Chipman,~D.~M.; Cramer,~C.~J.; Goddard,~W.~A.; Gordon,~M.~S.; Hehre,~W.~J.;
  Klamt,~A.; Schaefer,~H.~F.; Schmidt,~M.~W.; Sherrill,~C.~D.; Truhlar,~D.~G.;
  Warshel,~A.; Xu,~X.; Aspuru-Guzik,~A.; Baer,~R.; Bell,~A.~T.; Besley,~N.~A.;
  Chai,~J.~D.; Dreuw,~A.; Dunietz,~B.~D.; Furlani,~T.~R.; Gwaltney,~S.~R.;
  Hsu,~C.~P.; Jung,~Y.; Kong,~J.; Lambrecht,~D.~S.; Liang,~W.; Ochsenfeld,~C.;
  Rassolov,~V.~A.; Slipchenko,~L.~V.; Subotnik,~J.~E.; Van~Voorhis,~T.;
  Herbert,~J.~M.; Krylov,~A.~I.; Gill,~P.~M.; Head-Gordon,~M. {Advances in
  molecular quantum chemistry contained in the Q-Chem 4 program package}.
  \emph{Mol. Phys.} \textbf{2015}, \emph{113}, 184--215\relax
\mciteBstWouldAddEndPuncttrue
\mciteSetBstMidEndSepPunct{\mcitedefaultmidpunct}
{\mcitedefaultendpunct}{\mcitedefaultseppunct}\relax
\EndOfBibitem
\bibitem[Lee and Head-Gordon(2018)Lee, and Head-Gordon]{Lee2018}
Lee,~J.; Head-Gordon,~M. {Regularized Orbital-Optimized Second-Order
  M{\o}ller-Plesset Perturbation Theory: A Reliable Fifth-Order-Scaling
  Electron Correlation Model with Orbital Energy Dependent Regularizers}.
  \emph{J. Chem. Theory Comput.} \textbf{2018}, \emph{14}, 5203--5219\relax
\mciteBstWouldAddEndPuncttrue
\mciteSetBstMidEndSepPunct{\mcitedefaultmidpunct}
{\mcitedefaultendpunct}{\mcitedefaultseppunct}\relax
\EndOfBibitem
\bibitem[Chai and Head-Gordon(2008)Chai, and Head-Gordon]{Chai2008}
Chai,~J.-D.; Head-Gordon,~M. Long-range corrected hybrid density functionals
  with damped atom--atom dispersion corrections. \emph{Phys. Chem. Chem. Phys.}
  \textbf{2008}, \emph{10}, 6615--6620\relax
\mciteBstWouldAddEndPuncttrue
\mciteSetBstMidEndSepPunct{\mcitedefaultmidpunct}
{\mcitedefaultendpunct}{\mcitedefaultseppunct}\relax
\EndOfBibitem
\bibitem[Weigend and Ahlrichs(2005)Weigend, and Ahlrichs]{Weigend2005}
Weigend,~F.; Ahlrichs,~R. {Balanced basis sets of split valence, triple zeta
  valence and quadruple zeta valence quality for H to Rn: Design and assessment
  of accuracy}. \emph{Phys. Chem. Chem. Phys.} \textbf{2005}, \emph{7},
  3297--3305\relax
\mciteBstWouldAddEndPuncttrue
\mciteSetBstMidEndSepPunct{\mcitedefaultmidpunct}
{\mcitedefaultendpunct}{\mcitedefaultseppunct}\relax
\EndOfBibitem
\bibitem[Rappoport and Furche(2010)Rappoport, and Furche]{Rappoport2010a}
Rappoport,~D.; Furche,~F. {Property-optimized Gaussian basis sets for molecular
  response calculations}. \emph{J. Chem. Phys.} \textbf{2010}, \emph{133},
  134105\relax
\mciteBstWouldAddEndPuncttrue
\mciteSetBstMidEndSepPunct{\mcitedefaultmidpunct}
{\mcitedefaultendpunct}{\mcitedefaultseppunct}\relax
\EndOfBibitem
\bibitem[{Hunter}(2007)]{Hunter2007}
{Hunter},~J.~D. Matplotlib: A 2D Graphics Environment. \emph{Comput. Sci. Eng.}
  \textbf{2007}, \emph{9}, 90--95\relax
\mciteBstWouldAddEndPuncttrue
\mciteSetBstMidEndSepPunct{\mcitedefaultmidpunct}
{\mcitedefaultendpunct}{\mcitedefaultseppunct}\relax
\EndOfBibitem
\bibitem[Humphrey \latin{et~al.}(1996)Humphrey, Dalke, and Schulten]{HUMP96}
Humphrey,~W.; Dalke,~A.; Schulten,~K. {VMD} -- {V}isual {M}olecular {D}ynamics.
  \emph{J. Mol. Graphics} \textbf{1996}, \emph{14}, 33--38\relax
\mciteBstWouldAddEndPuncttrue
\mciteSetBstMidEndSepPunct{\mcitedefaultmidpunct}
{\mcitedefaultendpunct}{\mcitedefaultseppunct}\relax
\EndOfBibitem
\bibitem[Mao \latin{et~al.}(2016)Mao, Horn, Mardirossian, Head-Gordon,
  Skylaris, and Head-Gordon]{Mao2016approaching}
Mao,~Y.; Horn,~P.~R.; Mardirossian,~N.; Head-Gordon,~T.; Skylaris,~C.-K.;
  Head-Gordon,~M. Approaching the basis set limit for DFT calculations using an
  environment-adapted minimal basis with perturbation theory: Formulation,
  proof of concept, and a pilot implementation. \emph{J. Chem. Phys.}
  \textbf{2016}, \emph{145}, 044109\relax
\mciteBstWouldAddEndPuncttrue
\mciteSetBstMidEndSepPunct{\mcitedefaultmidpunct}
{\mcitedefaultendpunct}{\mcitedefaultseppunct}\relax
\EndOfBibitem
\bibitem[Owczarzy \latin{et~al.}(2008)Owczarzy, Moreira, You, Behlke, and
  Walder]{Owczarzy2008predicting}
Owczarzy,~R.; Moreira,~B.~G.; You,~Y.; Behlke,~M.~A.; Walder,~J.~A. Predicting
  stability of DNA duplexes in solutions containing magnesium and monovalent
  cations. \emph{Biochemistry} \textbf{2008}, \emph{47}, 5336--5353\relax
\mciteBstWouldAddEndPuncttrue
\mciteSetBstMidEndSepPunct{\mcitedefaultmidpunct}
{\mcitedefaultendpunct}{\mcitedefaultseppunct}\relax
\EndOfBibitem
\bibitem[Lippard and Berg(1994)Lippard, and Berg]{Lippard1994principles}
Lippard,~S.~J.; Berg,~J.~M. \emph{Principles of bioinorganic chemistry};
  University Science Books Mill Valley, CA, 1994; Vol.~70\relax
\mciteBstWouldAddEndPuncttrue
\mciteSetBstMidEndSepPunct{\mcitedefaultmidpunct}
{\mcitedefaultendpunct}{\mcitedefaultseppunct}\relax
\EndOfBibitem
\bibitem[Kaim \latin{et~al.}(2013)Kaim, Schwederski, and
  Klein]{Kaim2013bioinorganic}
Kaim,~W.; Schwederski,~B.; Klein,~A. \emph{Bioinorganic Chemistry--Inorganic
  Elements in the Chemistry of Life: An Introduction and Guide}; John Wiley \&
  Sons, 2013\relax
\mciteBstWouldAddEndPuncttrue
\mciteSetBstMidEndSepPunct{\mcitedefaultmidpunct}
{\mcitedefaultendpunct}{\mcitedefaultseppunct}\relax
\EndOfBibitem
\bibitem[Sigel(1993)]{Sigel1993interactions}
Sigel,~H. Interactions of metal ions with nucleotides and nucleic acids and
  their constituents. \emph{Chem. Soc. Rev.} \textbf{1993}, \emph{22},
  255--267\relax
\mciteBstWouldAddEndPuncttrue
\mciteSetBstMidEndSepPunct{\mcitedefaultmidpunct}
{\mcitedefaultendpunct}{\mcitedefaultseppunct}\relax
\EndOfBibitem
\bibitem[Lippert(2000)]{Lippert2000multiplicity}
Lippert,~B. Multiplicity of metal ion binding patterns to nucleobases.
  \emph{Coord. Chem. Rev.} \textbf{2000}, \emph{200}, 487--516\relax
\mciteBstWouldAddEndPuncttrue
\mciteSetBstMidEndSepPunct{\mcitedefaultmidpunct}
{\mcitedefaultendpunct}{\mcitedefaultseppunct}\relax
\EndOfBibitem
\bibitem[Valls \latin{et~al.}(2004)Valls, Us{\'o}n, Gouyette, and
  Subirana]{Valls2004cubic}
Valls,~N.; Us{\'o}n,~I.; Gouyette,~C.; Subirana,~J.~A. A cubic arrangement of
  DNA double helices based on nickel- guanine interactions. \emph{J. Am. Chem.
  Soc.} \textbf{2004}, \emph{126}, 7812--7816\relax
\mciteBstWouldAddEndPuncttrue
\mciteSetBstMidEndSepPunct{\mcitedefaultmidpunct}
{\mcitedefaultendpunct}{\mcitedefaultseppunct}\relax
\EndOfBibitem
\bibitem[Stasyuk \latin{et~al.}(2020)Stasyuk, Sol{\`a}, Swart, Fonseca~Guerra,
  Krygowski, and Szatylowicz]{Stasyuk2020effect}
Stasyuk,~O.~A.; Sol{\`a},~M.; Swart,~M.; Fonseca~Guerra,~C.; Krygowski,~T.~M.;
  Szatylowicz,~H. Effect of alkali metal cations on length and strength of
  hydrogen bonds in DNA base pairs. \emph{ChemPhysChem} \textbf{2020},
  \emph{21}, 2112--2126\relax
\mciteBstWouldAddEndPuncttrue
\mciteSetBstMidEndSepPunct{\mcitedefaultmidpunct}
{\mcitedefaultendpunct}{\mcitedefaultseppunct}\relax
\EndOfBibitem
\bibitem[Poater \latin{et~al.}(2005)Poater, Sodupe, Bertran*, and
  Sol{\`a}*]{Poater2005hydrogen}
Poater,~J.; Sodupe,~M.; Bertran*,~J.; Sol{\`a}*,~M. Hydrogen bonding and
  aromaticity in the guanine--cytosine base pair interacting with metal cations
  (M= Cu+, Ca2+ and Cu2+). \emph{Mol. Phys.} \textbf{2005}, \emph{103},
  163--173\relax
\mciteBstWouldAddEndPuncttrue
\mciteSetBstMidEndSepPunct{\mcitedefaultmidpunct}
{\mcitedefaultendpunct}{\mcitedefaultseppunct}\relax
\EndOfBibitem
\bibitem[Marder(2007)]{Marder2007will}
Marder,~T.~B. Will we soon be fueling our automobiles with ammonia--borane?
  \emph{Angew. Chem., Int. Ed.} \textbf{2007}, \emph{46}, 8116--8118\relax
\mciteBstWouldAddEndPuncttrue
\mciteSetBstMidEndSepPunct{\mcitedefaultmidpunct}
{\mcitedefaultendpunct}{\mcitedefaultseppunct}\relax
\EndOfBibitem
\bibitem[Peng and Chen(2008)Peng, and Chen]{Peng2008ammonia}
Peng,~B.; Chen,~J. Ammonia borane as an efficient and lightweight hydrogen
  storage medium. \emph{Energy Environ. Sci.} \textbf{2008}, \emph{1},
  479--483\relax
\mciteBstWouldAddEndPuncttrue
\mciteSetBstMidEndSepPunct{\mcitedefaultmidpunct}
{\mcitedefaultendpunct}{\mcitedefaultseppunct}\relax
\EndOfBibitem
\bibitem[Hirao \latin{et~al.}(1999)Hirao, Omoto, and Fujimoto]{Hirao1999lewis}
Hirao,~H.; Omoto,~K.; Fujimoto,~H. Lewis acidity of boron trihalides. \emph{J.
  Phys. Chem. A} \textbf{1999}, \emph{103}, 5807--5811\relax
\mciteBstWouldAddEndPuncttrue
\mciteSetBstMidEndSepPunct{\mcitedefaultmidpunct}
{\mcitedefaultendpunct}{\mcitedefaultseppunct}\relax
\EndOfBibitem
\bibitem[Plumley and Evanseck(2009)Plumley, and Evanseck]{Plumley2009periodic}
Plumley,~J.~A.; Evanseck,~J.~D. Periodic trends and index of boron Lewis
  acidity. \emph{J. Phys. Chem. A} \textbf{2009}, \emph{113}, 5985--5992\relax
\mciteBstWouldAddEndPuncttrue
\mciteSetBstMidEndSepPunct{\mcitedefaultmidpunct}
{\mcitedefaultendpunct}{\mcitedefaultseppunct}\relax
\EndOfBibitem
\bibitem[Branchadell and Oliva(1991)Branchadell, and
  Oliva]{Branchadell1991lewis}
Branchadell,~V.; Oliva,~A. The Lewis acidity scale of boron trihalides: an ab
  initio study. \emph{J. Mol. Struct.: THEOCHEM} \textbf{1991}, \emph{236},
  75--84\relax
\mciteBstWouldAddEndPuncttrue
\mciteSetBstMidEndSepPunct{\mcitedefaultmidpunct}
{\mcitedefaultendpunct}{\mcitedefaultseppunct}\relax
\EndOfBibitem
\bibitem[Bessac and Frenking(2003)Bessac, and Frenking]{Bessac2003bcl3}
Bessac,~F.; Frenking,~G. Why is BCl3 a stronger Lewis acid with respect to
  strong bases than BF3? \emph{Inorg. Chem.} \textbf{2003}, \emph{42},
  7990--7994\relax
\mciteBstWouldAddEndPuncttrue
\mciteSetBstMidEndSepPunct{\mcitedefaultmidpunct}
{\mcitedefaultendpunct}{\mcitedefaultseppunct}\relax
\EndOfBibitem
\bibitem[Staubitz \latin{et~al.}(2010)Staubitz, Robertson, Sloan, and
  Manners]{Staubitz2010amine}
Staubitz,~A.; Robertson,~A.~P.; Sloan,~M.~E.; Manners,~I. Amine- and phosphine-
  borane adducts: new interest in old molecules. \emph{Chem. Rev.}
  \textbf{2010}, \emph{110}, 4023--4078\relax
\mciteBstWouldAddEndPuncttrue
\mciteSetBstMidEndSepPunct{\mcitedefaultmidpunct}
{\mcitedefaultendpunct}{\mcitedefaultseppunct}\relax
\EndOfBibitem
\bibitem[Xu \latin{et~al.}(2000)Xu, Peng, Lam, Poon, Dong, Xu, Sun, Cheuk,
  Salhi, Lee, \latin{et~al.} others]{Xu2000transition}
others,, \latin{et~al.}  Transition metal carbonyl catalysts for
  polymerizations of substituted acetylenes. \emph{Macromolecules}
  \textbf{2000}, \emph{33}, 6918--6924\relax
\mciteBstWouldAddEndPuncttrue
\mciteSetBstMidEndSepPunct{\mcitedefaultmidpunct}
{\mcitedefaultendpunct}{\mcitedefaultseppunct}\relax
\EndOfBibitem
\bibitem[Sunley and Watson(2000)Sunley, and Watson]{Sunley2000high}
Sunley,~G.~J.; Watson,~D.~J. High productivity methanol carbonylation catalysis
  using iridium: the Cativa™ process for the manufacture of acetic acid.
  \emph{Catal. Today} \textbf{2000}, \emph{58}, 293--307\relax
\mciteBstWouldAddEndPuncttrue
\mciteSetBstMidEndSepPunct{\mcitedefaultmidpunct}
{\mcitedefaultendpunct}{\mcitedefaultseppunct}\relax
\EndOfBibitem
\bibitem[Rossomme \latin{et~al.}(2020)Rossomme, Lininger, Bell, Head-Gordon,
  and Head-Gordon]{Rossomme2020electronic}
Rossomme,~E.; Lininger,~C.~N.; Bell,~A.~T.; Head-Gordon,~T.; Head-Gordon,~M.
  Electronic structure calculations permit identification of the driving forces
  behind frequency shifts in transition metal monocarbonyls. \emph{Phys. Chem.
  Chem. Phys.} \textbf{2020}, \emph{22}, 781--798\relax
\mciteBstWouldAddEndPuncttrue
\mciteSetBstMidEndSepPunct{\mcitedefaultmidpunct}
{\mcitedefaultendpunct}{\mcitedefaultseppunct}\relax
\EndOfBibitem
\bibitem[Huber(2013)]{Huber2013molecular}
Huber,~K.-P. \emph{Molecular spectra and molecular structure: IV. Constants of
  diatomic molecules}; Springer Science \& Business Media, 2013\relax
\mciteBstWouldAddEndPuncttrue
\mciteSetBstMidEndSepPunct{\mcitedefaultmidpunct}
{\mcitedefaultendpunct}{\mcitedefaultseppunct}\relax
\EndOfBibitem
\bibitem[Abel \latin{et~al.}(1969)Abel, McLean, Tyfield, Braterman, Walker, and
  Hendra]{Abel1969vibrational}
Abel,~E.; McLean,~R.; Tyfield,~S.; Braterman,~P.; Walker,~A.; Hendra,~P.
  Vibrational and electronic spectra and bonding in ionic transition metal
  hexacarbonyls. \emph{J. Mol. Spectrosc.} \textbf{1969}, \emph{30},
  29--50\relax
\mciteBstWouldAddEndPuncttrue
\mciteSetBstMidEndSepPunct{\mcitedefaultmidpunct}
{\mcitedefaultendpunct}{\mcitedefaultseppunct}\relax
\EndOfBibitem
\bibitem[Reed and Duncan(2010)Reed, and Duncan]{Reed2010infrared}
Reed,~Z.~D.; Duncan,~M.~A. Infrared spectroscopy and structures of manganese
  carbonyl cations, Mn (CO) n+(n= 1- 9). \emph{J. Am. Soc. Mass Spectrom.}
  \textbf{2010}, \emph{21}, 739--749\relax
\mciteBstWouldAddEndPuncttrue
\mciteSetBstMidEndSepPunct{\mcitedefaultmidpunct}
{\mcitedefaultendpunct}{\mcitedefaultseppunct}\relax
\EndOfBibitem
\end{mcitethebibliography}
\end{document}

\section{Working basis}
In this section, we describe a method to construct an orthogonal working molecular orbital (MO) basis.
We will work with the occupied and unoccupied space defined by the polarized wavefunction throughout this section.
The polarized wavefunction $(\Phi_{\text{POL}})$ gives a set of non-orthogonal occupied orbitals spanning the occupied space.
In order to convert this set to an orthogonal basis, we symmetrically orthogonalize it as shown in Eq.~\eqref{eq::occ_sym_ortho} where the columns of the $\vb{T}^{\text{ortho}}$ and $\vb{T}$ contain the orthogonal and non-orthogonal occupied orbitals.
Symmetric orthogonalization is sufficient as there is no linear dependency in the occupied space.
\begin{align}
    \vb{T}^{\text{ortho}} &= \vb{T}\boldsymbol{\sigma}_{OO}^{-1/2} \label{eq::occ_sym_ortho} \\
    \boldsymbol{\sigma}_{OO} &= \vb{T}^T\vb{ST}  
\end{align}
$\Phi_{\text{POL}}$ also gives a set of non-orthogonal virtual orbitals defined by FERFs.
As these orbitals might have some over overlap with the occupied space, we will project them out first as shown in Eq.~\eqref{eq::Vproj} where $\vb{P} = \vb{T}\boldsymbol{\sigma}_{OO}^{-1}\vb{T}^T$ is the projector onto the occupied space. 
These orbitals are orthogonalized using standard canonical orthogonalization in order to account for linear dependence in the virtual space.
$\vb{U}$ is matrix of eigenvectors with eigenvalues greater than threshold and $\vb{s}$ is a diagonal matrix of eigenvalues greater than threshold.
\begin{align}
    \vb{V}^{\text{proj}} &= (\vb{I}-\vb{P}\vb{S})\vb{V} \label{eq::Vproj} \\
    \vb{V}^{\text{ortho}} &= \vb{V}^{\text{proj}}\vb{Us}^{-1/2} \label{eq::Vortho}
\end{align}
The space not spanned by these the occupied and the FERF virtual space is the void space.
An orthogonal basis for this space is formed by the eigenvectors of the projector ($\vb{R}_{\text{POL}}$) for this space (Eq.~\eqref{eq::R_pol}). and 
\begin{align}
    \vb{P}_{\text{POL}} &= \vb{C}_{\text{POL}}\boldsymbol{\sigma}\vb{C}_{\text{POL}}^T \label{eq::P_pol} \\
    \vb{R}_{\text{POL}} &= \vb{I} - \vb{P}_{\text{POL}} \label{eq::R_pol} \\
    \vb{R}^{\text{ortho}} &\leftarrow eigenvectors(\vb{R}_{\text{POL}})
\end{align}
An orthogonal basis for the entire Hilbert space is formed by the columns of $\vb{T}^{\text{ortho}}$, $\vb{V}^{\text{ortho}}$, and $\vb{R}^{\text{ortho}}$.

\section{Cost function and gradient}
In this section, we describe the cost function to be minimized to solve for $\vb{X}_{OV}^{\text{CT}}$ and derive its corresponding gradient.
Let $\vb{U}_{XY}$ be the cumulative unitary transformation generated by $\boldsymbol{\theta}_X$ and $\boldsymbol{\theta}_Y$ respectively where $\vb{U}_X$ (referred to as $\vb{U}_{\text{curr}}^{\text{CT}}$ in Eq.~(26)) denotes the current step taken and $\vb{U}_Y$ denotes the next step. 
\begin{align}
    \vb{U}_X &= e^{\boldsymbol{\theta}_X} \\
    \vb{U}_Y &= e^{\boldsymbol{\theta}_Y} \\
    \vb{U}_{XY} &= \vb{U}_X \vb{U}_Y
\end{align}
As we are trying to find the generator of unitary transformation that rotates the polarized density matrix ($\vb{P}^{\text{POL}}$) to the fully-relaxed density matrix ($\vb{P}^{\text{CT}}$), we are trying to find the $\boldsymbol{\theta}$ that satisfies Eq.~\eqref{eq::Pconnect_problem}
\begin{align}
    \vb{P^{\text{CT}}} &= e^{\boldsymbol{\theta}} \vb{P}^{\text{POL}} e^{-\boldsymbol{\theta}} \label{eq::Pconnect_problem} 
\end{align}
Hence, we define the cost function C as shown in Eq.~\eqref{eq::Pconnect_cost}
\begin{align}
    C &= || \vb{P}^{\text{CT}}-\vb{U}_{XY}\vb{P}^{\text{POL}}\vb{U}_{XY}^{\dagger} || _F^2 \label{eq::Pconnect_cost} 
\end{align}
Before minimizing the cost function $C$, let us form a working orthogonal basis.
Let $\mathbb{H}$ denote the Hilbert space of the whole complex.
For the polarized wavefunction $\Phi_{\text{POL}}$, $\mathbb{H}$ can be partitioned into occupied space ($\mathbb{T}$), virtual space spanned by FERFs ($\mathbb{Q}$), and the void space which is the unoccupied space not spanned by FERFs ($\mathbb{R}$).
$\mathbb{Q}$ and $\mathbb{R}$ can be clubbed together to form the unoccupied space $\mathbb{V}$.
These spaces can also be made strongly orthogonal to each other without loss of generality.
\begin{align}
    \mathbb{H} &= \mathbb{T} \oplus \mathbb{Q} \oplus \mathbb{R} \nonumber \\
    \mathbb{H} &= \mathbb{T} \oplus \mathbb{V} \quad \quad \quad \quad  \text{where } \mathbb{V} = \mathbb{Q} \oplus \mathbb{R}
\end{align}
Let the columns of $\vb{T}$ and $\vb{V}$ form a orthonormal basis for the occupied and unoccupied subspaces respectively.
Let $N_T$ and $N_V$ be the dimensions of these subspaces respectively.
In this orthonormal projectors, the density matrix which is the projector on the occupied space can be written as 
\begin{align}
    \vb{P}^{\text{POL}} &= \left[
\begin{array}{cC}
\vb{I}_{N_T \times N_T} & \vb{0} \\
\vb{0} & \vb{0} 
\end{array}
\right] \label{eq::Ppol}
\end{align}
Expressing the cost function in this orthonormal basis and taking its derivative with respect to the elements of the generator $\boldsymbol{\theta}_Y$,
\begin{align}
    \frac{\partial C}{\partial (\boldsymbol{\theta}_Y)_{ia}} &= \frac{\partial}{\partial (\boldsymbol{\theta}_Y)_{ia}} \sum_{p,q} \Big( \big(\vb{P}^{\text{CT}} - \vb{U}_{XY}\vb{P}^{\text{POL}}\vb{U}_{XY}^{\dagger}\big)_{pq}^2\Big)  \nonumber \\ 
    &= 2\sum_{p,q}\big(\vb{P}^{\text{CT}} - \vb{U}_{XY}\vb{P}^{\text{POL}}\vb{U}_{XY}^{\dagger}\big)_{pq}\frac{\partial}{\partial (\boldsymbol{\theta}_Y)_{ia}}\big(\vb{P}^{\text{CT}} - \vb{U}_{XY}\vb{P}^{\text{POL}}\vb{U}_{XY}^{\dagger}\big)_{pq}  \label{eq::cost_der}
\end{align}
Consider,
\begin{align}
    \frac{\partial}{\partial (\boldsymbol{\theta}_Y)_{ia}}\big(\vb{P}^{\text{CT}} - \vb{U}_{XY}\vb{P}^{\text{POL}}\vb{U}_{XY}^{\dagger}\big)_{pq} &= \frac{\partial}{\partial (\boldsymbol{\theta}_Y)_{ia}}\big(\vb{P}^{\text{CT}}_{pq} - \sum_{k,l,m,n}(\vb{U}_X)_{pk}(\vb{U}_Y)_{kl}P^{\text{POL}}_{lm}(\vb{U^{\dagger}}_Y)_{mn}(\vb{U}^{\dagger}_X)_{nq}\big) \nonumber \\
    &= -\sum_{k,l,m,n}(\vb{U}_X)_{pk} \frac{\partial \vb({U}_Y)_{kl}}{\partial(\boldsymbol{\theta}_Y)_{ia} }P^{\text{POL}}_{lm}(\vb{U^{\dagger}}_Y)_{mn}(\vb{U}^{\dagger}_X)_{nq} \nonumber \\
    & \quad - \sum_{k,l,m,n}(\vb{U}_X)_{pk}(\vb{U}_Y)_{kl}P^{\text{POL}}_{lm}\frac{\partial( \vb{U^{\dagger}}_Y)_{mn}}{\partial(\boldsymbol{\theta}_Y)_{ia}}(\vb{U}^{\dagger}_X)_{nq} \nonumber \\
\end{align}
We know that $\frac{\partial \vb({U}_Y)_{kl}}{\partial(\boldsymbol{\theta}_Y)_{ia}} = \delta_{ki}\delta_{la}-\delta_{ka}\delta_{li}$.
Substituting this above,
\begin{align}
    \frac{\partial}{\partial (\boldsymbol{\theta}_Y)_{ia}}\big(\vb{P}^{\text{CT}} - \vb{U}_{XY}\vb{P}^{\text{POL}}\vb{U}_{XY}^{\dagger}\big)_{pq} &= -\sum_{m,n}(\vb{U}_X)_{pi}P^{\text{POL}}_{am}(\vb{U}^{\dagger}_Y)_{mn}(\vb{U}^{\dagger}_X)_{nq}+\sum_{m,n}(\vb{U}_X)_{pa}P^{\text{POL}}_{im}(\vb{U}^{\dagger}_Y)_{mn}(\vb{U}^{\dagger}_X)_{nq}  \nonumber \\
    & \quad -\sum_{k,l}(\vb{U}_X)_{pk}(\vb{U}_Y)_{kl}P^{\text{POL}}_{la}(\vb{U}^{\dagger}_X)_{iq}
    +\sum_{k,l}(\vb{U}_X)_{pk}(\vb{U}_Y)_{kl}P^{\text{POL}}_{li}(\vb{U}^{\dagger}_X)_{aq} \nonumber
\end{align}
Evaluating the above expression at $\boldsymbol{\theta}_Y=\vb{0}$, we have $\vb{U}_Y=\vb{I}$ and $\vb{U}_{XY}=\vb{U}_X$.
\begin{align}
    \frac{\partial}{\partial (\boldsymbol{\theta}_Y)_{ia}}\big(\vb{P}^{\text{CT}} - \vb{U}_{XY}\vb{P}^{\text{POL}}\vb{U}_{XY}^{\dagger}\big)_{pq}\Big|_{\boldsymbol{\theta}_Y=\boldsymbol{0}} &= -\sum_{m}(\vb{U}_X)_{pi}P^{\text{POL}}_{am}(\vb{U}_X^{\dagger})_{mq}
    +\sum_{m}(\vb{U}_X)_{pa}P^{\text{POL}}_{im}(\vb{U}_X^{\dagger})_{mq} \nonumber \\
    \quad &-\sum_{l}(\vb{U}_X)_{pl}P^{\text{POL}}_{la}(\vb{U}_X^{\dagger})_{iq}
    +\sum_{l}(\vb{U}_X)_{pl}P^{\text{POL}}_{li}(\vb{U}_X^{\dagger})_{aq} \nonumber 
\end{align}
From Eq.~\eqref{eq::Ppol}, we know that $P^{\text{POL}}_{am} = P^{\text{POL}}_{la} = 0$ as $a$ is an index for the unoccupied space.
Thus we have,
\begin{align}
    \frac{\partial}{\partial (\boldsymbol{\theta}_Y)_{ia}}\big(\vb{P}^{\text{CT}} - \vb{U}_{XY}\vb{P}^{\text{POL}}\vb{U}_{XY}^{\dagger}\big)_{pq}\Big|_{\boldsymbol{\theta}_Y=\boldsymbol{0}} &= \sum_{m}(\vb{U}_X)_{pa}P^{\text{POL}}_{im}(\vb{U}_X^{\dagger})_{mq} +\sum_{l}(\vb{U}_X)_{pl}P^{\text{POL}}_{li}(\vb{U}_X^{\dagger})_{aq}  \nonumber \\
    &= (\vb{U}_X)_{pa}(\vb{U}^{\dagger}_X)_{iq} + (\vb{U}_X)_{pi}(\vb{U}^{\dagger}_X)_{aq} \label{eq::cost_der2}
\end{align}
Plugging Eq.~\eqref{eq::cost_der2} into Eq.~\eqref{eq::cost_der}, 
\begin{align}
    \frac{\partial C}{\partial (\boldsymbol{\theta}_Y)_{ia}}\Big|_{\boldsymbol{\theta}_Y=\boldsymbol{0}} &= 
    2\sum_{p,q}\Big(P^{\text{CT}}_{pq} - \sum_{j,k}^{N_T}(\vb{U}_X)_{pj}P^{\text{POL}}_{jk}(\vb{U}^{\dagger}_{kq})\Big)\Big((\vb{U}_X)_{pa}(\vb{U}^{\dagger}_X)_{iq} + (\vb{U}_X)_{pi}(\vb{U}^{\dagger}_X)_{aq} \Big) \nonumber \\
    &= 2\sum_{p,q} \Bigg( (\vb{U}^{\dagger}_X)_{iq}P^{\text{CT}}_{qp}(\vb{U}_X)_{pa} 
    + (\vb{U}^{\dagger}_X)_{ip}P^{\text{CT}}_{pq}(\vb{U}_X)_{qa} \nonumber \\ 
    &\quad -\sum_k^{N_T}(\vb{U}^{\dagger}_X)_{jq}(\vb{U}^{\dagger}_X)_{jp}(\vb{U}_X)_{pa}(\vb{U}_X^{\dagger})_{iq}  -\sum_k^{N_T}(\vb{U}_X)_{pj}(\vb{U}_X)_{pi}(\vb{U}_X^{\dagger})_{jq}(\vb{U}_X)_{qa}\Bigg) \label{eq::cost_der3}
\end{align}
From Eq.~\eqref{eq::cost_der3}, we can see that the first two terms of the RHS are equal and the third and fourth terms are $0$ as $\sum_p(\vb{U}_X^{\dagger})_{jp}(\vb{U}_X)_{pa}=0$ and $\sum_p(\vb{U}_X^{\dagger})_{jq}(\vb{U}_X)_{qa}=0$ as basis vectors spanning $\mathbb{T}$ and $\mathbb{V}$ are orthonormal to each other.
Finally, we have
\begin{align}
    \frac{\partial C}{\partial (\boldsymbol{\theta}_Y)_{ia}}\Big|_{\boldsymbol{\theta}_Y=\boldsymbol{0}} &= 4 \sum_{p,q}(\vb{U}^{\dagger}_X)_{iq}P^{\text{CT}}_{qp}(\vb{U}_X)_{pa} 
\end{align}
\begin{figure}[H]
    \centering
    \includegraphics[width=\textwidth]{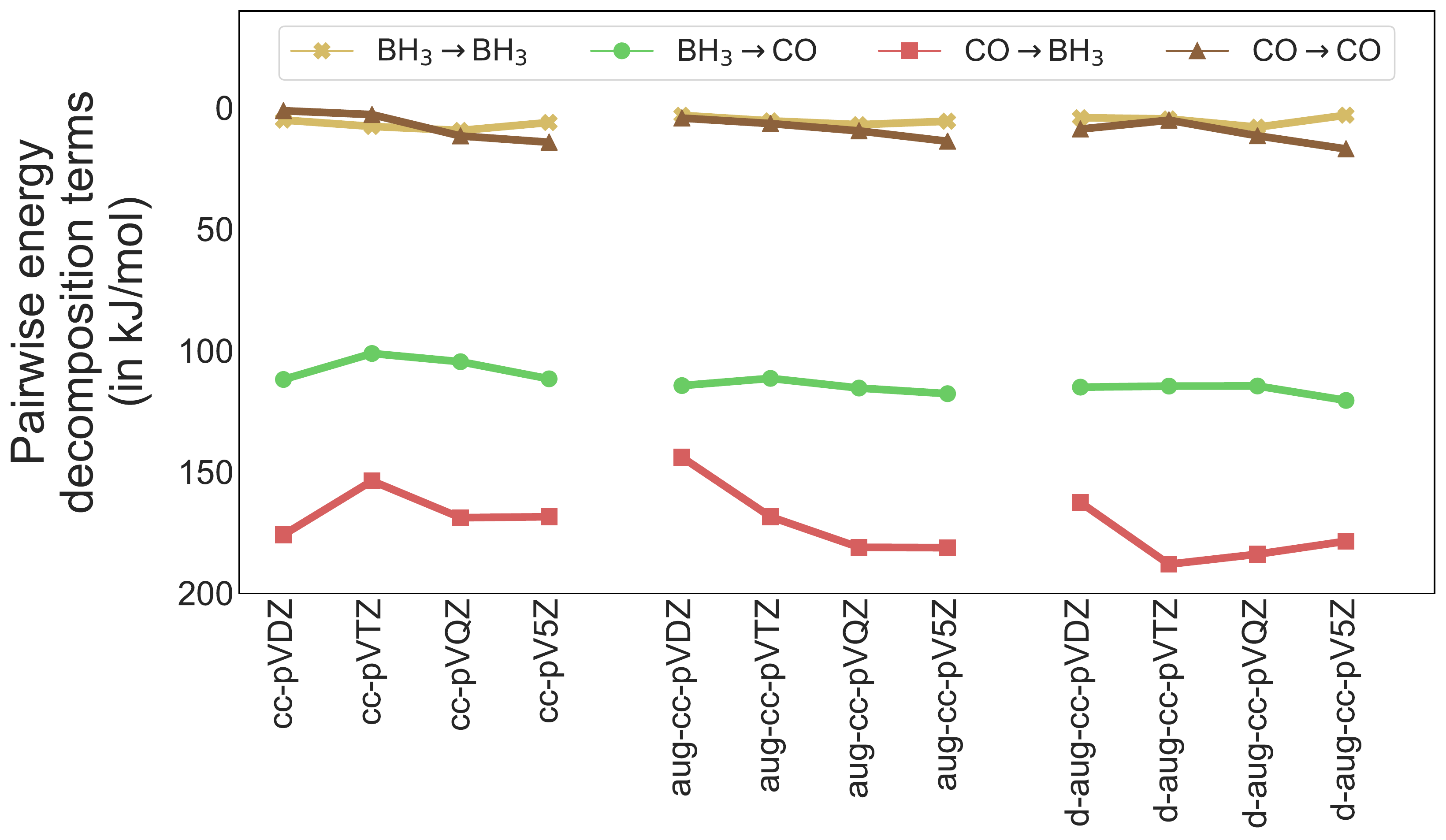}
    \caption{Convergence properties of the non-perturbative pairwise energy decomposition components of the total charge transfer energy with respect to increasing the highest angular momentum of the basis set for the Dunning basis set series: cc-pVXZ, aug-ccpVXZ, and d-aug-cc-pVXZ (X=D, T, Q, and 5) for the \ce{BH3}--\ce{CO} system at equilibrium geometry using $\omega$B97X-D. }
    \label{fig::basis_lim_energy_kJmol}
\end{figure}

\begin{figure}[H]
    \centering
    \includegraphics[width=\textwidth]{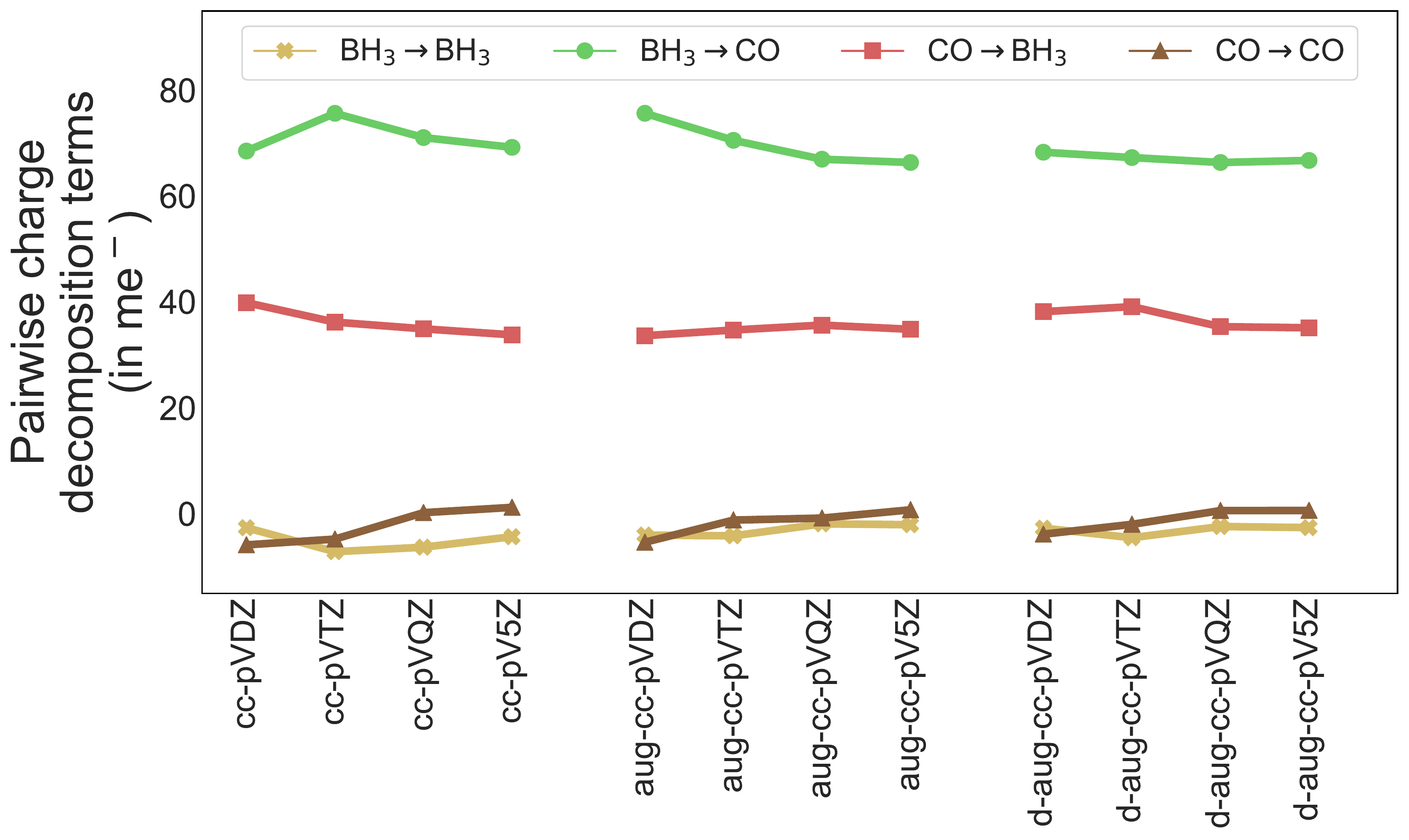}
    \caption{Convergence properties of the non-perturbative pairwise charge decomposition components of  total charge transfer with respect to increasing the highest angular momentum of the basis set for the Dunning basis set series: cc-pVXZ, aug-ccpVXZ, and d-aug-cc-pVXZ (X=D, T, Q, and 5) for the \ce{BH3}--\ce{CO} system at equilibrium geometry using $\omega$B97X-D.}
    \label{fig::basis_lim_charge_me}
\end{figure}

\begin{table}[H]
\label{tab::DNA_aEDA}
\caption{Adiabatic energy decomposition analysis for the DNA base pairs thymine(T)-adenine(A), guanine(G)-cytosine(C), and their corresponding metallated versions. All values are in kJ/mol.}
\begin{tabular}{ccccc}
\hline
    System        & $\Delta E_{\text{INT}}$   & $\Delta E_{\text{FRZ}}$    & $\Delta E_{\text{POL}}$   & $\Delta E_{\text{CT}}$    \\ \hline
T(1):A(2)   & -69.4  & -31.8  & -11.5 & -20.6 \\
G(1):C(2)    & -134.3 & -64.8  & -27.5 & -28.5 \\
\ce{Na+} G(1):C(2)  & -149.3 & -77.1  & -31.7 & -29.4 \\
\ce{Mg^2+} G(1):C(2) & -214.2 & -120.9 & -53.7 & -21.2 \\
\ce{Ca^2+} G(1):C(2) & -195.2 & -114.1 & -48.3 & -16.5 \\ \hline
\end{tabular}
\end{table}

\begin{table}[H]
\caption{Energy decomposition analysis and non-perturbative charge decomposition analysis for the adduct \ce{BH3}--\ce{NCl_pH_q} ($p+q=3$). All values are in kJ/mol.}
\resizebox{\textwidth}{!}{%
\begin{tabular}{cccccc|cccc}
\hline
          & \multicolumn{5}{c|}{Energy decomposition analysis}   & \multicolumn{4}{c}{non-perturbative charge decomposition}                          \\ \hline
          & $\Delta E_{\text{INT}}$ & $\Delta E_{\text{GD}}$   & $\Delta E_{\text{FRZ}}$   & $\Delta E_{\text{POL}}$    & $\Delta E_{\text{CT}}$        & \ce{BH3}$\rightarrow$\ce{BH3} & \ce{BH3}$\rightarrow$\ce{NCl_mH_n} & \ce{NCl_mH_n}$\rightarrow$\ce{BH3} & \ce{NCl_mH_n}$\rightarrow$\ce{NCl_pH_q} \\ \hline
\ce{BH3}-\ce{NH3}   & -133.1    & 56.3 & 116.5 & -148.1 & -157.7  & -2.7              & -16.4             & -138.1            & -0.5              \\
\ce{BH3}-\ce{NClH2} & -113.7    & 49.3 & 141.3 & -156.5 & -147.9 & -2.6              & -28.3             & -117.9            & 0.9               \\
\ce{BH3}-\ce{NCl2H} & -87.3     & 42.6 & 152.4 & -138.7 & -143.6  & -2.3              & -32.7             & -109.2            & 0.6               \\
\ce{BH3}-\ce{NCl3}  & -58.2     & 34.1 & 145.5 & -111.1 & -126.7  & -1.7              & -29.8             & -95.6             & 0.5              \\ \hline
\end{tabular}
}
\end{table}

\begin{table}[H]
\caption{Adiabatic energy decomposition analysis for the adduct \ce{BX3}--\ce{NH3} (\ce{X} = F, Cl, or, Br). All values are in kJ/mol.}
\begin{tabular}{ccccc}
\hline
    System        & $\Delta E_{\text{INT}}$   & $\Delta E_{\text{FRZ}}$    & $\Delta E_{\text{POL}}$   & $\Delta E_{\text{CT}}$    \\ \hline
\ce{BF3}--\ce{NH3}  & -90.8     & -20.4 & -6.6 & -63.8  \\
\ce{BCl3}--\ce{NH3} & -107.7    & -10.6 & -3.2 & -93.8  \\
\ce{BBr3}--\ce{NH3} & -119.2    & -9.9  & -3.0 & -106.3 \\ \hline
\end{tabular}
\end{table}

\begin{table}[H]
\caption{Adiabatic energy decomposition analysis for the adduct \ce{BH3}--\ce{NMe_pH_q} ($p+q=3$). All values are in kJ/mol.}
\label{tab::BH3_NMepHq_aEDA}
\begin{tabular}{ccccc}
\hline
    System        & $\Delta E_{\text{INT}}$   & $\Delta E_{\text{FRZ}}$    & $\Delta E_{\text{POL}}$   & $\Delta E_{\text{CT}}$    \\ \hline
\ce{BH3}--\ce{NH3}   & -133.1    & -16.7 & -14.5 & -101.9 \\
\ce{BH3}--\ce{NMeH2} & -153.2    & -18.3 & -28.2 & -106.8 \\
\ce{BH3}--\ce{NMe2H} & -163.1    & -19.7 & -37.4 & -106.0 \\
\ce{BH3}--\ce{NMe3}  & -163.8    & -21.1 & -40.4 & -102.3 \\ \hline
\end{tabular}
\end{table}

\begin{figure}[H]
     \centering
     \begin{subfigure}[b]{0.49\textwidth}
         \centering
         \includegraphics[width=\textwidth]{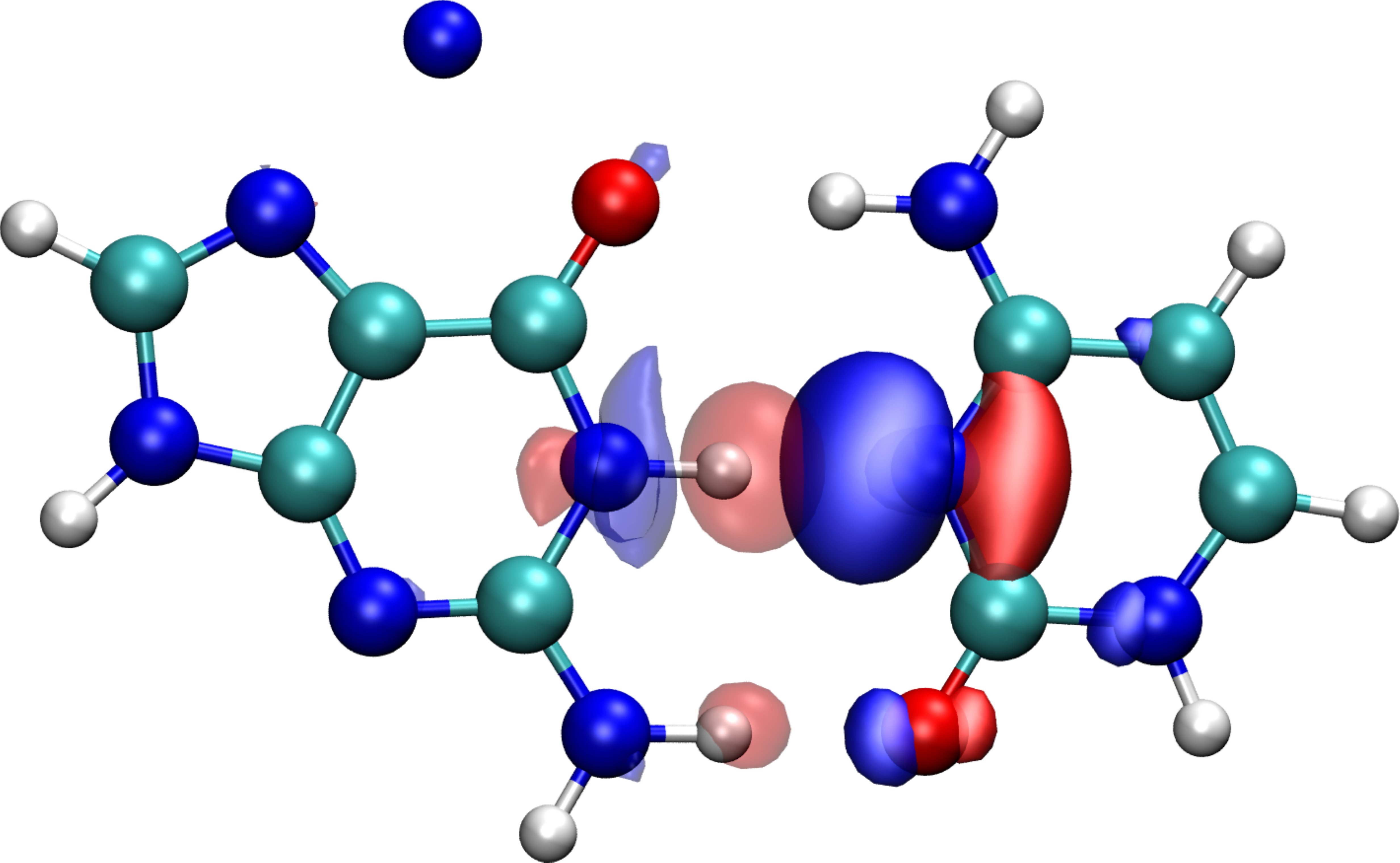}
         \caption{$\Delta E_{\text{CT}}^{\text{COVP1}}=-21.2$ kJ/mol}
         \label{fig:DNA_GC_Na_covp1}
     \end{subfigure}
     \begin{subfigure}[b]{0.49\textwidth}
         \centering
         \includegraphics[width=\textwidth]{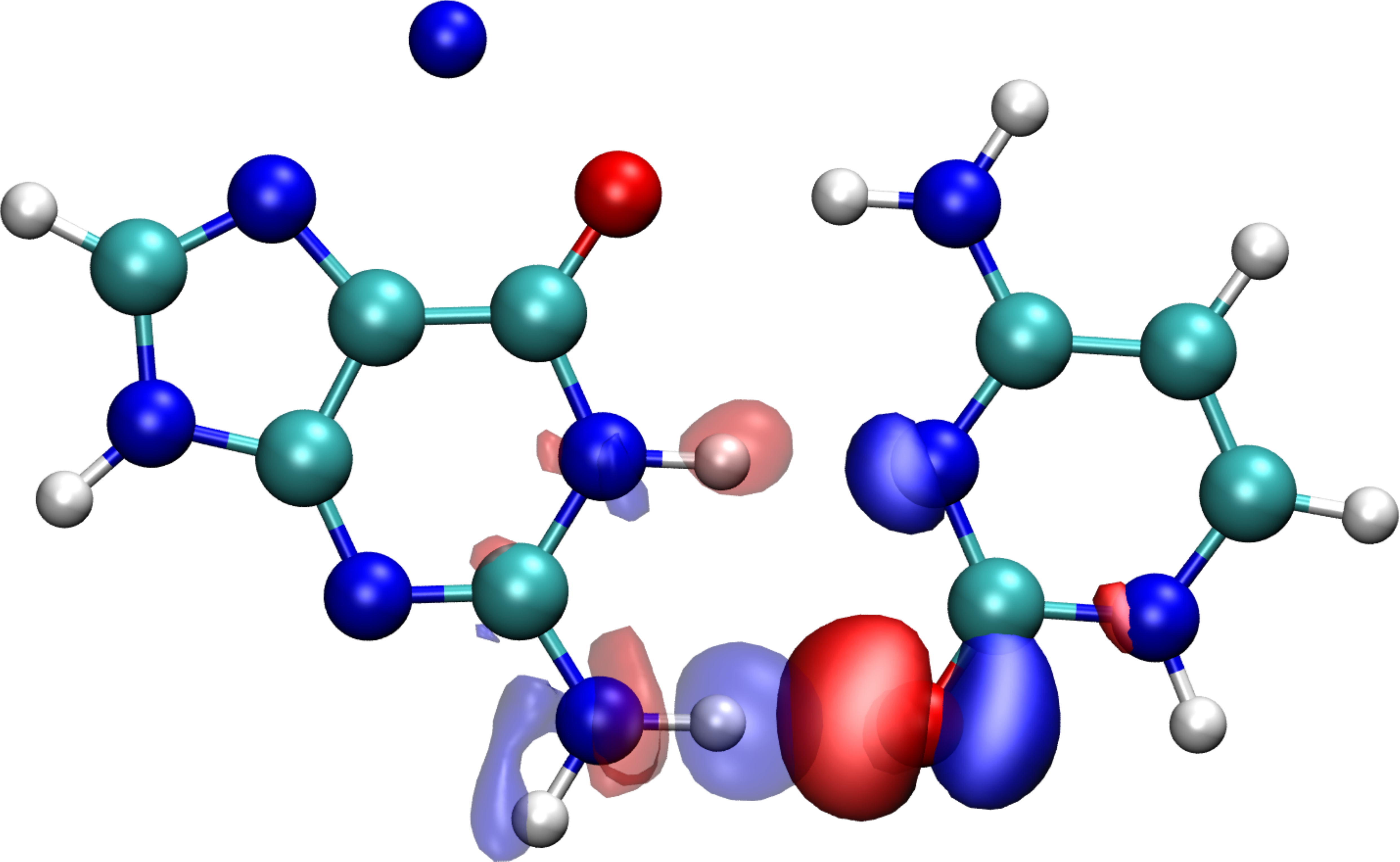}
         \caption{$\Delta E_{\text{CT}}^{\text{COVP2}}=-15.0$ kJ/mol}
         \label{fig:DNA_GC_Na_covp2}
     \end{subfigure}
     \begin{subfigure}[b]{0.49\textwidth}
         \centering
         \includegraphics[width=\textwidth]{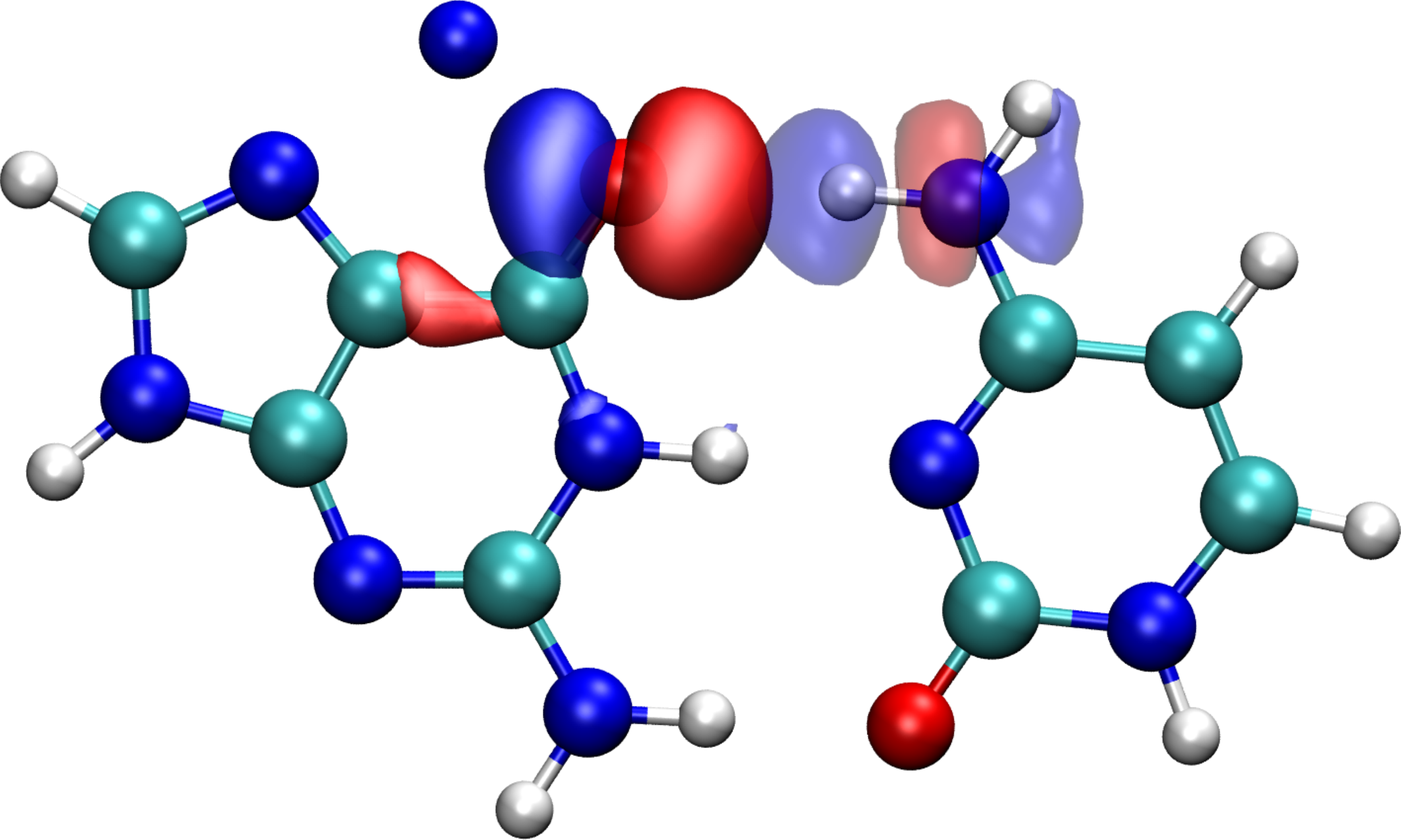}
         \caption{$\Delta E_{\text{CT}}^{\text{COVP3}}=-8.1$ kJ/mol}
         \label{fig:DNA_GC_Na_covp3}
     \end{subfigure}
        \caption{(a) Most significant COVP for \ce{Na+} guanine:cytosine complex (b) Second most significant COVP  (c) Third most significant COVP}
        \label{fig:DNA_GC_Na_covp}
\end{figure}

\begin{figure}[H]
     \centering
     \begin{subfigure}[b]{0.49\textwidth}
         \centering
         \includegraphics[width=\textwidth]{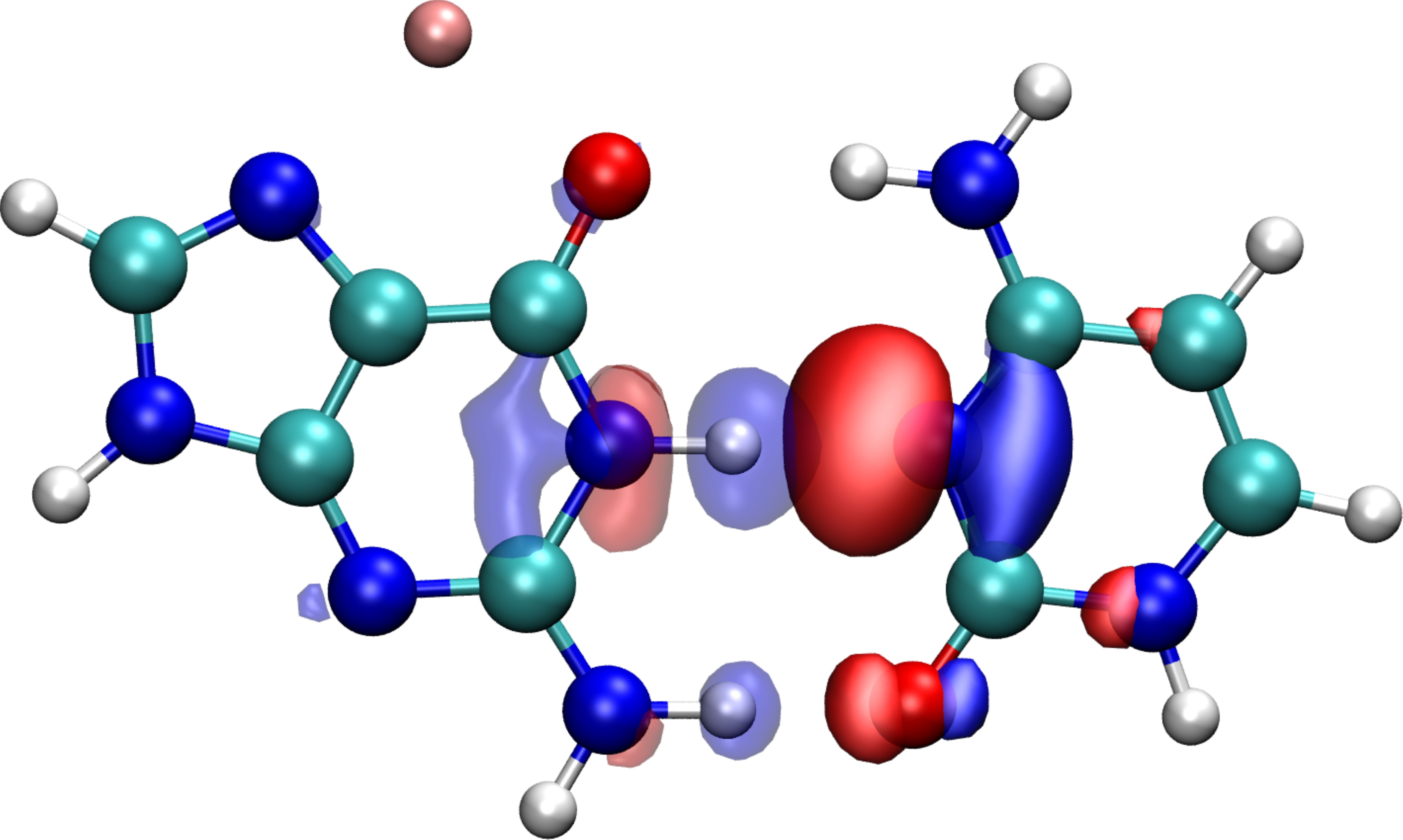}
         \caption{$\Delta E_{\text{CT}}^{\text{COVP1}}=-32.2$ kJ/mol}
         \label{fig:DNA_GC_Mg_covp1}
     \end{subfigure}
     \begin{subfigure}[b]{0.49\textwidth}
         \centering
         \includegraphics[width=\textwidth]{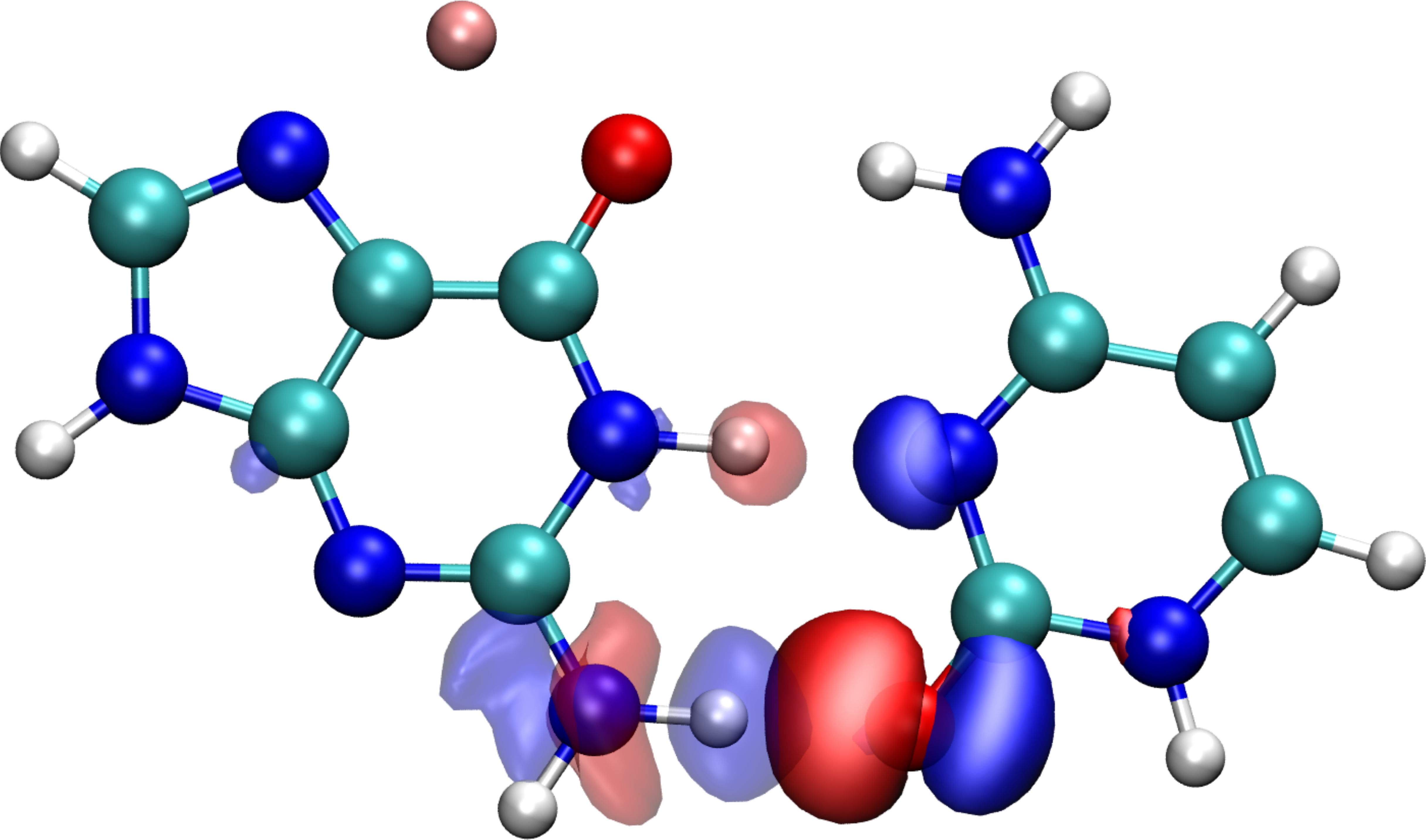}
         \caption{$\Delta E_{\text{CT}}^{\text{COVP2}}=-27.2$ kJ/mol}
         \label{fig:DNA_GC_Mg_covp2}
     \end{subfigure}
     \begin{subfigure}[b]{0.49\textwidth}
         \centering
         \includegraphics[width=\textwidth]{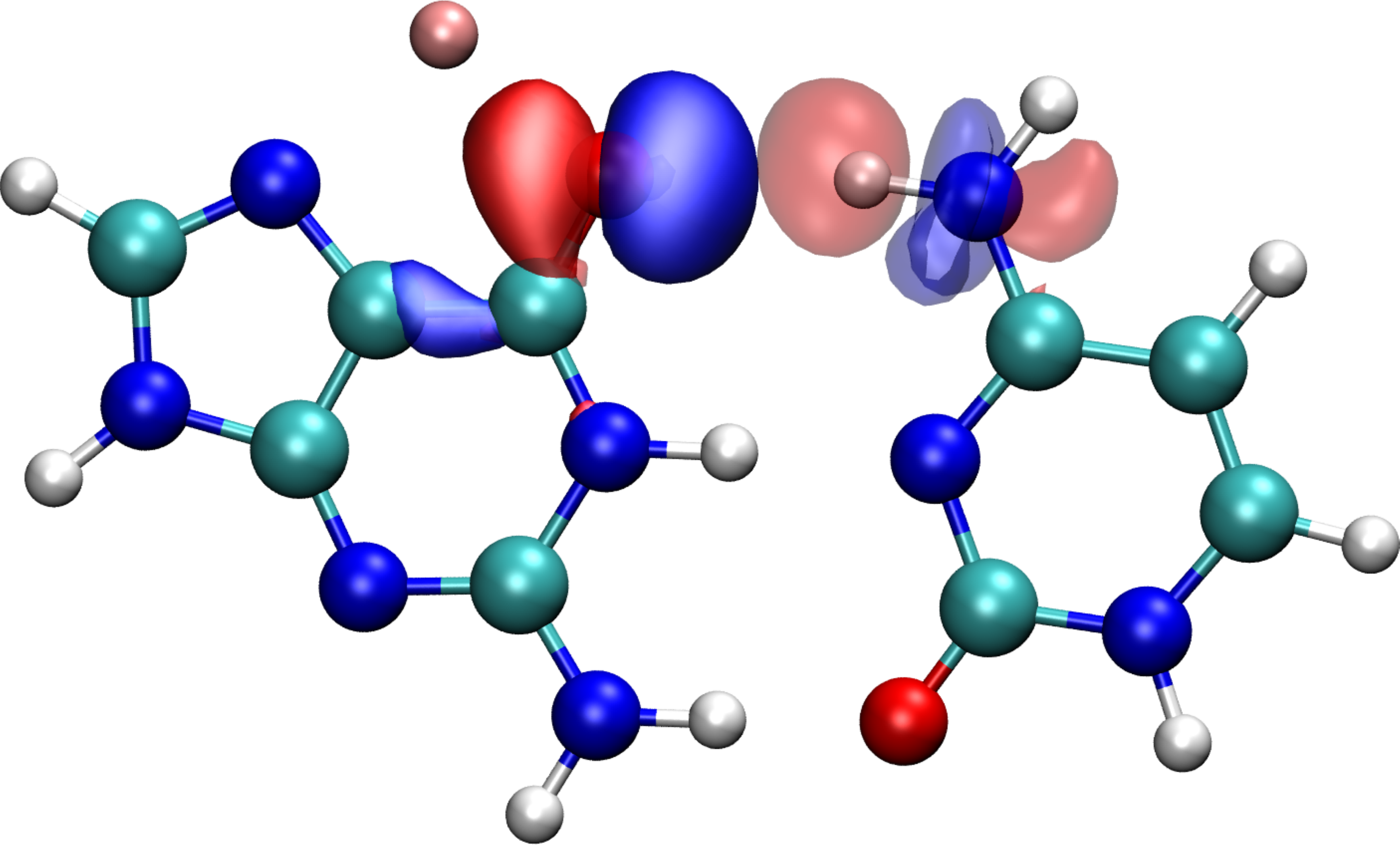}
         \caption{$\Delta E_{\text{CT}}^{\text{COVP3}}=-2.7$ kJ/mol}
         \label{fig:DNA_GC_Mg_covp3}
     \end{subfigure}
        \caption{(a) Most significant COVP for \ce{Mg^2+} guanine:cytosine complex (b) Second most significant COVP (c) Third most significant COVP.}
        \label{fig:DNA_GC_Mg_covp}
\end{figure}

\begin{figure}[H]
     \centering
     \begin{subfigure}[b]{0.49\textwidth}
         \centering
         \includegraphics[width=\textwidth]{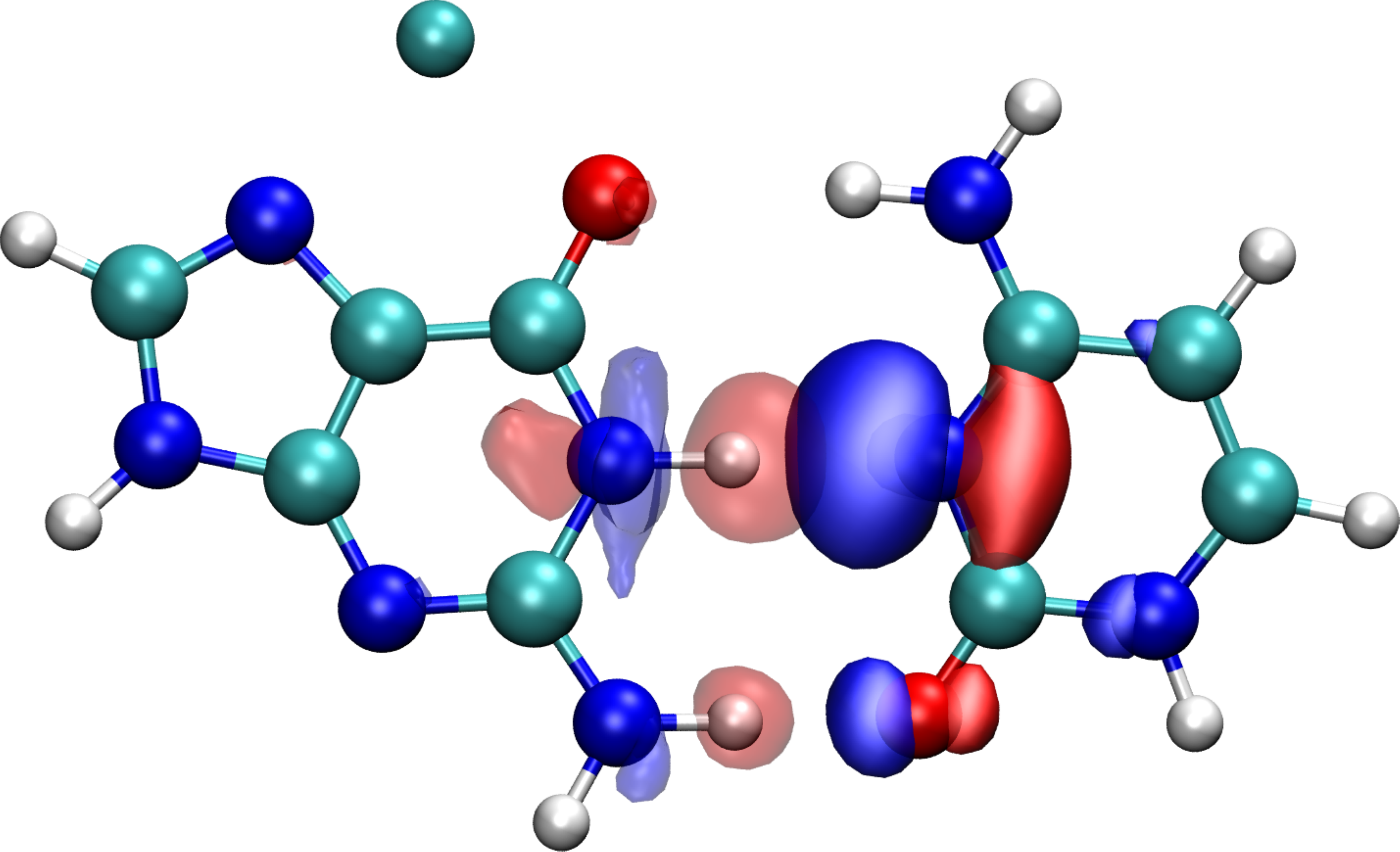}
         \caption{$\Delta E_{\text{CT}}^{\text{COVP1}}=-28.2$ kJ/mol}
         \label{fig:DNA_GC_Ca_covp1}
     \end{subfigure}
     \begin{subfigure}[b]{0.49\textwidth}
         \centering
         \includegraphics[width=\textwidth]{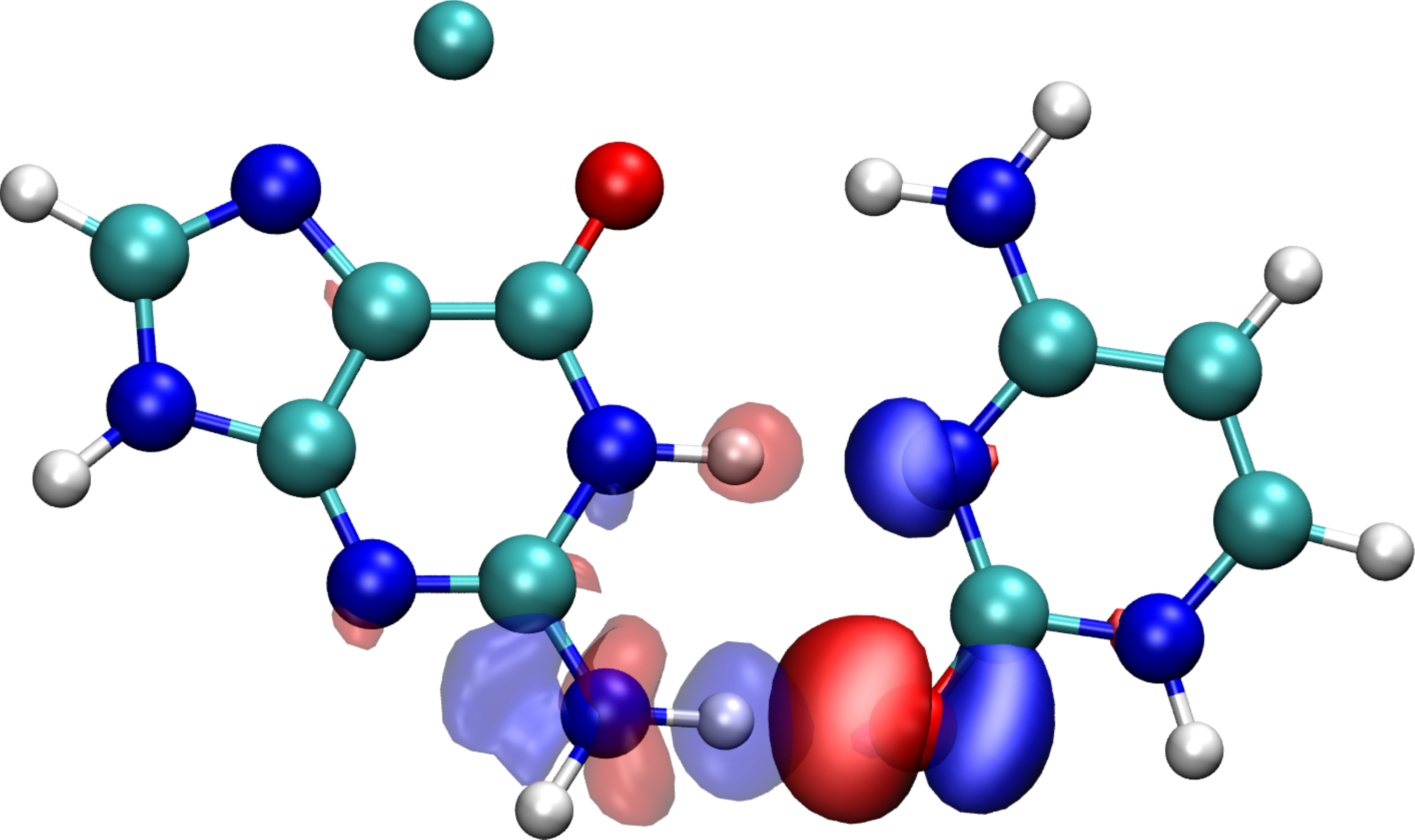}
         \caption{$\Delta E_{\text{CT}}^{\text{COVP2}}=-23.9$ kJ/mol}
         \label{fig:DNA_GC_Ca_covp2}
     \end{subfigure}
     \begin{subfigure}[b]{0.49\textwidth}
         \centering
         \includegraphics[width=\textwidth]{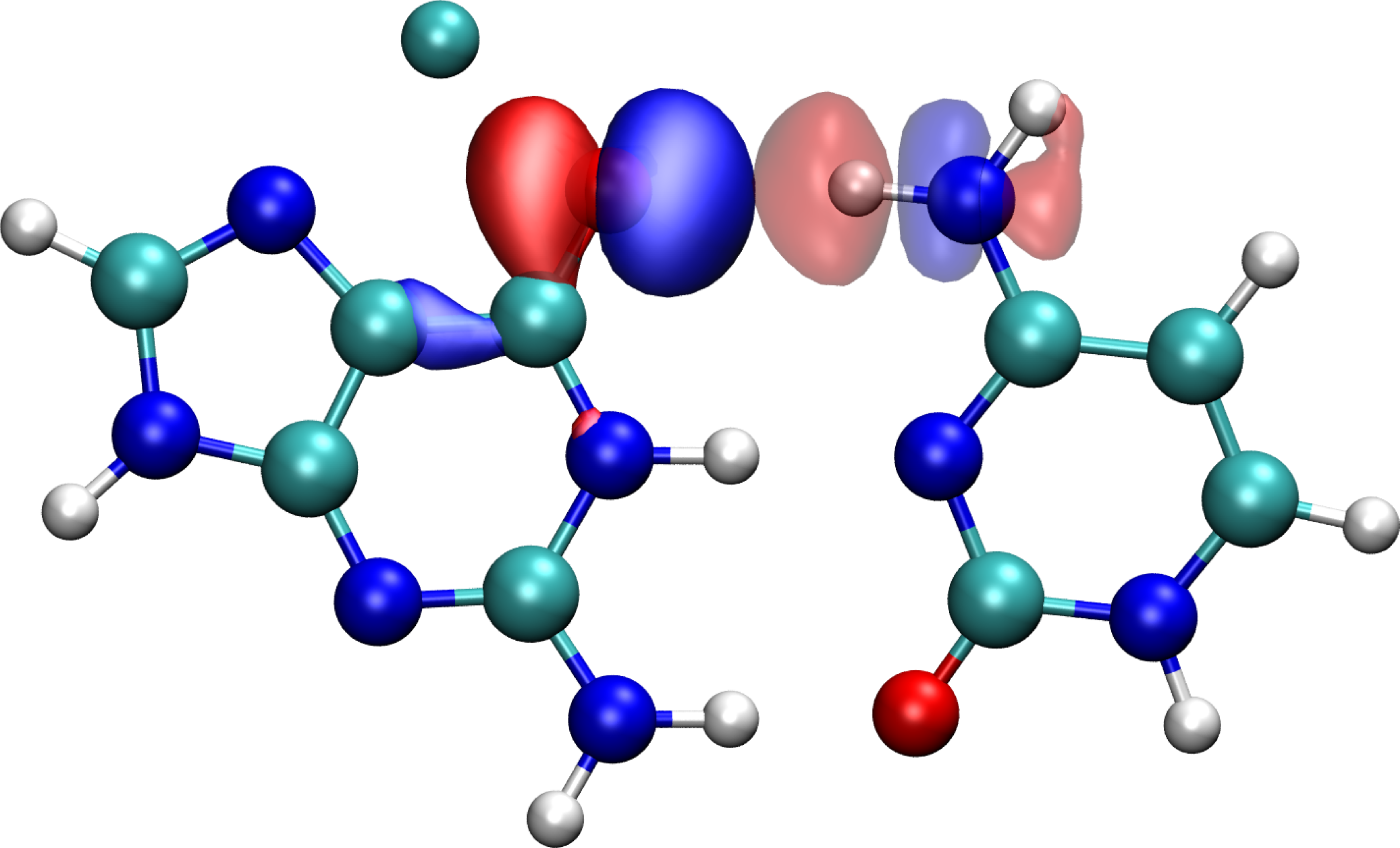}
         \caption{$\Delta E_{\text{CT}}^{\text{COVP3}}=-2.9$ kJ/mol}
         \label{fig:DNA_GC_Ca_covp3}
     \end{subfigure}
        \caption{(a) Most significant COVP for \ce{Ca^2+} guanine:cytosine complex (b) Second most significant COVP (c) Third most significant COVP.}
        \label{fig:DNA_GC_Ca_covp}
\end{figure}

\begin{table}[H]
\caption{Variational forward-backward charge transfer analysis for a series of 3d transition metal hexacarbonyls. All values are in kJ/mol.}
\begin{tabular}{ccc}
\hline
System        & \ce{M(CO)5}$\rightarrow$\ce{CO} & \ce{CO}$\rightarrow$\ce{M(CO)5} \\ \hline
\ce{V(CO)6-}  & -76.2      & -18.1       \\
\ce{Cr(CO)6} & -45.9      & -28.2       \\
\ce{Mn(CO)6+} & -22.6      & -39.7       \\
\hline
\end{tabular}
\end{table}

\begin{table}[H]
\caption{Residual energy not decomposed into pairwise additive terms using the non-perturbative CT analysis for some representative systems considered in this work using $\omega$B97X-D density functional and def2-TZVPD basis set. Residual energy is shown in both kJ/mol and as a percentage of total variational CT energy.}
\begin{tabular}{ccc}
\hline 
System    & Residual CT energy (kJ/mol) & Residual CT energy  \\ \hline
\ce{BF3-NH3}   & 6.86E-04       & 4.40E-04\% \\
\ce{BH3-NH3}   & 9.03E-04       & 5.78E-04\% \\
\ce{BH3-CO}    & 7.95E-04       & 2.70E-04\% \\
\ce{Mg^2+} guanine : cytosine & 7.95E-04       & 1.18E-03\% \\
\ce{Cr(CO)6}     & 1.13E-06       & 5.33E-07\% \\ \hline
\end{tabular}
\end{table}

\section{Geometries}
\textbf{\ce{BH3-CO}} \\
Charge=0, Multiplicity=1
\vskip-15pt
\begin{table}[H]
\hskip-7.0cm\begin{tabular}{llll}
B & -1.39961 & 0.70844 & 0.18735 \\
H & -0.23322 & 0.93106 & -0.03460 \\
H & -1.99575 & 0.22613 & -0.74595 \\
H & -1.98125 & 1.61908 & 0.72730 \\
C & -1.37842 & -0.39863 & 1.23375 \\
O & -1.36206 & -1.21576 & 2.00602 \\
    \end{tabular}
\end{table}

\textbf{Adenine $-$ thymine} \\
Charge = 0, Multiplicity=1
\vskip-15pt
\begin{table}[H]
\hskip-7.0cm\begin{tabular}{llll}
N & 0.93525 & 0.00327 & -0.37856 \\
C & 1.67122 & 0.00447 & 0.74242 \\
C & 3.06945 & 0.00430 & 0.59464 \\
C & 3.55409 & 0.00299 & -0.70204 \\
N & 2.83563 & 0.00192 & -1.82718 \\
C & 1.54067 & 0.00208 & -1.57294 \\
N & 4.09798 & 0.00504 & 1.51249 \\
C & 5.17274 & 0.00449 & 0.78245 \\
N & 4.91459 & 0.00351 & -0.56354 \\
N & 1.06445 & 0.00576 & 1.93073 \\
H & 0.86880 & 0.00126 & -2.42521 \\
H & 6.18223 & 0.00493 & 1.16396 \\
H & 5.58348 & 0.00302 & -1.31239 \\
H & 0.04879 & 0.00524 & 1.99236 \\
H & 1.62641 & 0.00609 & 2.76146 \\
N & -3.90108 & 0.00699 & -1.51169 \\
C & -4.59567 & 0.00885 & -0.33183 \\
C & -3.98779 & 0.00849 & 0.86365 \\
C & -2.52733 & 0.00612 & 0.87694 \\
N & -1.90988 & 0.00427 & -0.35524 \\
C & -2.52226 & 0.00443 & -1.58215 \\
C & -4.71384 & 0.01056 & 2.16875 \\
O & -1.86472 & 0.00574 & 1.90468 \\
O & -1.92589 & 0.00347 & -2.63504 \\
H & -4.37932 & 0.00746 & -2.39553 \\
H & -0.86725 & 0.00300 & -0.36549 \\
H & -5.67377 & 0.01071 & -0.42975 \\
H & -4.44606 & -0.86505 & 2.76243 \\
H & -4.44328 & 0.88616 & 2.76118 \\
H & -5.79328 & 0.01219 & 2.01585 \\
    \end{tabular}
\end{table}

\textbf{guanine $-$ cytosine} \\
Charge=0, Multiplicity=1
\vskip-15pt
\begin{table}[H]
\hskip-7.0cm\begin{tabular}{llll}
N & -2.71342 & -0.03847 & 0.00000 \\
O & 1.20960 & 2.30520 & 0.00000 \\
N & -0.79313 & 1.23201 & 0.00000 \\
C & -2.15789 & 1.18326 & 0.00000 \\
N & -2.91822 & 2.25468 & 0.00000 \\
C & -2.20356 & 3.39259 & 0.00000 \\
C & -0.82822 & 3.56058 & 0.00000 \\
C & -0.01516 & 2.39072 & 0.00000 \\
N & -2.71003 & 4.65671 & 0.00000 \\
C & -1.63428 & 5.51622 & 0.00000 \\
N & -0.49613 & 4.89892 & 0.00000 \\
H & -2.16332 & -0.89459 & 0.00000 \\
H & -3.71436 & -0.08261 & 0.00000 \\
H & -0.26632 & 0.34822 & 0.00000 \\
H & -3.68517 & 4.89551 & 0.00000 \\
H & -1.76809 & 6.58698 & 0.00000 \\
O & -1.11321 & -2.47931 & 0.00000 \\
N & 2.72139 & -0.02967 & 0.00000 \\
N & 0.78997 & -1.23187 & 0.00000 \\
C & 2.89188 & -2.42337 & 0.00000 \\
C & 0.10811 & -2.39792 & 0.00000 \\
N & 0.85582 & -3.57753 & 0.00000 \\
C & 2.20860 & -3.58117 & 0.00000 \\
C & 2.11745 & -1.21115 & 0.00000 \\
H & 3.97064 & -2.40651 & 0.00000 \\
H & 0.32769 & -4.43351 & 0.00000 \\
H & 2.69059 & -4.54922 & 0.00000 \\
H & 2.16866 & 0.84189 & 0.00000 \\
H & 3.72260 & 0.02130 & 0.00000 \\
    \end{tabular}
\end{table}

\textbf{\ce{Na+} guanine $-$ cytosine} \\
Charge=+1, Multiplicity=1
\vskip-15pt
\begin{table}[H]
\hskip-7.0cm\begin{tabular}{llll}
N & -2.65427 & -0.03416 & 0.03627 \\
O & 1.23315 & 2.41597 & -0.06545 \\
N & -0.73630 & 1.24166 & -0.01668 \\
C & -2.11519 & 1.17712 & 0.00788 \\
N & -2.89187 & 2.24861 & 0.00465 \\
C & -2.21123 & 3.38907 & -0.02376 \\
C & -0.84074 & 3.55313 & -0.04865 \\
C & -0.01845 & 2.40865 & -0.04524 \\
N & -2.70715 & 4.66311 & -0.03474 \\
C & -1.64204 & 5.51877 & -0.06468 \\
N & -0.49939 & 4.89434 & -0.07407 \\
H & -2.09506 & -0.89372 & 0.03818 \\
H & -3.65671 & -0.08599 & 0.05442 \\
H & -0.18946 & 0.35722 & -0.01308 \\
H & -3.68093 & 4.91639 & -0.02263 \\
H & -1.77192 & 6.58931 & -0.07877 \\
Na & 1.84056 & 4.62888 & -0.10276 \\
O & -1.11024 & -2.36217 & 0.03702 \\
N & 2.82195 & -0.07269 & -0.05055 \\
N & 0.84004 & -1.19096 & -0.00644 \\
C & 2.88894 & -2.46889 & -0.02441 \\
C & 0.11296 & -2.33555 & 0.01748 \\
N & 0.80910 & -3.53840 & 0.01934 \\
C & 2.15829 & -3.59976 & -0.00106 \\
C & 2.16524 & -1.23419 & -0.02692 \\
H & 3.96712 & -2.49681 & -0.04085 \\
H & 0.24667 & -4.37374 & 0.03659 \\
H & 2.60010 & -4.58655 & 0.00255 \\
H & 2.30412 & 0.80227 & -0.05347 \\
H & 3.82405 & -0.05981 & -0.06808 \\
    \end{tabular}
\end{table}

\textbf{\ce{Mg^2+} guanine $-$ cytosine} \\
Charge=+2, Multiplicity=1
\vskip-15pt
\begin{table}[H]
\hskip-7.0cm\begin{tabular}{llll}
N & -2.61052 & -0.03711 & 0.02913 \\
O & 1.23629 & 2.56983 & -0.07040 \\
N & -0.68532 & 1.23801 & -0.02048 \\
C & -2.08767 & 1.15804 & 0.00365 \\
N & -2.88581 & 2.23779 & 0.00206 \\
C & -2.24699 & 3.37849 & -0.02531 \\
C & -0.88101 & 3.52903 & -0.05089 \\
C & -0.05249 & 2.41972 & -0.04797 \\
N & -2.71709 & 4.67819 & -0.03546 \\
C & -1.66054 & 5.52027 & -0.06499 \\
N & -0.51309 & 4.87276 & -0.07549 \\
H & -2.03211 & -0.90577 & 0.02630 \\
H & -3.61605 & -0.09801 & 0.04688 \\
H & -0.10135 & 0.34587 & -0.01622 \\
H & -3.68792 & 4.95986 & -0.02287 \\
H & -1.77273 & 6.59371 & -0.07830 \\
Mg & 1.45076 & 4.48589 & -0.10116 \\
O & -1.10289 & -2.22319 & 0.01710 \\
N & 2.94599 & -0.12510 & -0.03382 \\
N & 0.89565 & -1.13327 & -0.00904 \\
C & 2.88131 & -2.51154 & -0.01200 \\
C & 0.12312 & -2.25302 & 0.00700 \\
N & 0.75563 & -3.48006 & 0.01205 \\
C & 2.09895 & -3.60922 & 0.00302 \\
C & 2.22167 & -1.24973 & -0.01835 \\
H & 3.95711 & -2.58991 & -0.01938 \\
H & 0.15265 & -4.28890 & 0.02365 \\
H & 2.49260 & -4.61628 & 0.00852 \\
H & 2.48475 & 0.76840 & -0.04205 \\
H & 3.94843 & -0.16363 & -0.04308 \\
    \end{tabular}
\end{table}

\textbf{\ce{Ca^2+} guanine$-$cytosine} \\
Charge=+2, Multiplicity=1
\vskip-15pt
\begin{table}[H]
\hskip-7.0cm\begin{tabular}{llll}
N & -2.58298 & -0.02635 & 0.12588 \\
O & 1.24827 & 2.53708 & -0.29794 \\
N & -0.67352 & 1.25761 & -0.06985 \\
C & -2.06212 & 1.17282 & 0.06706 \\
N & -2.85085 & 2.24999 & 0.14043 \\
C & -2.21180 & 3.39024 & 0.04367 \\
C & -0.85156 & 3.55722 & -0.11985 \\
C & -0.02982 & 2.43714 & -0.16788 \\
N & -2.70333 & 4.67550 & 0.07637 \\
C & -1.66509 & 5.53215 & -0.06648 \\
N & -0.51361 & 4.91001 & -0.18933 \\
H & -2.01933 & -0.89045 & 0.00183 \\
H & -3.58350 & -0.08686 & 0.22211 \\
H & -0.09705 & 0.37090 & -0.08356 \\
Ca & 1.76233 & 4.63674 & -0.55969 \\
H & -3.67270 & 4.93583 & 0.18876 \\
H & -1.79983 & 6.60293 & -0.07610 \\
O & -1.09576 & -2.22619 & -0.24098 \\
N & 2.93860 & -0.14267 & 0.13480 \\
N & 0.89972 & -1.14774 & -0.05726 \\
C & 2.87318 & -2.53252 & 0.10355 \\
C & 0.12503 & -2.26268 & -0.14235 \\
N & 0.75440 & -3.49334 & -0.11493 \\
C & 2.09120 & -3.62701 & 0.00868 \\
C & 2.21776 & -1.26905 & 0.05960 \\
H & 3.94410 & -2.61578 & 0.20225 \\
H & 0.15213 & -4.29975 & -0.18306 \\
H & 2.48143 & -4.63531 & 0.02510 \\
H & 2.47403 & 0.74792 & 0.07500 \\
H & 3.93527 & -0.18024 & 0.24002 \\
    \end{tabular}
\end{table}

\textbf{\ce{BF3-NH3}} \\
Charge=0, Multiplicity=1
\vskip-15pt
\begin{table}[H]
\hskip-7.0cm\begin{tabular}{llll}
B & -1.59518 & 0.54532 & 0.32606 \\
F & -0.23695 & 0.73407 & 0.37665 \\
F & -2.04606 & 0.02288 & -0.85972 \\
F & -2.32812 & 1.61046 & 0.78543 \\
N & -1.88362 & -0.66855 & 1.44789 \\
H & -1.37027 & -1.50567 & 1.19306 \\
H & -1.58449 & -0.36790 & 2.36960 \\
H & -2.87397 & -0.88713 & 1.47898 \\
    \end{tabular}
\end{table}

\textbf{\ce{BCl3-NH3}} \\
Charge=0, Multiplicity=1
\vskip-15pt
\begin{table}[H]
\hskip-7.0cm\begin{tabular}{llll}
B & -1.59730 & 0.47668 & 0.37371 \\
Cl & 0.21507 & 0.75918 & 0.43351 \\
Cl & -2.17485 & -0.19770 & -1.23203 \\
Cl & -2.57784 & 1.91478 & 0.95444 \\
N & -1.87918 & -0.69508 & 1.45460 \\
H & -1.36490 & -1.53304 & 1.19591 \\
H & -1.57836 & -0.39161 & 2.37738 \\
H & -2.87293 & -0.90943 & 1.48346 \\
    \end{tabular}
\end{table}

\textbf{\ce{BBr3-NH3}} \\
Charge=0, Multiplicity=1
\vskip-15pt
\begin{table}[H]
\hskip-7.0cm\begin{tabular}{llll}
B & -1.62062 & 0.49314 & 0.42622 \\
Br & 0.36052 & 0.77141 & 0.45700 \\
Br & -2.28711 & -0.20549 & -1.32786 \\
Br & -2.65735 & 2.06939 & 1.09335 \\
N & -1.90097 & -0.68153 & 1.48934 \\
H & -1.41915 & -1.53020 & 1.20189 \\
H & -1.56429 & -0.40332 & 2.40867 \\
H & -2.89964 & -0.86928 & 1.54658 \\
    \end{tabular}
\end{table}

\textbf{\ce{BH3-NH3}} \\
Charge=0, Multiplicity=1
\vskip-15pt
\begin{table}[H]
\hskip-7.0cm\begin{tabular}{llll}
B & -1.58381 & 0.55528 & 0.34974 \\
H & -0.38328 & 0.72225 & 0.38092 \\
H & -1.98319 & 0.12157 & -0.70965 \\
H & -2.21409 & 1.51121 & 0.74819 \\
N & -1.87902 & -0.64048 & 1.44259 \\
H & -1.38608 & -1.48859 & 1.18744 \\
H & -1.57317 & -0.36299 & 2.36835 \\
H & -2.86964 & -0.85142 & 1.48661 \\
    \end{tabular}
\end{table}

\textbf{\ce{BH3-NMeH2}} \\
Charge=0, Multiplicity=1
\vskip-15pt
\begin{table}[H]
\hskip-7.0cm\begin{tabular}{llll}
B & -1.59029 & 0.54147 & 0.39945 \\
H & -0.39079 & 0.72711 & 0.43604 \\
H & -1.97430 & 0.09320 & -0.66147 \\
H & -2.23445 & 1.49911 & 0.77674 \\
N & -1.87937 & -0.63349 & 1.49587 \\
H & -1.38938 & -1.46918 & 1.19558 \\
C & -1.48404 & -0.28508 & 2.87007 \\
H & -2.86906 & -0.85414 & 1.46563 \\
H & -0.42001 & -0.05929 & 2.87101 \\
H & -1.69613 & -1.09524 & 3.56844 \\
H & -2.02592 & 0.61129 & 3.16307 \\
    \end{tabular}
\end{table}

\textbf{\ce{BH3-NMe2H}} \\
Charge=0, Multiplicity=1
\vskip-15pt
\begin{table}[H]
\hskip-7.0cm\begin{tabular}{llll}
B & -1.65861 & 0.49965 & 0.45298 \\
H & -0.45988 & 0.69270 & 0.47355 \\
H & -2.05038 & 0.05252 & -0.60590 \\
H & -2.30362 & 1.45632 & 0.83556 \\
N & -1.93468 & -0.67196 & 1.55441 \\
H & -1.39656 & -1.47217 & 1.23838 \\
C & -1.44483 & -0.29306 & 2.88935 \\
C & -3.35278 & -1.06310 & 1.59141 \\
H & -0.38434 & -0.06317 & 2.82100 \\
H & -1.61396 & -1.09433 & 3.61145 \\
H & -1.97644 & 0.60373 & 3.20305 \\
H & -3.65413 & -1.38649 & 0.59816 \\
H & -3.93824 & -0.18679 & 1.86471 \\
H & -3.52118 & -1.86102 & 2.31737 \\
    \end{tabular}
\end{table}

\textbf{\ce{BH3-NMe3}} \\
Charge=0, Multiplicity=1
\vskip-15pt
\begin{table}[H]
\hskip-7.0cm\begin{tabular}{llll}
B & -1.64959 & 0.48063 & 0.45236 \\
H & -0.45143 & 0.67820 & 0.46539 \\
H & -2.05248 & 0.07545 & -0.61921 \\
H & -2.29029 & 1.42127 & 0.87602 \\
N & -1.90758 & -0.74278 & 1.51306 \\
C & -1.16084 & -1.93826 & 1.07956 \\
C & -1.44677 & -0.33716 & 2.85395 \\
C & -3.34795 & -1.05555 & 1.56332 \\
H & -0.38655 & -0.10000 & 2.80202 \\
H & -1.61644 & -1.14081 & 3.57503 \\
H & -1.99170 & 0.55415 & 3.15628 \\
H & -3.67753 & -1.34798 & 0.56899 \\
H & -3.88925 & -0.16164 & 1.86463 \\
H & -3.53785 & -1.86403 & 2.27386 \\
H & -1.49956 & -2.22155 & 0.08562 \\
H & -1.32323 & -2.76194 & 1.77935 \\
H & -0.10253 & -1.69175 & 1.03217 \\
    \end{tabular}
\end{table}

\textbf{\ce{BH3-NClH2}} \\
Charge=0, Multiplicity=1
\vskip-15pt
\begin{table}[H]
\hskip-7.0cm\begin{tabular}{llll}
B & -1.53814 & 0.56757 & 0.34596 \\
H & -0.32860 & 0.71229 & 0.34544 \\
H & -1.97360 & 0.10722 & -0.68119 \\
H & -2.14127 & 1.51869 & 0.77989 \\
N & -1.68261 & -0.60847 & 1.46811 \\
H & -1.16707 & -1.43357 & 1.17646 \\
H & -1.29749 & -0.29016 & 2.35227 \\
Cl & -3.29323 & -1.16311 & 1.82886 \\
    \end{tabular}
\end{table}

\textbf{\ce{BH3-NCl2H}} \\
Charge=0, Multiplicity=1
\vskip-15pt
\begin{table}[H]
\hskip-7.0cm\begin{tabular}{llll}
B & -1.61591 & 0.60936 & 0.30585 \\
H & -0.43193 & 0.81038 & 0.22872 \\
H & -2.12018 & 0.10704 & -0.67266 \\
H & -2.26226 & 1.50726 & 0.79531 \\
N & -1.80504 & -0.58507 & 1.42578 \\
Cl & -1.07122 & -2.10160 & 0.98489 \\
Cl & -1.31931 & -0.13703 & 3.03652 \\
H & -2.79617 & -0.79988 & 1.51139 \\
    \end{tabular}
\end{table}

\textbf{\ce{BH3-NCl3}} \\
Charge=0, Multiplicity=1
\vskip-15pt
\begin{table}[H]
\hskip-7.0cm\begin{tabular}{llll}
B & -1.57622 & 0.64506 & 0.27283 \\
H & -0.38144 & 0.79375 & 0.24696 \\
H & -2.06443 & 0.13474 & -0.70278 \\
H & -2.20824 & 1.54234 & 0.76857 \\
N & -1.75093 & -0.58737 & 1.43594 \\
Cl & -0.89748 & -2.03312 & 0.95071 \\
Cl & -1.11081 & -0.08456 & 2.98367 \\
Cl & -3.43245 & -1.00039 & 1.65990 \\
    \end{tabular}
\end{table}

\textbf{\ce{V(CO)5-CO}} \\
Charge=-1, Multiplicity=1
\vskip-15pt
\begin{table}[H]
\hskip-7.0cm\begin{tabular}{llll}
V & -1.05210 & 1.59272 & -0.03670 \\
C & 0.90067 & 1.57110 & -0.03405 \\
C & -3.00482 & 1.61663 & -0.04089 \\
C & -1.06987 & -0.36021 & -0.05396 \\
C & -1.05834 & 1.58186 & 1.91658 \\
C & -1.04625 & 1.60436 & -1.98915 \\
O & -1.06337 & 1.57707 & 3.06784 \\
O & -1.04170 & 1.61032 & -3.14041 \\
O & 2.05187 & 1.55644 & -0.03428 \\
O & -4.15598 & 1.63404 & -0.04307 \\
O & -1.07863 & -1.51140 & -0.06736 \\
C & -1.03419 & 3.54531 & -0.02414 \\
O & -1.02535 & 4.69655 & -0.01610 \\
    \end{tabular}
\end{table}

\textbf{\ce{Cr(CO)5-CO}} \\
Charge=0, Multiplicity=1
\vskip-15pt
\begin{table}[H]
\hskip-7.0cm\begin{tabular}{llll}
Cr & -1.05211 & 1.59284 & -0.03692 \\
C & 0.85796 & 1.57173 & -0.03412 \\
C & -2.96212 & 1.61590 & -0.04103 \\
C & -1.06970 & -0.31747 & -0.05341 \\
C & -1.05809 & 1.58204 & 1.87381 \\
C & -1.04646 & 1.60430 & -1.94662 \\
O & -1.06308 & 1.57722 & 3.00875 \\
O & -1.04195 & 1.61002 & -3.08156 \\
O & 1.99283 & 1.55718 & -0.03416 \\
O & -4.09695 & 1.63306 & -0.04290 \\
O & -1.07824 & -1.45232 & -0.06656 \\
C & -1.03442 & 3.50269 & -0.02454 \\
O & -1.02574 & 4.63759 & -0.01643 \\
    \end{tabular}
\end{table}

\textbf{\ce{Mn(CO)5-CO}} \\
Charge=+1, Multiplicity=1
\vskip-15pt
\begin{table}[H]
\hskip-7.0cm\begin{tabular}{llll}
Mn & -1.05217 & 1.59359 & -0.03842 \\
C & 0.85778 & 1.57339 & -0.03352 \\
C & -2.96213 & 1.61388 & -0.04247 \\
C & -1.07189 & -0.31631 & -0.05159 \\
C & -1.05710 & 1.58062 & 1.87154 \\
C & -1.04733 & 1.60611 & -1.94849 \\
O & -1.05998 & 1.57267 & 2.99262 \\
O & -1.04449 & 1.61318 & -3.06958 \\
O & 1.97881 & 1.56117 & -0.02988 \\
O & -4.08317 & 1.62573 & -0.04420 \\
O & -1.08351 & -1.43732 & -0.05906 \\
C & -1.03233 & 3.50355 & -0.02536 \\
O & -1.02056 & 4.62455 & -0.01727 \\
    \end{tabular}
\end{table}

% \bibliography{references}